\documentclass{vldb}
\usepackage{balance}  
\usepackage{comment}
\usepackage{graphicx}
\usepackage{subcaption}
\usepackage{multirow}
\usepackage{tabularx}
\usepackage{xcolor}
\usepackage{colortbl}
\usepackage[hidelinks]{hyperref}
\usepackage{array}
\usepackage{dblfloatfix}

\newcolumntype{L}[1]{>{\raggedright\let\newline\\\arraybackslash\hspace{0pt}}m{#1}}
\newcolumntype{C}[1]{>{\centering\let\newline\\\arraybackslash\hspace{0pt}}m{#1}}
\newcolumntype{R}[1]{>{\raggedleft\let\newline\\\arraybackslash\hspace{0pt}}m{#1}}

\makeatletter
\def\@opargbegintheorem#1#2#3{\trivlist
\item[\hskip\dimexpr\labelsep+10pt\relax{\scshape #1\ #2}](\textsc{#3}).\ \itshape}
\makeatother
\newtheorem{obs}{Observation}
\newtheorem{subobs}{Observation}[obs]

\newcommand{\subfigsizea}{0.45\textwidth}
\newcommand{\subfigsizeb}{0.315\textwidth}
\newcommand{\subfigsizec}{0.475\textwidth}
\newcommand{\subfigspace}{~~~}
\newcommand{\figvspace}{\vspace{0pt}}

\definecolor{grad_0}{RGB}{235, 120, 120}
\definecolor{grad_1}{RGB}{240, 185, 147}
\definecolor{grad_2}{RGB}{245, 245, 150}
\definecolor{grad_3}{RGB}{145, 235, 185}
\definecolor{grad_4}{RGB}{120, 235, 120}

\begin{document}

\title{Blockchain Goes Green? An Analysis of \\ Blockchain on Low-Power Nodes}

\numberofauthors{1} 

\author{
%
%
\alignauthor
Dumitrel Loghin\footnotemark[1]~, Gang
Chen\footnotemark[3]~, Tien Tuan Anh Dinh\footnotemark[1]~, Beng Chin Ooi\footnotemark[1]~, Yong Meng Teo\footnotemark[1]~\\
\affaddr{\footnotemark[1]~National University of Singapore,
\footnotemark[3]~Zhejiang University}\\
\email{\footnotemark[1]~[dumitrel,dinhtta,ooibc,teoym]@comp.nus.edu.sg,
\footnotemark[3]~cg@zju.edu.cn}
}

\maketitle

\begin{abstract}

Motivated by the massive energy usage of blockchain, on the one hand, and by
significant performance improvements in low-power, wimpy systems, on the other
hand, we perform an in-depth time-energy analysis of blockchain systems on
low-power nodes in comparison to high-performance nodes. We use three low-power
systems to represent a wide range of the performance-power spectrum, while
covering both x86/64 and ARM architectures. We show that low-end wimpy nodes are
struggling to run full-fledged blockchains mainly due to their small and
low-bandwidth memory. On the other hand, wimpy systems with balanced
performance-to-power ratio achieve reasonable performance while saving
significant amounts of energy. For example, Jetson TX2 nodes achieve around 80\%
and 30\% of the throughput of Parity and Hyperledger, respectively, while using
$18\times$ and $23\times$ less energy compared to traditional brawny servers
with Intel Xeon CPU.

\end{abstract}	

\section{Introduction}

Recent years have seen the increasing adoption of blockchain systems by
enthusiasts, industry and the academia. Starting with
Bitcoin~\cite{Bitcon_2008}, and followed by a myriad of other networks, among
which Ethereum~\cite{Ethereum_2013} is one of the most popular, these public
blockchain systems are mainly used for cryptocurrencies. More recently,
permissioned blockchains, such as Hyperledger~\cite{Hyperledger_2018}, are used
to facilitate asset management among mutually distrusting entities.

With an estimated annual energy usage of 52~TWh in 2018, Bitcoin network is
using more energy than developed countries such as Portugal or
Singapore~\cite{Bitcoin_Energy_18}. Even Ethereum would enter top 100 countries
with an estimated energy usage of almost 10~TWh~\cite{Ethereum_Energy_18}. When
accounting for the other hundreds or thousands of blockchain systems, we obtain
a worrying figure for energy usage. While other IT domains, such as cloud and
high-performance computing~\cite{Rajovic_13}, have been optimizing to reduce
power consumption and to increase energy efficiency, we believe it is time for
blockchain systems to follow this trend.

Energy represents the electricity used in a period of time. In computer systems,
high energy is the result of (i) long execution time and/or (ii) high power
usage of active subsystems, such as the main processor (CPU), graphics processor
(GPU) or other accelerators. In public blockchain platforms, Proof-of-Work (PoW)
consensus protocols are considered the Achilles' heel in terms of time-energy
performance~\cite{Bitcoin_Energy_18,Ethereum_Energy_18}. This is because PoW
consists of a compute-intensive mining phase running cryptographic algorithms.
In addition to the high cost of the consensus phase, we show in this paper that
newer versions of Ethereum exhibit long transaction execution time due to a
design choice that leads to many transactions being restarted multiple times.

Other proposed consensus protocols, including proof-of-authority (PoA),
proof-of-stake (PoS) and proof-of-elapsed-time (PoET), among others, are
promising in terms of lowering the time-energy costs because they do not include
resource-intensive algorithms. In this paper, we analyze three blockchain
systems with different consensus protocols, namely Ethereum, Parity and
Hyperledger which are using PoW, PoA and Practical Byzantine Fault Tolerance
(PBFT) \cite{PBFT_99} consensus protocols, respectively.

The high energy consumption of well-known blockchain systems is a result of
using traditional high-performance hardware in the mining process. The ASICs
used in mining Bitcoin or the high-end GPUs used by Ethereum have power
characteristics in the order of hundreds or thousands of Watts (W). Even
traditional high-performance CPU-only systems used by many blockchains are using
a significant quantity of energy. On the other hand, emerging low-power devices
based on ARM architecture are showing significant performance improvements that
promote them as alternative to traditional x86/64
servers~\cite{Microsoft_ARM_17}. While the majority of low-power, wimpy
nodes~\cite{BrawnyWimpy_13} are deployed at the edge, in Internet of Things
(IoT) setups, higher-end devices are targeting the server market. In this paper,
we seek answer to the question of whether these low-power devices are able to
run blockchain and at what performance-to-power cost.

The performance of modern blockchain systems on traditional servers was analysed
using BLOCKBENCH in~\cite{Dinh_SIGMOD_2017}, among other works. In this paper,
we extend BLOCKBENCH in two significant directions. First, we extend the
original time-performance analysis performed on Xeon CPUs to low-power, wimpy
nodes. Second, we provide in-depth analysis of energy cost on both brawny and
wimpy systems. For brawny nodes, we use the same system specified
in~\cite{Dinh_SIGMOD_2017}. For wimpy nodes, we use three systems covering a
wide performance-to-power spectrum, as well as both x86/64 and ARM
architectures.

We make the following contributions in this paper:
\begin{itemize}
  \item We provide the first extensive time-energy performance study of
  state-of-the-art blockchain systems on both high-performance and low-power
  nodes acting as miners or validators. We examine Hyperledger, Ethereum and
  Parity running on Intel Xeon, NUC~\cite{NUC_Specs}, NVIDIA Jetson
  TX2~\cite{Jetson_TX2_17}, and Raspberry Pi 3~\cite{Pi3_Specs}. We share not
  only the analysis, but our experience of running blockchain on various system
  architectures.
  
  \item We show that low-power ARM-based systems struggle to run full-fledged
  blockchain workloads mainly due to insufficient memory size and bandwidth. For
  example, the low-end Raspberry Pi 3 wimpy node is unable to run Ethereum, and
  it requires non-trivial code modifications and special configuration to run
  Hyperledger.
  
  \item We show that systems with the lowest power profile do not necessarily
  achieve the best energy efficiency. For example, Jetson TX2 is more
  energy-efficient than Raspberry Pi 3, even if the latter has a lower power
  profile.
  
  \item We show that wimpy nodes can achieve reasonable performance while saving
  significant amounts of energy. For example, eight Jetson TX2 nodes trade 17\%
  and 72\% of Parity and Hyperledger throughput, respectively, for $18\times$
  and $23\times$ lower energy consumption compared to eight Xeon nodes.
  
  \item Our analysis of Ethereum performance leads to an insight into the design
  trade-off in newer Ethereum releases compared to older ones that are used
  in~\cite{Dinh_SIGMOD_2017}. In particular, the new design has lower throughput
  due to the cost of many transaction execution restarts.
\end{itemize}

The remainder of this paper is organized as follows. In
Section~\ref{sec:rel_work} we present background and related work on blockchain
systems. In Section~\ref{sec:setup} we describe the hardware systems and
blockchain workloads used in this study. We also provide a detailed
characterization of the hardware systems in this section.  In the next two
sections, we analyze the time and energy performance at single-node and cluster
level. We conclude in Section~\ref{sec:concl}.

\section{Background and Related Work}
\label{sec:rel_work}

In this section, we provide a background on blockchain systems and survey the
related work on time and energy performance analysis of blockchains. 

\subsection{Blockchain Systems}

A blockchain is a distributed ledger running on a network of mutually
distrusting nodes (or peers). The ledger is stored as a linked list (or chain)
of blocks of transactions. The links in the chain are built using cryptographic
pointers to ensure that no one can tamper with the chain or with the data inside
a block.

Blockchains are most famous for being the underlying technology of
cryptocurrencies, but many blockchains are able to support general-purpose
applications. This ability is determined by the execution engine and data model.
For example, Bitcoin~\cite{Bitcon_2008} supports only operations related to
cryptocurrency (or token) manipulation. On the other hand,
Ethereum~\cite{Ethereum_2013} can run arbitrary computations on its
Turing-complete Ethereum Virtual Machine (EVM). At data model level, there are
at least three alternatives used in practice. The \textit{Unspent Transaction
Output} (\textit{UTXO}) model, used by Bitcoin, represents the ledger states as
transaction ids and associated unspent amounts which are the input of future
transactions. The \textit{account/balance} model resembles a classic banking
ledger. A more generic model used by Hyperledger consists of \textit{key-value
states}. On top of the data model, ones can write general applications that
operate on the blockchains states. Such applications are called~\textit{smart
contracts}. In this paper, we use BLOCKBENCH benchmarks which provides a set of
smart contracts for Hyperledger, Ethereum and Parity.

Depending on how nodes can join the network, the blockchain is \textit{public}
or \textit{private} (or \textit{permissioned}). In public networks, anybody can
join or leave and, thus, the security risks are high. Most of the cryptocurrency
blockchains are public, such as Bitcoin~\cite{Bitcon_2008} and
Ethereum~\cite{Ethereum_2013}. On the other hand, private blockchains allow only
authenticated peers to join the network. Typically, private blockchains, such as
Hyperledger~\cite{Hyperledger_2018} and Parity~\cite{Parity_2018}, are deployed
inside or across big organizations.

Blockchains operate in a network of mutually distrusting peers, where some peers
may not be just faulty but malicious. Hence, they assume a Byzantine
environment, in contrast to the crash-failure model used by the majority of
distributed systems. To ensure consistency among honest peers, most private
blockchains use Byzantine fault-tolerant consensus protocols such as
PBFT~\cite{PBFT_99}, whereas most public blockchains use proof-of-work (PoW)
consensus protocols.

In PoW, participating nodes, called miners, need to solve a difficult
cryptographic puzzle. The node that solves the puzzle first has the right to
append transactions to the ledger. On the other hand, PBFT consists of
exchanging $O(n^2)$ messages among the nodes to reach agreement on the
transactions to be appended to the blockchain. These consensus protocols are
considered the Achilles' heel of blockchain due to poor time-energy performance.
While PoW is scalable since it can run in parallel on all nodes, it is
compute-intensive and, thus, it is both slow and power-hungry on traditional
brawny servers. PBFT exhibits quadratic time growth with the number of nodes in
the network, leading to energy wastage.

Our analysis in the next sections confirms that a PoW-based blockchain, such as
Ethereum, uses more power compared to a PBFT- or PoA-based blockchain.
A PBFT-based blockchain, such as Hyperledger, uses almost the same power as a
PoA-based blockchains, such as Parity, on small networks of up to eight nodes.

\begin{table*}[t]
\centering
\caption{Systems characterization}
\label{table:sys_char}
\resizebox{0.96\textwidth}{!} {
\begin{tabularx}{\textwidth}{|X|l|r|r|r|r|}
\hline
& & \bf{Xeon} & \bf{NUC} & \bf{TX2} & \bf{RP3} \\
\hline
\hline
\multirow{9}{*}{Specs} & ISA & x86-64 & x86-64 & AARCH64 & ARMv7l \\
& Cores & 6(12) & 2(4) & 6 & 4 \\
& Frequency & 3.50~GHz & 2.40~GHz & 2.04~GHz & 1.20~GHz \\
& L1 Data Cache & 32~kB & 32~kB & 32-128~kB & 32~kB\\
& L2 Cache & 256~kB (core) & 256~kB (core) & 2~MB & 512~kB\\
& L3 Cache & 12~MB & 3~MB & N/A & N/A \\ 
& Memory & 32~GB DDR3 & 32~GB DDR4 & 8~GB LPDDR4 & 1~GB LPDDR2 \\
& Storage & 2~TB HDD & 256~GB SSD & 64~GB SD card & 64~GB SD card \\
& Networking & Gbit & Gbit & Gbit & 100 Mbit \\
\hline
\multirow{9}{*}{CPU} & CoreMark (one core) [IPS] & 25201.6 & 18022.3 & 12019.7
& 3591.8 \\
& System power [W] & 70.6 & 13.8 & 8.8 & 3.0 \\
\cline{2-6}
& CoreMark (all cores) [IPS] & 186924.9 & 50582.4 & 67345.9 & 11031.3 \\
& System power [W] & 127.7 & 18.6 & 11.7 & 4.9 \\
\cline{2-6}
& Keccak256 [MBPS] & 314.2 & 217.4 & 119.0 & 1.0 \\
& System power [W] & 77.1 & 14.7 & 4.9 & 2.4 \\
\cline{2-6}
& Keccak512 [MBPS] & 169.9 & 116.2 & 65.2 & 0.9 \\
& System power [W] & 74.1 & 14.7 & 4.8 & 2.3 \\
\cline{2-6}
& Idle system power [W] & 50.8 & 9 & 2.4 & 1.9 \\
\hline
\multirow{5}{*}{Storage} & Write throughput [MB/s] & 160.0 & 409.0 & 16.3 & 12.5
\\
& Read throughput [MB/s] & 172.0 & 551.0 & 88.9 & 22.6 \\
& Buffered read throughput [GB/s] & 8.1 & 6.6 & 2.7 & 0.8 \\
& Write latency [ms] & 9.3 & 1.0 & 17.1 & 14.0 \\
& Read latency [ms] & 2.5 & 0.2 & 2.8 & 25.5 \\
\hline
\multirow{3}{*}{Network} & TCP bandwidth [Mbits/s] & 941 & 839 & 943 & 94 \\
& UDP bandwidth [Mbits/s] & 810 & 813 & 755 & 96 \\
& Ping latency [ms] & 0.14 & 0.14 & 0.3 & 0.35 \\
\hline
\end{tabularx}
}
\vspace{10pt}
\end{table*}

\subsection{The Time-Energy Analysis of Blockchains}

There are a number of related works that analyze performance of
blockchains~\cite{Dinh_SIGMOD_2017, Pongnumkul_ICCCN_17}. However, only a few
include energy analysis~\cite{Sankaran_ICDCS_2018, Suankaewmanee_ICNC_18}, and
the analysis is of limited depth.
 
BLOCKBENCH~\cite{Dinh_SIGMOD_2017} is a benchmarking suite comprising both
simple (micro) benchmarks and complex (macro) benchmarks. The micro benchmarks,
namely \textit{CPUHeavy}, \textit{IOHeavy} and \textit{Analytics}, stress
different subsystems such as the CPU, memory and IO. On the other hand,
\textit{YCSB} macro benchmark implements a key-value storage, while
\textit{Smallbank} represents OLTP and simulates banking operations. These
benchmarks are implemented as smart contracts in Ethereum, Parity and
Hyperledger. Their performance in terms of throughput and latency is evaluated
on traditional high-performance servers with Intel Xeon CPU. In this paper, we
extend BLOCKBENCH to include time-energy analysis of a wider range of systems,
with focus on low-power nodes.

Sankaran et al.~\cite{Sankaran_ICDCS_2018} analyze the time and energy
performance of an in-house Ethereum network consisting of high-performance
mining servers and low-power Raspberry Pi clients. These low-power systems
cannot run Ethereum mining due to their limited memory size, hence, they only
take the role of clients. In this paper, we run the Ethereum full nodes on
low-power devices with higher performance, such as Intel NUC and Jetson TX2. To
the best of our knowledge, we are the first to run and analyze the time-energy
performance of full-fledged blockchains on low-power systems.
 
MobiChain~\cite{Suankaewmanee_ICNC_18} is an approach that allows mining on
mobile devices running Android OS, in the context of mobile commerce. While
containing analysis of both time and energy performance, MobiChain has no
comparison to other blockchains. In terms of energy analysis, the authors show
that it is more energy-efficient to group multiple transactions in a single
block since there is less mining work and therefore less time and power wasted
in this process. However, larger blocks increase latency and result in poor user
experience.

Jupiter~\cite{Jupiter_ICDE_18} is a blockchain designed for mobile devices. It
aims to address the problem storing large ledger on mobile devices with limited
storage capacity. However, there is no time or energy performance evaluation.

To the best of our knowledge, we provide the first extensive time-energy
performance analysis of blockchain systems on low-power, wimpy nodes in
comparison with high-performance server systems.

\begin{figure*}[!t]
\centering
\begin{subfigure}{\subfigsizea}
\centering
\includegraphics[width=0.5\textwidth,angle=270]{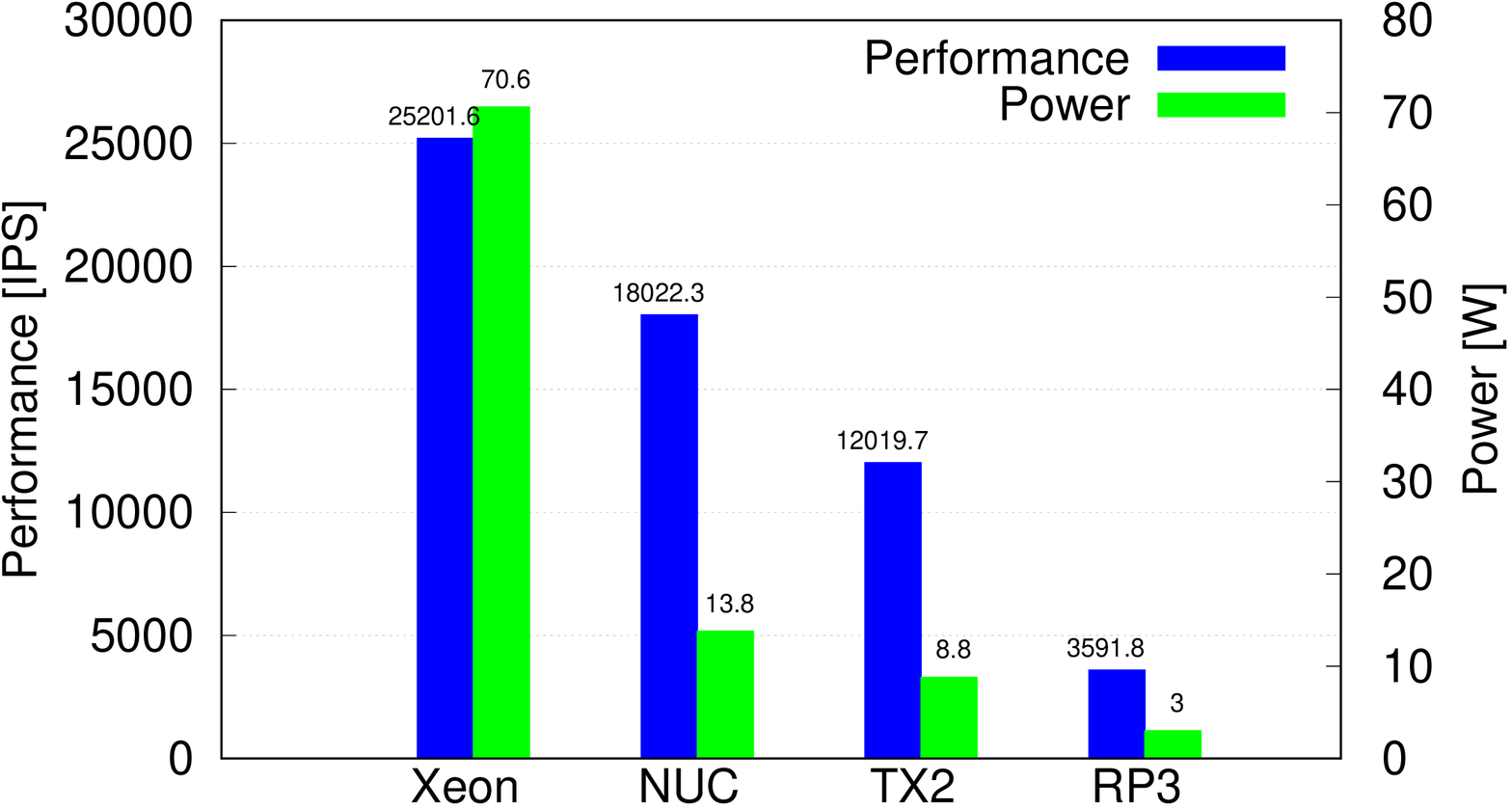}
\caption{CoreMark on one core}
\label{fig:coremark_one_core}
\end{subfigure}
\subfigspace
\begin{subfigure}{\subfigsizea}
\centering
\includegraphics[width=0.53\textwidth,angle=270]{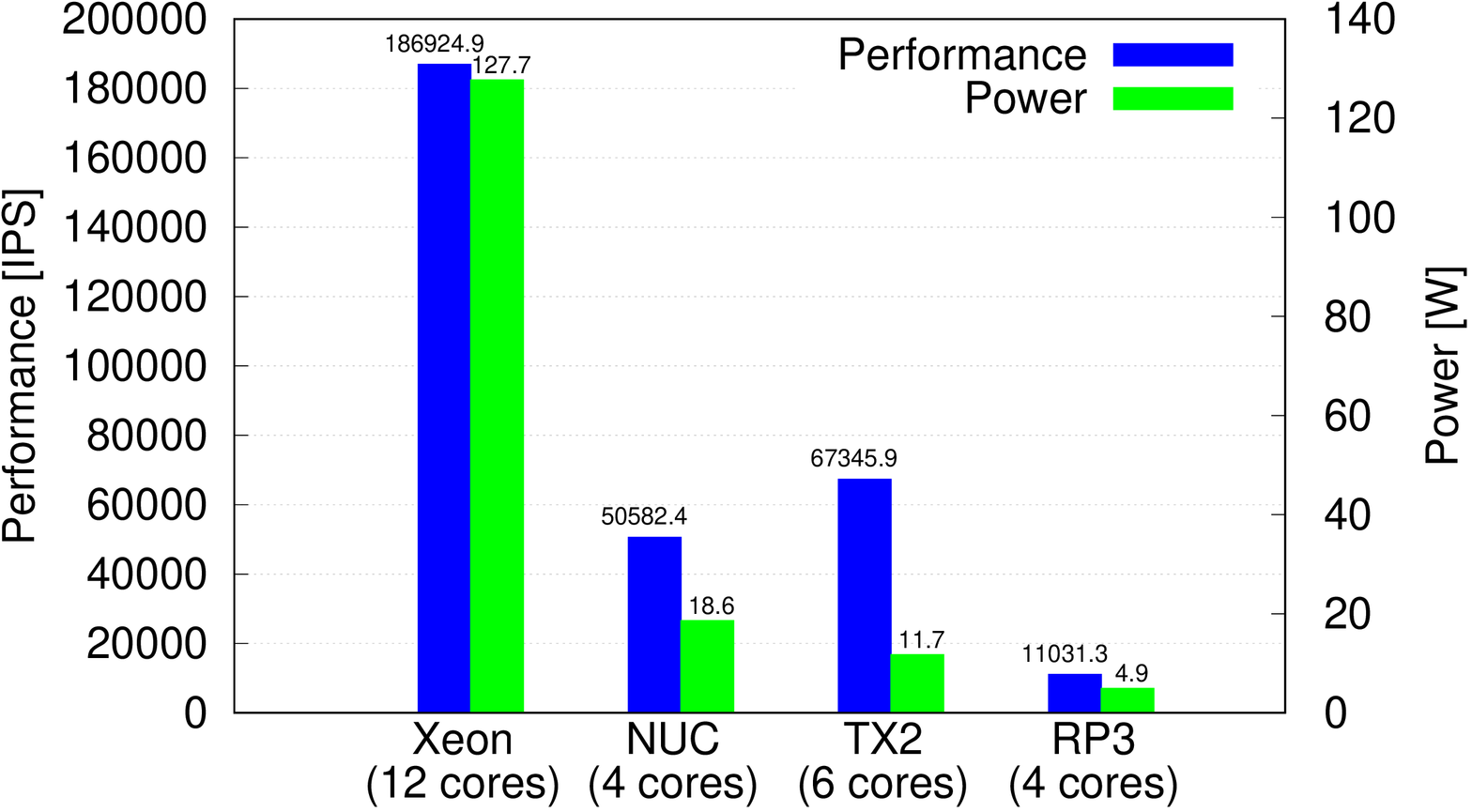}
\caption{CoreMark on all cores}
\label{fig:coremark_all_cores}
\end{subfigure}
\begin{subfigure}{\subfigsizea}
\centering
\includegraphics[width=0.5\textwidth,angle=270]{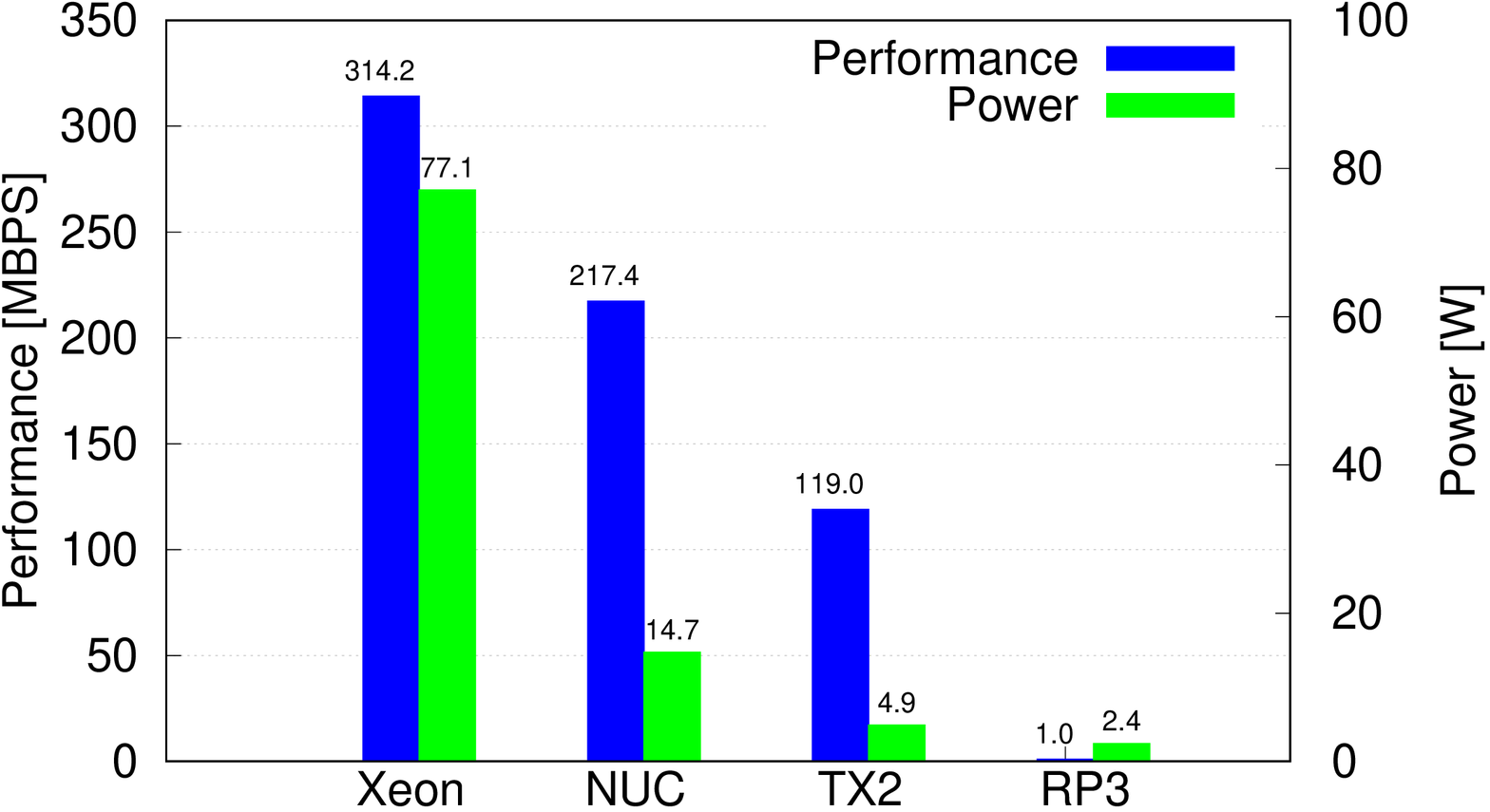}
\caption{Keccak256 on one core}
\label{fig:keccak256}
\end{subfigure}
\subfigspace
\begin{subfigure}{\subfigsizea}
\centering
\includegraphics[width=0.53\textwidth,angle=270]{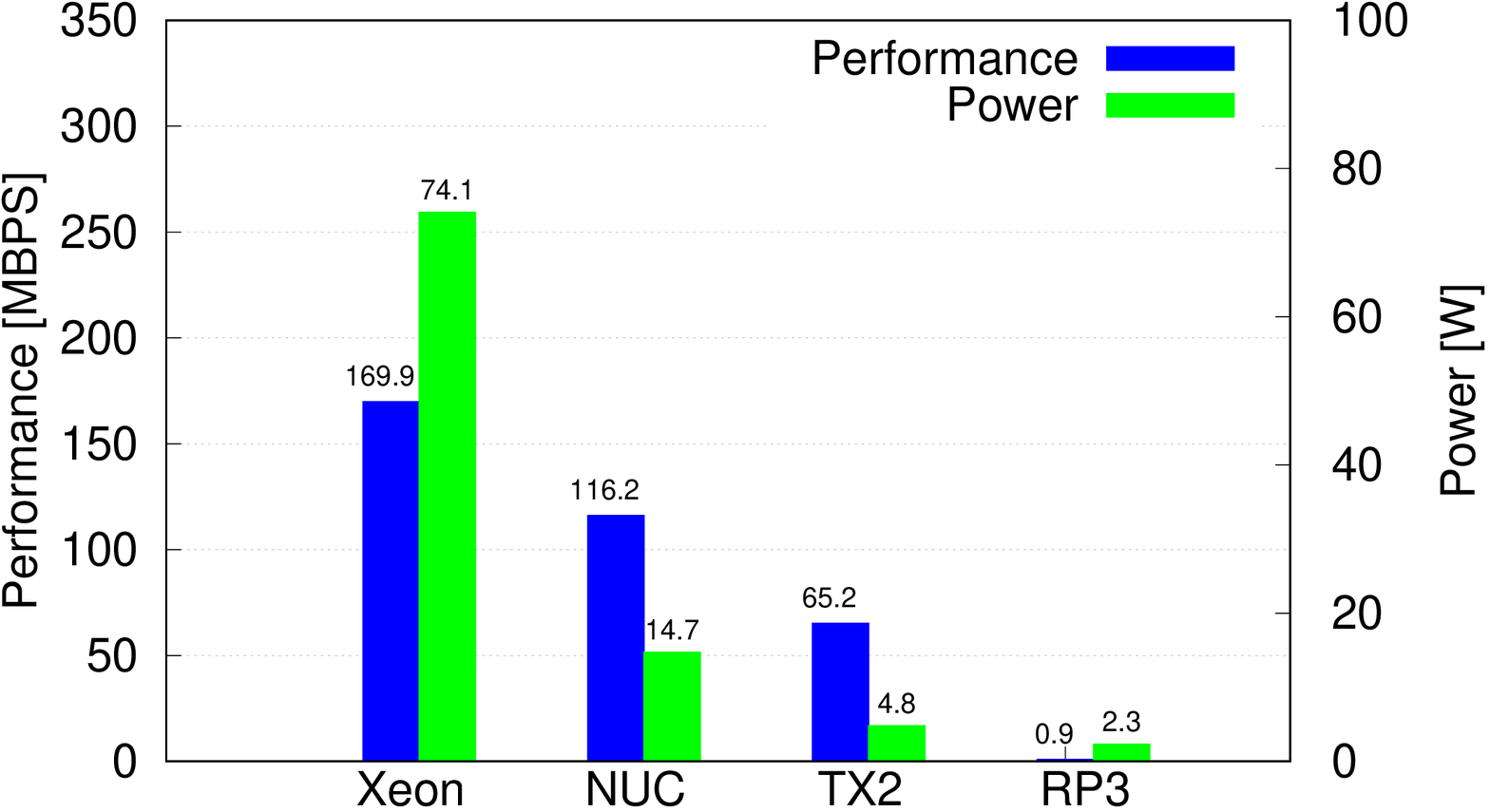}
\caption{Keccak512 on one core}
\label{fig:keccak512}
\end{subfigure}
\caption{Performance and power at CPU level}
\label{fig:sys_cpu}
\end{figure*}

\section{Experimental Setup}
\label{sec:setup}

In this section, we describe our experimental setup, starting with the systems
and ending with the workloads. We present a detailed characterization of the
selected systems at CPU, memory, storage and networking levels. The results of
this detailed characterization are summarized in Table~\ref{table:sys_char}.

\subsection{Systems}
\label{sec:sys_specs}

We compare the time and energy performance of low-power systems against a
high-performance traditional server system. This server system is based on a
x86/64 Intel \textbf{Xeon} E5-1650 v3 CPU clocked at 3.5GHz, and has 32GB DDR3
memory, 2TB hard-disk (HDD) and 1Gbps networking interface card (NIC). It runs
Ubuntu 14.04 with Linux kernel 3.13.0-95.

The low-power systems used for the analysis are (i) Intel
\textbf{NUC}~\cite{NUC_Specs}, (ii) NVIDIA Jetson
\textbf{TX2}~\cite{Jetson_TX2_17} and (iii) Raspberry Pi 3 model B
(\textbf{RP3})~\cite{Pi3_Specs}. The NUC system is based on a x86/64 Intel Core
i3 CPU with two physical cores that support Hyperthreading and run at 2.4GHz. 
This system has 32GB DDR4, 256GB solid-state drive (SSD) and 1Gbps NIC.  It runs
Ubuntu 16.04 with Linux kernel 4.15.0-34.

The TX2 system is based on a heterogeneous 6-core 64-bit CPU with two NVIDIA  
Denver cores and four ARM Cortex-A57 cores clocked at more than 2GHz. The system
has 8GB LPDDR4, a 32GB SD card and 1Gbps NIC. TX2 is running Ubuntu 16.04
with Linux kernel 4.4.38-tegra of \textit{aarch64} (64-bit ARM) architecture.

The RP3 has a 4-core ARM Cortex-A53 CPU of 64-bit ARM architecture and 1GB of
LPDDR2 memory. This system has a 64GB SD card that acts as storage and 100Mbps
NIC. It runs  Debian 9 (stretch) with Linux kernel 4.9.80-v7+ (32-bit ARM).

We measure power and energy consumption of these systems with a Yokogawa power meter
connected to the AC lines.  We report only AC power and energy values in
this paper. We believe that these values are more useful compared to DC
measurements since they reflect the final billable energy.

\subsection{Systems Characterization}
\label{sec:sys_char}
Before analyzing the time and energy of blockchains on the selected systems, we
evaluate the hardware at CPU, memory, storage and networking level to understand
their relative performance. The measured values and system characteristics are
summarized in Table~\ref{table:sys_char}.

We first measure idle system power when the hardware is running only the OS. We
obtain 50W, 9W, 2.4W and 1.9W for Xeon, NUC, TX2 and RP3, respectively. These
values clearly show the power efficiency gap between brawny nodes used in the majority
of datacenters and supercomputers, and wimpy nodes used at the edge.

To assess CPU performance, we use CoreMark benchmark which is increasingly used
by the industry, including vendors that equip their systems with ARM
CPUs~\cite{ARM_CoreMark_09}. CoreMark measures CPU performance in terms of
iterations per second (IPS). We present the performance and average power usage
in Figure~\ref{fig:coremark_one_core} and Figure~\ref{fig:coremark_all_cores}
for CoreMark running on a single core and all cores, respectively. For
multi-core analysis, we enable all available cores, including virtual cores in
systems that support Hyperthreading. For example, we use twelve and four virtual
cores on Xeon and NUC, respectively.

At single-core level, the performance of Xeon is 1.4, 2.1 and 7 times higher
compared to NUC, TX2 and RP3, respectively. But this performance comes at the
cost of $5.1\times$, $8\times$ and $23.5\times$ higher power consumption.
However, we note that this is the power used by the entire system which includes
other components beside the CPU. We then estimate the power of CPU by
subtracting the values for idle system power. One Xeon core uses almost 20W,
while one ARM core from RP3 uses only 1.1W. Hence, the performance-to-power
ratio (PPR) of the RP3 is superior to that of the Xeon.

At multi-core level, TX2 exhibits better performance than NUC, mainly because of
its six real cores compared to only two real cores on NUC. Moreover, TX2 uses
less power than NUC to deliver higher performance. Therefore, it is expected
that TX2 has a better time-energy performance for multi-threaded workloads. We
also observe that the performance is not scaling perfectly with the number of
cores.  For example, Xeon exhibits only 7.4 times performance boost when 12
cores are used.  TX2 is performing better, with a 5.6 performance increase when
6 cores are used. This sub-linearity is due to resource contention, both in-core
and off-core~\cite{Tudor_2011}.

Blockchain systems rely heavily on cryptographic operations that are
CPU-intensive. We evaluate the CPU on running this type of workload by measuring
the performance and average power of Keccak secure hash algorithm from
\textit{go-ethereum v1.8.15}, compiled with \textit{go 1.11}. We run both
Keccak256 and Keccak512 on a random input of one billion bytes. The throughput
measured in MB per second (MBPS) represents the performance of these
cryptographic algorithms on the selected systems. As shown in
Figure~\ref{fig:keccak256} and Figure~\ref{fig:keccak512}, the performance
trends are similar to CoreMark. RP3 exhibits much lower performance: almost
$320\times$ and $190\times$ lower throughput compared to Xeon on Keccak256 and
Keccak512, respectively. The lower system power of RP3 running these
cryptographic operations compared to CoreMark suggests  that the core is not
fully utilized. In fact, it is often stuck in memory operations that use less
power compared to arithmetic operations. As we shall see in the next paragraph,
RP3's memory has significantly lower bandwidth than the other three systems.

\begin{figure}[tbp]
\centering
\includegraphics[width=0.38\textwidth]{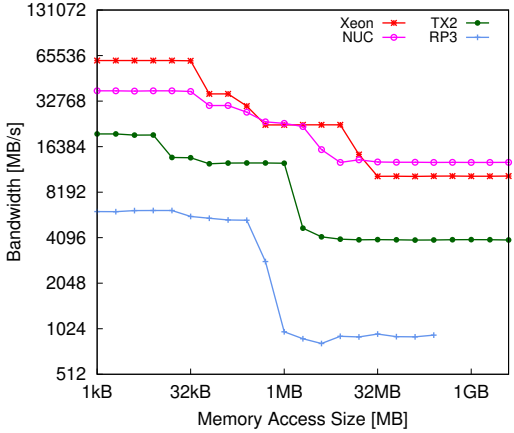}
\caption{Memory bandwidth comparison}
\label{fig:sys_mem}
\figvspace
\end{figure}

We analyze the performance of the memory subsystem in terms of bandwidth.
We use \textit{lmbench}~\cite{lmbench} to get the read-write bandwidth and plot
the results in Figure~\ref{fig:sys_mem}. At level one cache (L1), Xeon has the
highest bandwidth, which is almost 60GB/s, while NUC, TX2 and RP3 exhibit
bandwidths of 37GB/s, 19GB/s and 6GB/s, respectively. This is expected since
server-class processors, such as Xeon, have optimized caches. However, at the
main memory level, NUC leads with a bandwidth of 12.5GB/s, followed closely by
Xeon with 10GB/s. This lower performance of Xeon is attributed to the older DDR3
memory generation. TX2 and RP3 exhibit main memory bandwidths of less than 4GB/s
and 1GB/s, respectively. This low bandwidth, together with small memory size
hinder the execution of modern workloads on wimpy systems.

At storage level, there is a mixed performance profile since the systems are
equipped with different types of storage mediums. To assess throughput and
latency, we use \textit{dd} and \textit{ioping} Linux commands, respectively. As
expected, the SSD of NUC exhibits the highest throughput and the lowest latency.
On the other hand, the SD cards used by TX2 and RP3 exhibit low throughput, high
latency and significant read/write asymmetry. Since modern operating systems are
caching files or chunks in memory, we also measure the buffered read throughput.
We observe that this throughput follows the memory bandwidth trend, except for
NUC where the buffered throughput of 6.6GB/s is half of the memory bandwidth.

At networking level, we measure the bandwidth and latency using \textit{iperf}
and \textit{ping} Linux commands, respectively. As expected, RP3 exhibits lower
TCP and UDP bandwidths since it is equipped with 100Mbps NIC, compared to the
Gigabit Ethernet NICs of the other systems. The slightly higher latency of TX2
and RP3 can be attributed to the lower clock frequency of the wimpy systems. To
validate this hypothesis, we have measured the networking latency while setting
the clock frequency to a fixed step. TX2 supports twelve frequency steps in the
range 346MHz-2.04GHz. We obtained a Pearson correlation coefficient of -0.93
between the twelve frequency steps and corresponding latencies, suggesting
strong inverse proportionality. For example, the networking latency at 346MHz is
0.33ms, while at 2.04GHz is 0.25ms. On RP3, there are only two available
frequency steps, but we obtained similar results. While setting the frequency to
600MHz and 1.2GHz, we obtained networking latencies of 0.35ms and 0.29ms,
respectively.

\begin{obs}
\normalfont{In summary, the hardware systems have the following characteristics.}
\begin{subobs}
\label{obs:sys_char_nuc}
\textit{Low-power x86/64 devices, such as Intel NUC, can match the performance
of server-class systems at memory and storage level while using $5\times$ less 
power. However, CPU performance is lower when running multi-threaded
workloads due to the small number of cores.}
\end{subobs}
\begin{subobs}
\label{obs:sys_char_tx2}
\textit{
High-end ARM-based wimpy devices, such as Jetson TX2, have potential to
achieve high PPR at the cost of lower time performance compared to x86/64
systems.}
\end{subobs}
\begin{subobs}
\label{obs:sys_char_rp3}
\textit{
Low-end ARM-based devices, such as Raspberry Pi 3, suffer from low core clock
frequency, small and low-bandwidth memory. These systems may not
be able to run modern server-class workloads, including blockchains.}
\end{subobs}
\end{obs}

\subsection{Workloads}

We use BLOCKBENCH~\cite{Dinh_SIGMOD_2017} with minor changes\footnote{\small{The
updated source code of BLOCKBENCH is available at
\url{https://github.com/dloghin/blockbench}}} to assess blockchain performance.
We were not able to compile \textit{go-ethereum v1.4.18} evaluted in the
original BLOCKBENCH paper~\cite{Dinh_SIGMOD_2017} on TX2 due to issues with
older versions of \textit{go} toolchain on \textit{aarch64} architecture. We
also encountered issues with the compilation of \textit{parity-ethereum v1.6.0}
on all systems due to broken Rust packages.
Hence, we use \textit{go-ethereum v1.8.15} compiled with \textit{go 1.11} and
\textit{parity-ethereum v2.1.6} compiled with \textit{cargo 1.30.0} on all
systems. For Hyperledger experiments we use version \textit{v0.6} which supports
PBFT consensus.

The micro-benchmarks in BLOCKBENCH assess the performance of different
subsystems. CPUHeavy uses quicksort to sort an array of integers, while IOHeavy
implements Write and Scan operations that touch key-value pairs to stress the
memory and IO subsystems. The analytics benchmark simulates typical OLAP
workloads as found in traditional databases. It implements three queries. The
first query (Q1) computes the total value of transactions between two blocks.
The second (Q2) and third (Q3) computes the maximum transaction and the maximum
account balance, respectively, between two blocks for a given account. This
benchmark requires an initialization step that creates 120,000 accounts and
generates over 100,000 blocks with an average of three transactions per block.

The macro-benchmarks in BLOCKBENCH are complex database applications stressing
all key subsystems. For example, YCSB evaluates the performance of a key-value
store with configurable read-write ratios, while Smallbank represents OLTP
workloads by simulating banking transactions. \textit{Donothing} benchmark
estimates the overhead of consensus protocols since it performs no computations
and no IO operations inside the smart contract. In this paper, the
macro-benchmarks are run on clusters of nodes.

All workloads are run at least three times. We report the average values and standard deviations.

\subsection{Raspberry Pi 3 (RP3) Setup}

RP3 is unable to run \textit{go-ethereum} since it has only 1GB of RAM while
Ethereum requires more than 4GB. Modifying \textit{go-ethereum} to run on
low-end wimpy devices is left to future work. In this paper, we only report the
performance and energy of Ethereum on Xeon, NUC and TX2.

Running Hyperledger on ARM-based devices is challenging and requires non-trivial
 engineering work to patch the existing code\footnote{\small{The modified Fabric
code is available on GitHub at \\
\url{https://github.com/dloghin/fabric/tree/v0.6_raspberrypi}}}.  Firstly, we
need to recompile the Linux kernel to support Docker, since Hyperledger is
running the chaincode inside Docker containers. Secondly, we need to pre-compile
several tools, such as \textit{protoc} and \textit{grpc}, used in these
containers for \textit{armv7l} (32-bit ARM) and \textit{aarch64} (64-bit ARM)
architectures.
Thirdly, we need to decrease the size of some buffers and increase timeouts for
the execution on RP3. For example, we decreased \texttt{cNameArr} buffer size
from 256MB (which is one fourth of the available memory on RP3) to 1MB, and we
increased request execution timeout from 30s to 10m.

Even with all these changes, populating the blockchain for Analytics queries on
RP3 leads to system crashes. We discovered that the swap space on the default
Debian 9 OS of RP3 was 100MB which is too small given the main memory size of
1GB. We increased this swap size to 2GB. We also found through profiling with Go
\textit{pprof} that much of the memory is used in encoding and decoding
\textit{protobuf} objects. Many of these operations are redundant and can be
avoided by keeping extra fields in blockchain data structures.
One option to save memory is to de-allocate unused memory space. To this end, we
insert \texttt{debug.FreeOSMemory()} in Hyperledger's code after
memory-intensive routines to make sure the garbage collector (GC) is
de-allocating memory faster.  We also decreased client's transaction rate during
the initialization step of Analytics to allow GC to free more memory. With all
these changes in place, we were able to run Hyperledger without problems.

\begin{figure}[!t]
\centering
\begin{subfigure}{\subfigsizec}
\centering
\includegraphics[width=0.98\textwidth]{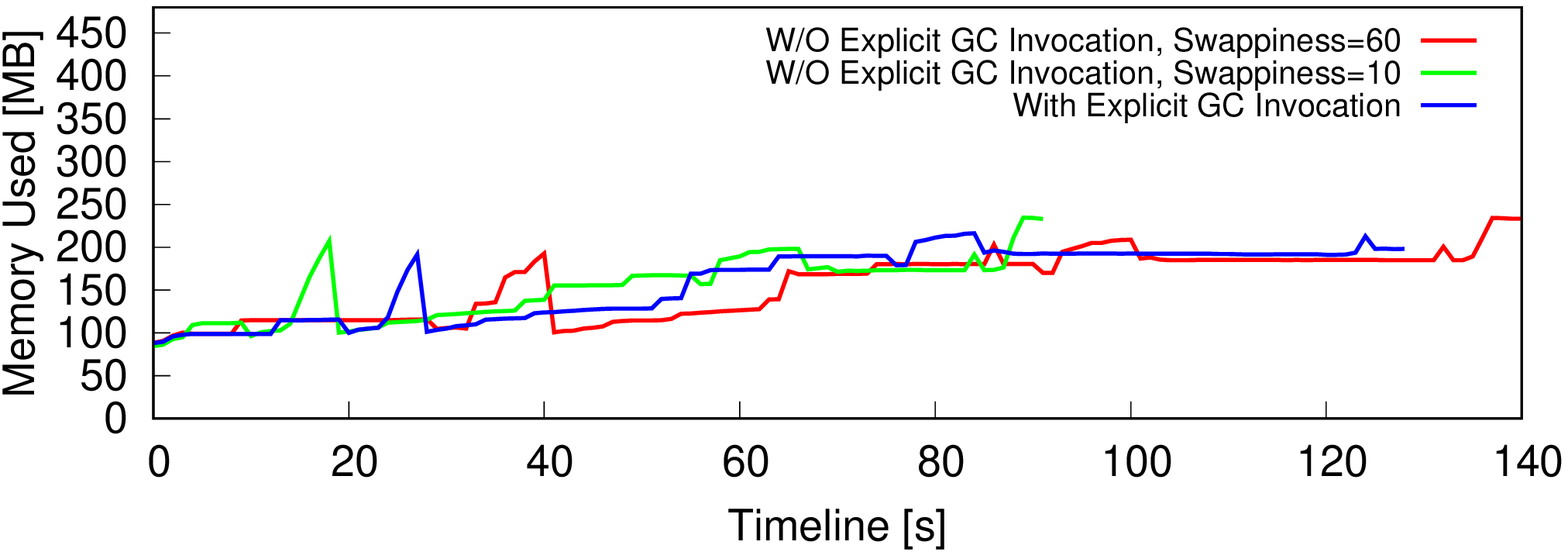}
\caption{Phase 1 (CPUHeavy Deploy)}
\end{subfigure}
\begin{subfigure}{\subfigsizec}
\centering
\includegraphics[width=0.98\textwidth]{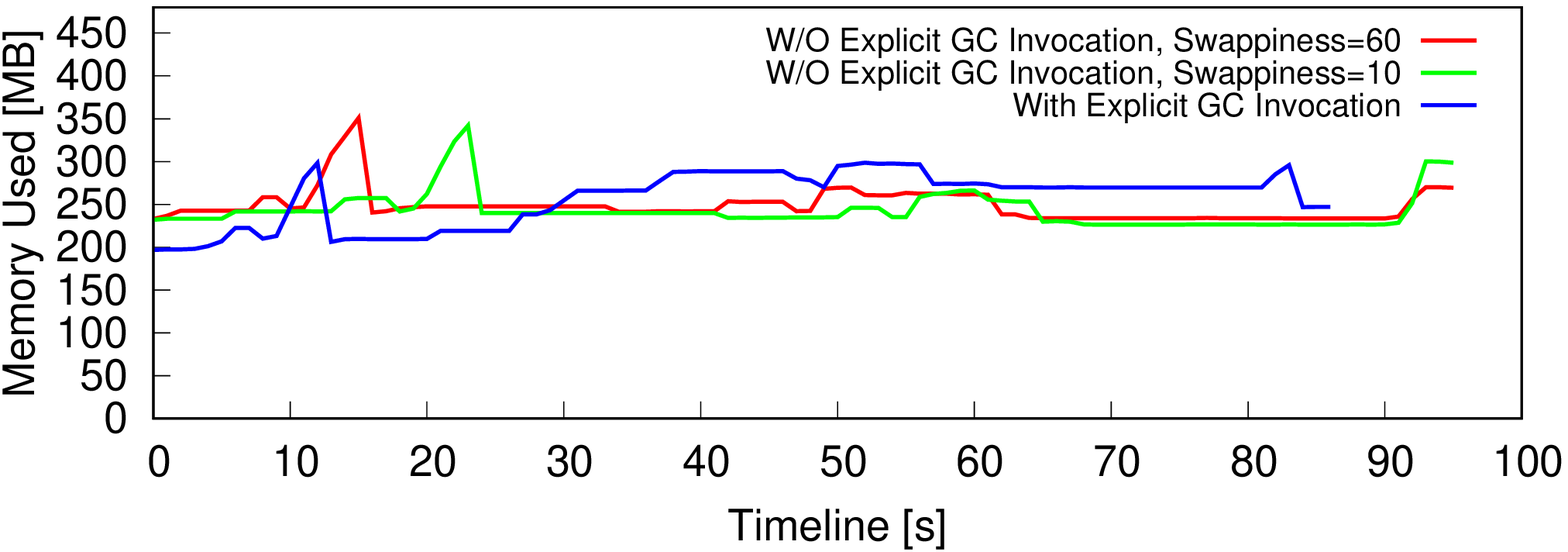}
\caption{Phase 2 (IOHeavy Deploy)}
\label{fig:gc_mem_rp3_phase2}
\end{subfigure}
\begin{subfigure}{\subfigsizec}
\centering
\includegraphics[width=0.98\textwidth]{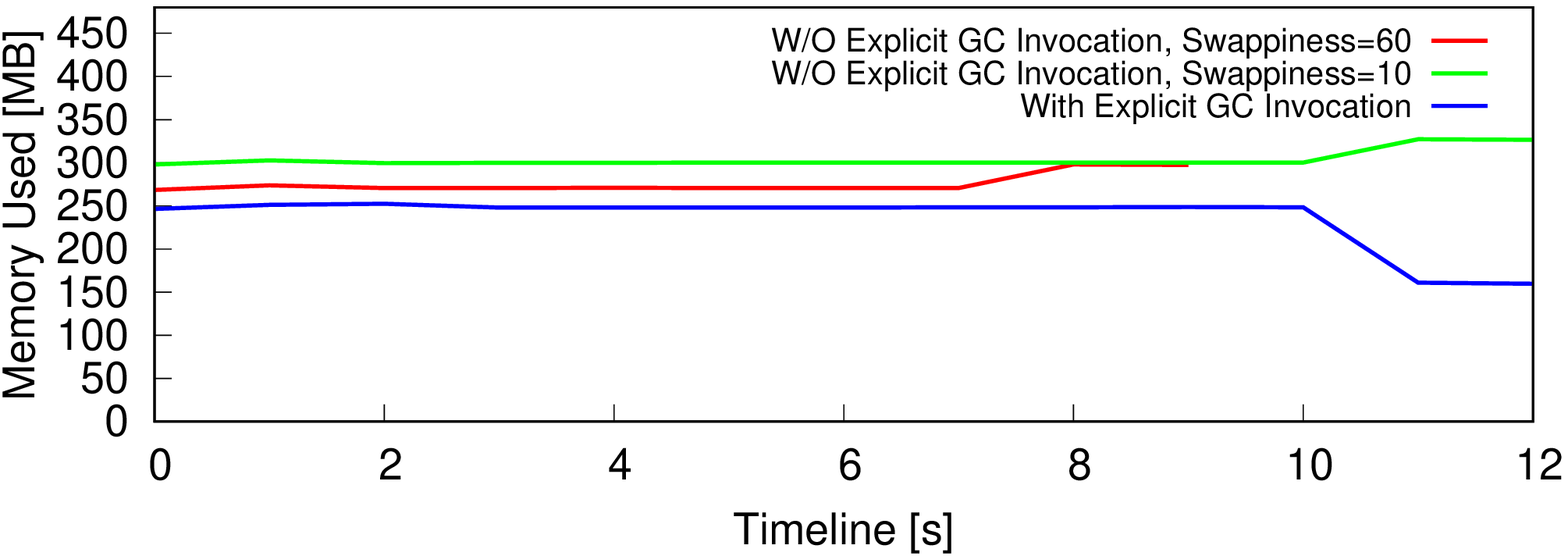}
\caption{Phase 3 (CPUHeavy)}
\label{fig:gc_mem_rp3_phase3}
\end{subfigure}
\caption{Hyperledger memory usage on RP3}
\label{fig:gc_mem_rp3}
\figvspace
\end{figure}

While explicitly invoking Go's GC is not necessary for some workloads, such as
CPUHeavy, we observed that IOheavy Write/Scan and Analytics almost always crash
without this change. We evaluate the effects of this change on a sequence of
operations consisting of CPUHeavy deploy (phase 1), IOHeavy deploy (phase 2),
and CPUHeavy (phase 3) executions. The memory usage, depicted in
Figure~\ref{fig:gc_mem_rp3}, shows that when the GC is called explicitly (i)
peak memory usage is smaller, as expected, but interestingly (ii) the execution
becomes faster than without explicit GC invocation.
Figures~\ref{fig:gc_mem_rp3_phase2} and~\ref{fig:gc_mem_rp3_phase3} illustrates
clearly that explicit GC invocation decreases the memory footprint at the end of
phase execution.

Intuitively, explicit GC should incur overheads and delay the execution.  But
freeing memory aggressively may reduce pressure on the Linux kernel swapping
mechanism, and consequently lead to faster execution. In Debian/Ubuntu OS, the
swapping is controlled by a parameter called \textit{swappiness} which is by
default 60, but can take values from 0 to 100. A value of 0 indicates that the
kernel avoids swapping pages out of the physical memory, while a value of 100
indicates an aggressive swapping mechanism~\cite{swappiness}. To test our
hypothesis, we lower the swappiness value to 10 for a lower swapping overhead.
With this change, the overall execution is faster, but the memory footprint is
higher during execution,  as shown in Figure~\ref{fig:gc_mem_rp3_phase3}.
However, we note that decreasing swappiness value does not help with
Hyperledger's crashes on heavy workloads.

\begin{table*}[t]
\caption{Time, Power and PPR of
Hyperledger\scriptsize{$^{\ref{ftn:perf}}$}}
\label{table:hl-ppr}
\resizebox{1.0\textwidth}{!} {
\begin{tabular}{|c|r||r|r|r|r||r|r|r|r||r|r|r|r||r|r|r|r||R{48pt}|R{48pt}|R{48pt}|R{48pt}|}
\hline
\multirow{3}{*}{\textbf{Workload}} &
\multicolumn{1}{c|}{\multirow{3}{*}{\textbf{Size}}} &
\multicolumn{8}{c||}{\textbf{Execution Time [s]}}
& \multicolumn{8}{c||}{\textbf{Power [W]}} &
\multicolumn{4}{c|}{\multirow{2}{*}{\textbf{Performance-to-Power Ratio
[ops/J]}}} \\
\cline{3-18} & & \multicolumn{4}{c||}{\textbf{Average}} &
\multicolumn{4}{c||}{\textbf{Std. dev.}} &
\multicolumn{4}{c||}{\textbf{Average}} & \multicolumn{4}{c||}{\textbf{Std.
dev.}} & \multicolumn{4}{c|}{} \\
\cline{3-22} & & Xeon & NUC & TX2 & RP3 & Xeon & NUC & TX2 & RP3 & Xeon & NUC &
TX2 & RP3 & Xeon & NUC & TX2 & RP3 & Xeon & NUC & TX2 & RP3 \\
\hline
\hline\multirow{3}{*}{CPUHeavy}
& 1000000
& \cellcolor{grad_2} 1.0
& \cellcolor{grad_2} 1.0
& \cellcolor{grad_1} 1.1
& \cellcolor{grad_0} 2.4
& 0.0
& 0.0
& 0.0
& 1.7
& \cellcolor{grad_0} 50.6
& \cellcolor{grad_1} 9.0
& \cellcolor{grad_2} 2.4
& \cellcolor{grad_3} 2.1
& 1.7
& 0.2
& 0.0
& 0.1
& \cellcolor{grad_0} 19,459.2
& \cellcolor{grad_1} 106,885.5
& \cellcolor{grad_3} 383,004.3
& \cellcolor{grad_2} 308,875.7
\\
\cline{2-22}
& 10000000
& \cellcolor{grad_2} 1.2
& \cellcolor{grad_2} 1.2
& \cellcolor{grad_1} 1.7
& \cellcolor{grad_0} 2.5
& 0.0
& 0.0
& 0.0
& 0.0
& \cellcolor{grad_0} 52.0
& \cellcolor{grad_1} 10.9
& \cellcolor{grad_2} 3.8
& \cellcolor{grad_3} 2.7
& 1.0
& 0.7
& 0.6
& 0.2
& \cellcolor{grad_0} 165,735.7
& \cellcolor{grad_1} 739,181.8
& \cellcolor{grad_3} 1,597,910.5
& \cellcolor{grad_2} 1,494,017.5
\\
\cline{2-22}
& 100000000
& \cellcolor{grad_3} 2.8
& \cellcolor{grad_2} 3.9
& \cellcolor{grad_1} 8.3
& \cellcolor{grad_0} 17.7
& 0.0
& 0.2
& 0.2
& 1.1
& \cellcolor{grad_0} 72.9
& \cellcolor{grad_1} 14.8
& \cellcolor{grad_2} 4.6
& \cellcolor{grad_3} 2.9
& 1.0
& 1.2
& 0.1
& 0.1
& \cellcolor{grad_0} 485,530.3
& \cellcolor{grad_1} 1,720,378.6
& \cellcolor{grad_3} 2,607,959.9
& \cellcolor{grad_2} 1,978,696.9
\\
\cline{2-22}
\hline
\multirow{3}{*}{IOHeavy Write}
& 3200000
& \cellcolor{grad_3} 1,055.2
& \cellcolor{grad_2} 1,721.1
& \cellcolor{grad_1} 7,365.3
& \cellcolor{grad_0} 11,911.7
& 52.3
& 88.9
& 78.3
& 142.1
& \cellcolor{grad_0} 83.1
& \cellcolor{grad_1} 17.9
& \cellcolor{grad_2} 3.7
& \cellcolor{grad_3} 3.3
& 0.1
& 0.0
& 0.0
& 0.0
& \cellcolor{grad_0} 36.6
& \cellcolor{grad_2} 104.4
& \cellcolor{grad_3} 117.2
& \cellcolor{grad_1} 82.3
\\
\cline{2-22}
& 6400000
& \cellcolor{grad_3} 2,125.1
& \cellcolor{grad_2} 3,473.2
& \cellcolor{grad_1} 14,675.3
& \cellcolor{grad_0} 27,246.0
& 85.3
& 139.9
& 114.0
& 309.6
& \cellcolor{grad_0} 83.0
& \cellcolor{grad_1} 17.9
& \cellcolor{grad_2} 4.5
& \cellcolor{grad_3} 3.1
& 0.1
& 0.1
& 1.1
& 0.0
& \cellcolor{grad_0} 36.3
& \cellcolor{grad_3} 103.1
& \cellcolor{grad_2} 102.9
& \cellcolor{grad_1} 74.9
\\
\cline{2-22}
& 12800000
& \cellcolor{grad_3} 4,299.0
& \cellcolor{grad_2} 7,025.0
& \cellcolor{grad_1} 28,957.0
& \cellcolor{grad_0} 63,891.3
& 102.5
& 162.7
& 751.9
& 917.2
& \cellcolor{grad_0} 83.0
& \cellcolor{grad_1} 17.9
& \cellcolor{grad_2} 3.7
& \cellcolor{grad_3} 3.0
& 0.1
& 0.0
& 0.1
& 0.0
& \cellcolor{grad_0} 35.9
& \cellcolor{grad_2} 101.6
& \cellcolor{grad_3} 120.1
& \cellcolor{grad_1} 66.2
\\
\cline{2-22}
\hline
\multirow{3}{*}{IOHeavy Scan}
& 3200000
& \cellcolor{grad_3} 744.1
& \cellcolor{grad_2} 1,442.0
& \cellcolor{grad_1} 6,191.7
& \cellcolor{grad_0} 8,915.3
& 5.9
& 7.2
& 79.3
& 42.6
& \cellcolor{grad_0} 83.7
& \cellcolor{grad_1} 15.1
& \cellcolor{grad_3} 3.1
& \cellcolor{grad_2} 3.2
& 0.1
& 0.0
& 0.0
& 0.0
& \cellcolor{grad_0} 51.4
& \cellcolor{grad_2} 147.1
& \cellcolor{grad_3} 169.3
& \cellcolor{grad_1} 112.4
\\
\cline{2-22}
& 6400000
& \cellcolor{grad_3} 1,487.1
& \cellcolor{grad_2} 2,871.4
& \cellcolor{grad_1} 10,960.3
& \cellcolor{grad_0} 17,296.3
& 11.5
& 26.5
& 2195.8
& 118.8
& \cellcolor{grad_0} 83.6
& \cellcolor{grad_1} 15.1
& \cellcolor{grad_2} 3.6
& \cellcolor{grad_3} 3.2
& 0.0
& 0.0
& 0.8
& 0.0
& \cellcolor{grad_0} 51.5
& \cellcolor{grad_2} 147.4
& \cellcolor{grad_3} 169.4
& \cellcolor{grad_1} 114.6
\\
\cline{2-22}
& 12800000
& \cellcolor{grad_3} 2,966.1
& \cellcolor{grad_2} 5,768.3
& \cellcolor{grad_1} 25,049.0
& \cellcolor{grad_0} 34,274.7
& 20.2
& 84.4
& 257.7
& 702.6
& \cellcolor{grad_0} 83.6
& \cellcolor{grad_1} 15.1
& \cellcolor{grad_3} 3.0
& \cellcolor{grad_2} 3.2
& 0.1
& 0.1
& 0.0
& 0.0
& \cellcolor{grad_0} 51.6
& \cellcolor{grad_2} 147.3
& \cellcolor{grad_3} 167.7
& \cellcolor{grad_1} 115.3
\\
\cline{2-22}
\hline
\multirow{1}{*}{Analytics Q1}
& 10000
& \cellcolor{grad_3} 9.7
& \cellcolor{grad_2} 18.8
& \cellcolor{grad_0} 157.6
& \cellcolor{grad_1} 103.8
& 0.0
& 0.0
& 0.3
& 5.0
& \cellcolor{grad_0} 90.3
& \cellcolor{grad_1} 18.1
& \cellcolor{grad_3} 2.9
& \cellcolor{grad_2} 3.2
& 0.5
& 0.1
& 0.0
& 0.0
& \cellcolor{grad_0} 11.4
& \cellcolor{grad_2} 29.5
& \cellcolor{grad_1} 21.8
& \cellcolor{grad_3} 30.6
\\
\cline{2-22}
\hline
\end{tabular}
}
\end{table*}

\section[Single Node Analysis]{Single Node Analysis}
\label{sec:analysis_sn}

\subsection{Hyperledger}
\label{sec:hl}

The time-power performance\footnote{Background colors represent relative
performance, with green and red being the best and the worst,
respectively.\label{ftn:perf}} of Hyperledger workloads is shown in
Table~\ref{table:hl-ppr}, across four systems under evaluation. Lower values
are better for the execution time and power, while for the performance-to-power
ratio (PPR) higher values are better. We define the PPR as the ratio between
useful work and power. For simplicity, we consider useful work to be
proportional with the input size. For example, for Xeon system, sorting one million numbers
using the CPUHeavy smart contract performs one million operations and
consumes $50.6$W, therefore its PPR is $19,459.2$ ops/J.

\begin{figure*}[t]
\centering
\begin{subfigure}{\subfigsizeb}
\includegraphics[width=0.65\textwidth,angle=270]{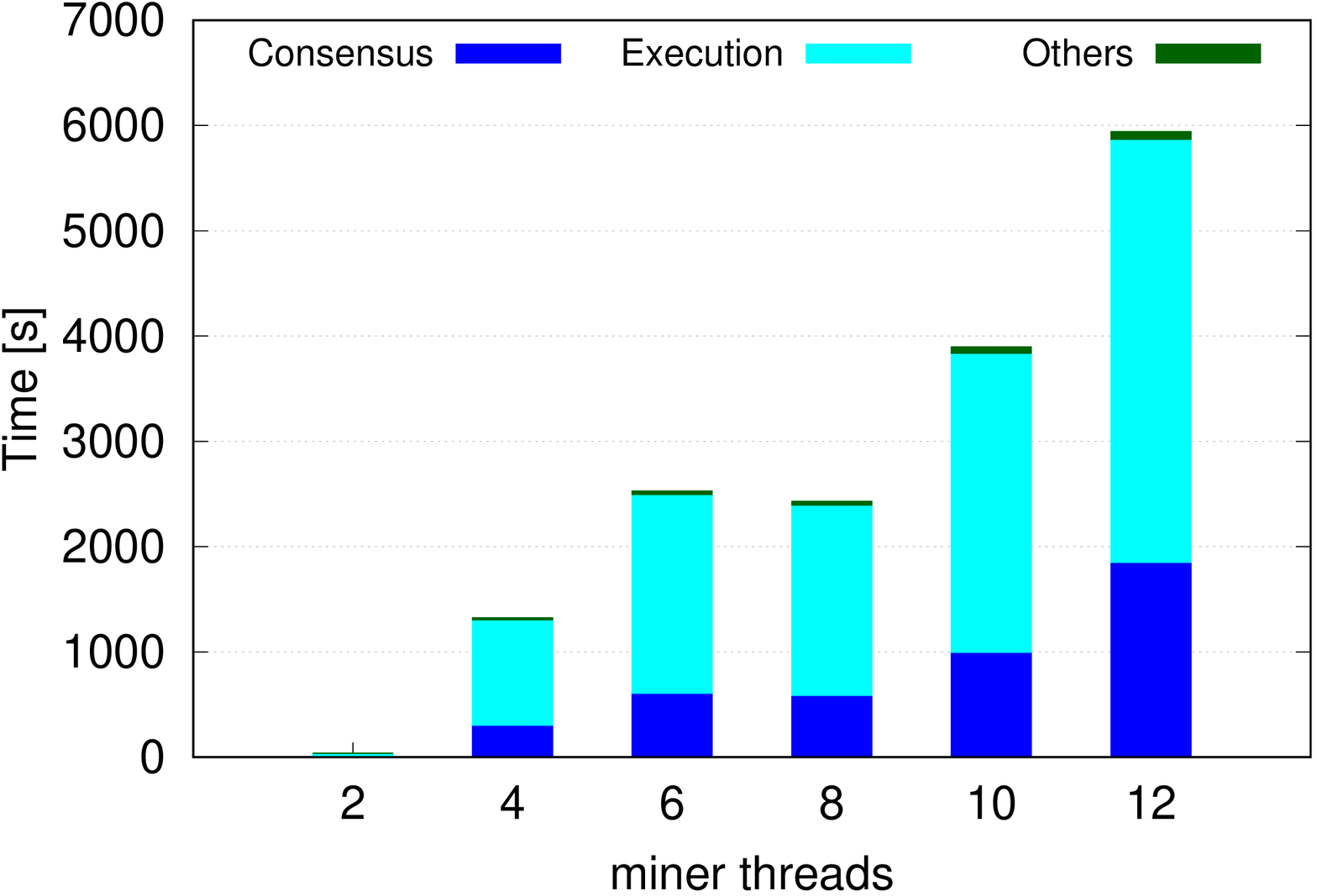}
\caption{On Xeon}
\end{subfigure}
\begin{subfigure}{\subfigsizeb}
\includegraphics[width=0.65\textwidth,angle=270]{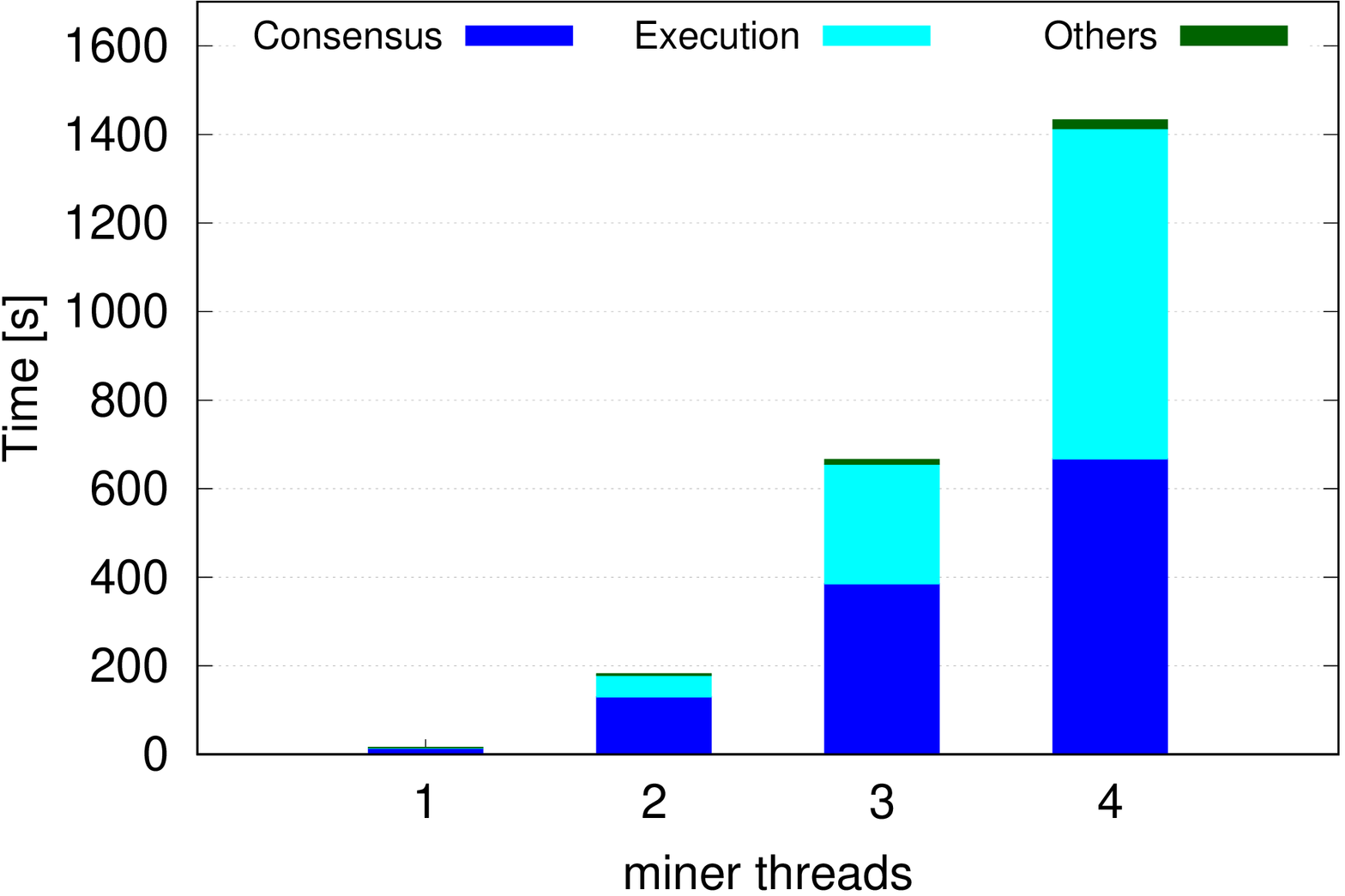}
\caption{On NUC}
\end{subfigure}
\begin{subfigure}{\subfigsizeb}
\includegraphics[width=0.65\textwidth,angle=270]{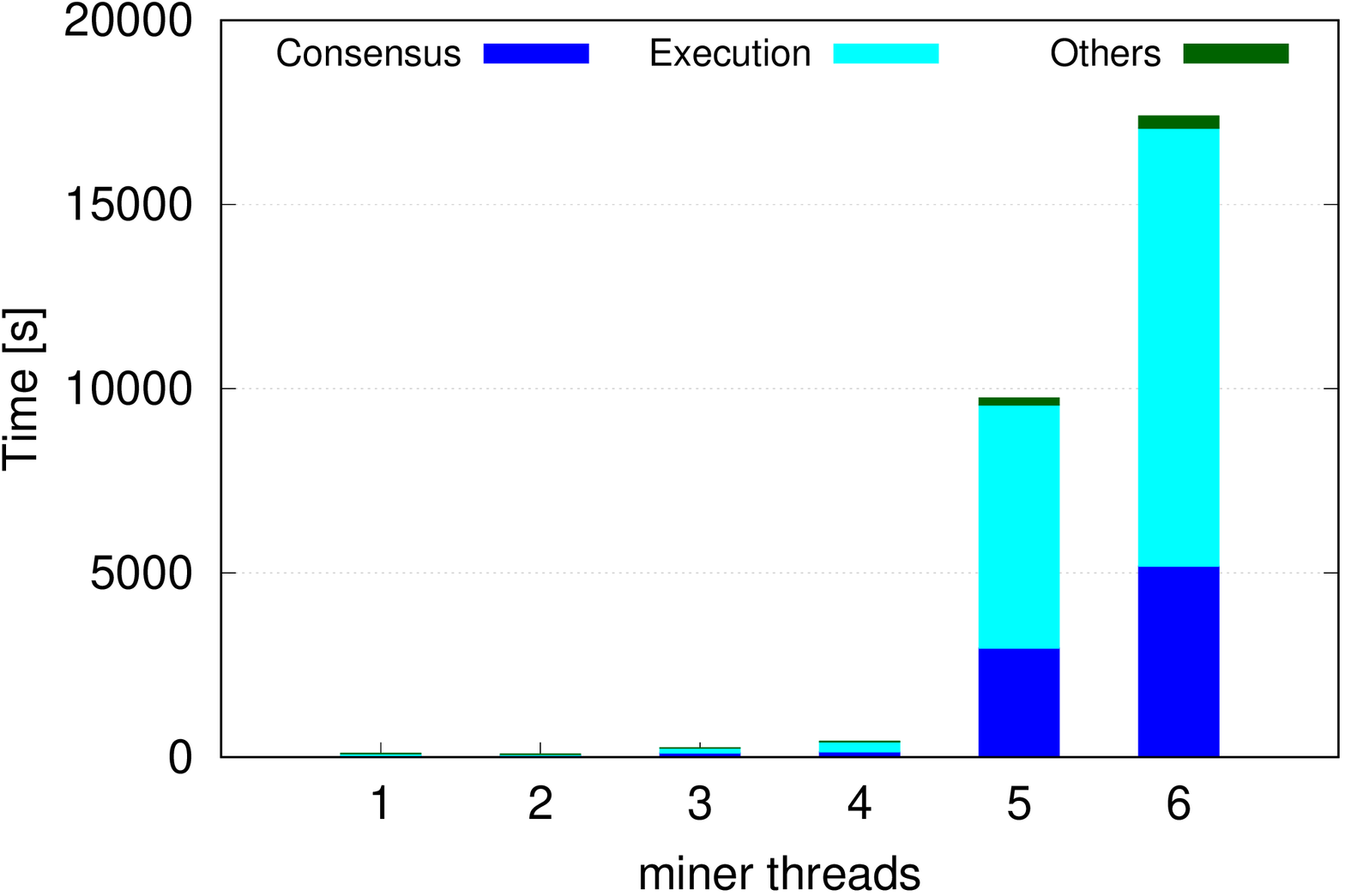}
\caption{On TX2}
\end{subfigure}
\caption{Ethereum execution breakdown}
\label{fig:geth_bd}
\end{figure*}

\begin{table*}[t]
\centering
\caption{Comparison of five CPUHeavy(1M) executions with four miner threads on
NUC with different versions of Ethereum}
\label{table:eth_runs_apply}
\resizebox{0.9\textwidth}{!} {
\begin{tabular}{|c|r|r|r|r|r|r|r|r|r|r|r|}
\hline
\textbf{geth} & \multicolumn{5}{c|}{\textbf{Execution Time [s]}} &
\multicolumn{5}{c|}{\textbf{Apply Transaction Count}} \\
\cline{2-11}
\textbf{version} & Run \#1 & Run \#2 & Run \#3 & Run \#4 & Run \#5 & Run \#1 &
Run \#2 & Run \#3 & Run \#4 & Run \#5 \\
\hline
\hline
1.4.18 & 146.7 & 97.7 & 120.9 & 98.9 & 99.2 & 6 & 4 & 5 & 4 & 4 \\
1.8.13 & 5.5 & 5.0 & 4.5 & 7.5 & 5.0 & 1 & 1 & 1 & 1 & 1 \\
1.8.14 & 1369.3 & 2152.8 & 561.7 & 197.3 & 72.6 & 314 & 481 & 126 & 42 & 16 \\
1.8.15 & 62.08 & 125.45 & 27.68 & 141.64 & 186.4 & 13 & 27 & 4 & 31 & 42 \\
\hline
\end{tabular}
}
\end{table*}

As expected, the fastest system is Xeon, followed by NUC, TX2 and RP3, in this
order. For example, RP3 is up to $15\times$ slower than Xeon for IOHeavy Write,
while TX2 is more than $8\times$ slower than Xeon for IOHeavy Scan. These
performance patterns are in accordance with the characterization in
Section~\ref{sec:sys_char}.

The average power profile is the opposite of time performance, with Xeon being
the most power-hungry and RP3 the most power-efficient. In particular, Xeon
uses 50W and up to 84W for CPUHeavy and IOHeavy, respectively.
For the CPUHeavy benchmark there is a gradual increase in power usage with the
growing input size since the CPU is doing more work. The IOHeavy workloads incur
more power compared to CPUHeavy because they stress the memory and storage, in
addition to using the CPU. At the same time, this high-power usage of IOHeavy
remains constant as input size increases because (i) the CPU utilization is low
but roughly constant, and (ii) the memory and storage have lower dynamic power
fluctuations~\cite{GoogleDatacenter_13}. For example, the average CPU
utilization is 9.5\% (3.5\% standard deviation) and 8.1\% (3.6\% standard
deviation) during the IOHeavy Write and Scan, respectively, for 3.2 million
key-value pairs.

The energy is the product of execution time and average power usage. Xeon and
RP3 exhibit the highest energy cost due to high power usage for the former and
long execution time for the latter. On the other hand, TX2 and NUC use almost
always the lowest energy. TX2 is almost always the most efficient because of its
lower power profile compared to NUC, and higher performance compared to RP3. We
note that even if RP3 has a very low power profile, its memory and CPU
limitation translate to larger energy cost than systems with higher power
profile.

In summary, we make the
following observation concerning Hyperledger execution.  \begin{obs}
\label{obs:hl}
\textit{The highest energy efficiency is achieved by low-power systems with
balanced performance-to-power profile, rather than systems with low power
profile but also low performance.}
\end{obs}

\subsection{Ethereum}
\label{sec:eth}

\begin{table*}[t]
\caption{Time, Power and PPR of Ethereum}
\label{table:eth-ppr}
\resizebox{1.0\textwidth}{!} {
\begin{tabular}{|c|r||r|r|r||r|r|r||r|r|r||r|r|r||R{50pt}|R{50pt}|R{50pt}|}
\hline
\multirow{3}{*}{\textbf{Workload}} &
\multicolumn{1}{c|}{\multirow{3}{*}{\textbf{Size}}} &
\multicolumn{6}{c||}{\textbf{Execution Time [s]}}
& \multicolumn{6}{c||}{\textbf{Power [W]}} &
\multicolumn{3}{c|}{\multirow{2}{*}{\textbf{Performance-to-Power Ratio
[ops/J]}}} \\
\cline{3-14} & & \multicolumn{3}{c||}{\textbf{Average}} &
\multicolumn{3}{c||}{\textbf{Std. dev.}} & \multicolumn{3}{c||}{\textbf{Average}} &
\multicolumn{3}{c||}{\textbf{Std. dev.}} & \multicolumn{3}{c|}{} \\
\cline{3-17} & & Xeon & NUC & TX2 & Xeon & NUC & TX2 & Xeon & NUC & TX2 & Xeon &
NUC & TX2 & Xeon & NUC & TX2 \\
\hline
\hline
\multirow{1}{*}{CPUHeavy}
& 1000000
& \cellcolor{grad_3} 6.2
& \cellcolor{grad_1} 40.6
& \cellcolor{grad_2} 31.4
& 1.4
& 12.3
& 5.5
& \cellcolor{grad_1} 80.9
& \cellcolor{grad_2} 17.1
& \cellcolor{grad_3} 5.0
& 1.4
& 0.2
& 0.1
& \cellcolor{grad_2} 2,112.5
& \cellcolor{grad_1} 1,586.8
& \cellcolor{grad_3} 4,111.3
\\
\cline{2-17}
\hline
\multirow{3}{*}{IOHeavy Write}
& 100
& \cellcolor{grad_2} 3.5
& \cellcolor{grad_3} 2.9
& \cellcolor{grad_1} 9.2
& 0.8
& 1.4
& 7.7
& \cellcolor{grad_1} 76.7
& \cellcolor{grad_2} 16.4
& \cellcolor{grad_3} 4.8
& 0.1
& 0.5
& 0.0
& \cellcolor{grad_1} 0.5
& \cellcolor{grad_3} 3.1
& \cellcolor{grad_2} 2.3
\\
\cline{2-17}
& 1000
& \cellcolor{grad_3} 4.4
& \cellcolor{grad_2} 10.5
& \cellcolor{grad_1} 12.6
& 2.5
& 5.0
& 7.2
& \cellcolor{grad_1} 81.4
& \cellcolor{grad_2} 16.1
& \cellcolor{grad_3} 5.1
& 4.0
& 0.2
& 0.3
& \cellcolor{grad_1} 4.1
& \cellcolor{grad_2} 7.5
& \cellcolor{grad_3} 15.5
\\
\cline{2-17}
& 10000
& \cellcolor{grad_1} 329.9
& \cellcolor{grad_3} 142.4
& \cellcolor{grad_2} 194.4
& 48.9
& 118.6
& 173.7
& \cellcolor{grad_1} 80.5
& \cellcolor{grad_2} 17.5
& \cellcolor{grad_3} 5.2
& 0.5
& 0.4
& 0.2
& \cellcolor{grad_1} 0.4
& \cellcolor{grad_2} 10.4
& \cellcolor{grad_3} 24.9
\\
\cline{2-17}
\hline
\multirow{3}{*}{IOHeavy Scan}
& 100
& \cellcolor{grad_2} 3.9
& \cellcolor{grad_3} 3.4
& \cellcolor{grad_1} 5.2
& 1.7
& 0.9
& 3.3
& \cellcolor{grad_1} 78.0
& \cellcolor{grad_2} 16.7
& \cellcolor{grad_3} 5.0
& 1.1
& 0.4
& 0.1
& \cellcolor{grad_1} 0.4
& \cellcolor{grad_2} 1.9
& \cellcolor{grad_3} 3.0
\\
\cline{2-17}
& 1000
& \cellcolor{grad_3} 1.8
& \cellcolor{grad_2} 6.5
& \cellcolor{grad_1} 29.6
& 1.0
& 2.3
& 10.7
& \cellcolor{grad_1} 77.4
& \cellcolor{grad_2} 16.0
& \cellcolor{grad_3} 4.7
& 3.2
& 0.5
& 0.0
& \cellcolor{grad_3} 12.3
& \cellcolor{grad_2} 10.7
& \cellcolor{grad_1} 7.1
\\
\cline{2-17}
& 10000
& \cellcolor{grad_3} 1.8
& \cellcolor{grad_2} 2.4
& \cellcolor{grad_1} 11.1
& 0.9
& 0.6
& 1.4
& \cellcolor{grad_1} 79.7
& \cellcolor{grad_2} 17.0
& \cellcolor{grad_3} 5.0
& 2.8
& 0.3
& 0.2
& \cellcolor{grad_1} 81.7
& \cellcolor{grad_3} 270.6
& \cellcolor{grad_2} 140.7
\\
\cline{2-17}
\hline
\multirow{1}{*}{Analytics Q1}
& 1000
& \cellcolor{grad_3} 0.7
& \cellcolor{grad_2} 8.1
& \cellcolor{grad_1} 19.5
& 0.0
& 0.0
& 0.1
& \cellcolor{grad_1} 70.4
& \cellcolor{grad_2} 19.4
& \cellcolor{grad_3} 5.5
& 4.1
& 0.1
& 0.1
& \cellcolor{grad_1} 20.4
& \cellcolor{grad_2} 63.9
& \cellcolor{grad_3} 91.9
\\
\cline{2-17}
\hline
\multirow{1}{*}{Analytics Q2}
& 1000
& \cellcolor{grad_3} 0.7
& \cellcolor{grad_2} 8.1
& \cellcolor{grad_1} 19.6
& 0.0
& 0.0
& 0.3
& \cellcolor{grad_1} 73.5
& \cellcolor{grad_2} 19.4
& \cellcolor{grad_3} 5.5
& 0.5
& 0.2
& 0.0
& \cellcolor{grad_1} 19.2
& \cellcolor{grad_2} 63.8
& \cellcolor{grad_3} 92.9
\\
\cline{2-17}
\hline
\multirow{1}{*}{Analytics Q3}
& 1000
& \cellcolor{grad_3} 0.4
& \cellcolor{grad_2} 6.6
& \cellcolor{grad_1} 16.9
& 0.0
& 0.0
& 0.1
& \cellcolor{grad_1} 71.7
& \cellcolor{grad_2} 19.6
& \cellcolor{grad_3} 5.5
& 0.3
& 0.2
& 0.1
& \cellcolor{grad_1} 32.0
& \cellcolor{grad_2} 76.9
& \cellcolor{grad_3} 105.8
\\
\cline{2-17}
\hline
\end{tabular}
}
\end{table*}

\begin{table*}[t]
\caption{Time, Power and PPR of Parity}
\label{table:parity-ppr}
\resizebox{1.0\textwidth}{!} {
\begin{tabular}{|c|r||r|r|r|r||r|r|r|r||r|r|r|r||r|r|r|r||r|r|r|r|}
\hline
\multirow{3}{*}{\textbf{Workload}} &
\multicolumn{1}{c|}{\multirow{3}{*}{\textbf{Size}}} &
\multicolumn{8}{c||}{\textbf{Execution Time [s]}}
& \multicolumn{8}{c||}{\textbf{Power [W]}} &
\multicolumn{4}{c|}{\multirow{2}{*}{\textbf{Performance-to-Power Ratio
[ops/J]}}} \\
\cline{3-18} & & \multicolumn{4}{c||}{\textbf{Average}} &
\multicolumn{4}{c||}{\textbf{Std. dev.}} &
\multicolumn{4}{c||}{\textbf{Average}} & \multicolumn{4}{c||}{\textbf{Std.
dev.}} & \multicolumn{4}{c|}{} \\
\cline{3-22} & & Xeon & NUC & TX2 & RP3 & Xeon & NUC & TX2 & RP3 & Xeon & NUC &
TX2 & RP3 & Xeon & NUC & TX2 & RP3 & Xeon & NUC & TX2 & RP3 \\
\hline
\hline
\multirow{2}{*}{CPUHeavy}
& 1000000
& \cellcolor{grad_3} 64.9
& \cellcolor{grad_2} 71.1
& \cellcolor{grad_1} 147.1
& \cellcolor{grad_0} 1,205.5
& 26.0
& 0.0
& 3.7
& 6.5
& \cellcolor{grad_0} 57.6
& \cellcolor{grad_1} 12.7
& \cellcolor{grad_2} 3.6
& \cellcolor{grad_3} 2.6
& 1.2
& 0.3
& 0.1
& 0.0
& \cellcolor{grad_1} 324.2
& \cellcolor{grad_2} 1,106.6
& \cellcolor{grad_3} 1,910.1
& \cellcolor{grad_0} 316.0
\\
\cline{2-22}
& 10000000
& \cellcolor{grad_3} 469.7
& \cellcolor{grad_2} 705.1
& \cellcolor{grad_1} 1,371.0
& \cellcolor{grad_0} 12,205.4
& 4.3
& 0.2
& 83.0
& 367.2
& \cellcolor{grad_0} 71.5
& \cellcolor{grad_1} 14.7
& \cellcolor{grad_2} 4.5
& \cellcolor{grad_3} 2.7
& 1.2
& 0.4
& 0.0
& 0.0
& \cellcolor{grad_0} 298.1
& \cellcolor{grad_2} 967.4
& \cellcolor{grad_3} 1,634.5
& \cellcolor{grad_1} 302.0
\\
\cline{2-22}
\hline
\multirow{3}{*}{IOHeavy Write}
& 100
& \cellcolor{grad_0} 84.8
& \cellcolor{grad_2} 42.1
& \cellcolor{grad_3} 30.8
& \cellcolor{grad_1} 62.2
& 15.3
& 15.2
& 0.2
& 45.3
& \cellcolor{grad_0} 50.7
& \cellcolor{grad_1} 8.8
& \cellcolor{grad_2} 2.5
& \cellcolor{grad_3} 2.1
& 0.8
& 0.0
& 0.0
& 0.0
& \cellcolor{grad_0} 0.0
& \cellcolor{grad_1} 0.3
& \cellcolor{grad_3} 1.3
& \cellcolor{grad_2} 1.2
\\
\cline{2-22}
& 1000
& \cellcolor{grad_1} 170.4
& \cellcolor{grad_3} 96.0
& \cellcolor{grad_2} 106.7
& \cellcolor{grad_0} 287.0
& 84.1
& 26.0
& 40.0
& 5.6
& \cellcolor{grad_0} 52.8
& \cellcolor{grad_1} 10.5
& \cellcolor{grad_2} 3.0
& \cellcolor{grad_3} 2.5
& 0.1
& 0.2
& 0.1
& 0.1
& \cellcolor{grad_0} 0.1
& \cellcolor{grad_1} 1.1
& \cellcolor{grad_3} 3.6
& \cellcolor{grad_2} 1.4
\\
\cline{2-22}
& 10000
& \cellcolor{grad_3} 124.7
& \cellcolor{grad_2} 186.9
& \cellcolor{grad_1} 380.7
& \cellcolor{grad_0} 2,996.5
& 0.4
& 5.8
& 1.1
& 25.8
& \cellcolor{grad_0} 71.0
& \cellcolor{grad_1} 14.2
& \cellcolor{grad_2} 4.2
& \cellcolor{grad_3} 2.7
& 1.8
& 0.1
& 0.1
& 0.0
& \cellcolor{grad_0} 1.1
& \cellcolor{grad_2} 3.7
& \cellcolor{grad_3} 6.3
& \cellcolor{grad_1} 1.2
\\
\cline{2-22}
\hline
\multirow{3}{*}{IOHeavy Scan}
& 100
& \cellcolor{grad_1} 63.1
& \cellcolor{grad_2} 52.6
& \cellcolor{grad_3} 30.7
& \cellcolor{grad_0} 82.5
& 0.0
& 15.2
& 0.0
& 40.0
& \cellcolor{grad_0} 50.0
& \cellcolor{grad_1} 8.8
& \cellcolor{grad_2} 2.5
& \cellcolor{grad_3} 1.9
& 0.8
& 0.1
& 0.0
& 0.0
& \cellcolor{grad_0} 0.0
& \cellcolor{grad_1} 0.3
& \cellcolor{grad_3} 1.3
& \cellcolor{grad_2} 0.9
\\
\cline{2-22}
& 1000
& \cellcolor{grad_3} 42.2
& \cellcolor{grad_0} 149.0
& \cellcolor{grad_2} 51.9
& \cellcolor{grad_1} 72.2
& 15.1
& 30.0
& 30.0
& 40.0
& \cellcolor{grad_0} 50.1
& \cellcolor{grad_1} 8.7
& \cellcolor{grad_2} 2.5
& \cellcolor{grad_3} 2.0
& 0.9
& 0.0
& 0.1
& 0.0
& \cellcolor{grad_0} 0.5
& \cellcolor{grad_1} 0.8
& \cellcolor{grad_3} 10.1
& \cellcolor{grad_2} 9.7
\\
\cline{2-22}
& 10000
& \cellcolor{grad_2} 52.6
& \cellcolor{grad_0} 191.7
& \cellcolor{grad_3} 30.3
& \cellcolor{grad_1} 112.7
& 14.9
& 78.3
& 1.2
& 2.8
& \cellcolor{grad_0} 51.0
& \cellcolor{grad_1} 9.5
& \cellcolor{grad_2} 2.8
& \cellcolor{grad_3} 2.5
& 0.9
& 0.1
& 0.1
& 0.0
& \cellcolor{grad_0} 4.1
& \cellcolor{grad_1} 6.7
& \cellcolor{grad_3} 119.6
& \cellcolor{grad_2} 35.5
\\
\cline{2-22}
\hline
\multirow{1}{*}{Analytics Q1}
& 1000
& \cellcolor{grad_3} 1.2
& \cellcolor{grad_2} 2.0
& \cellcolor{grad_1} 10.5
& \cellcolor{grad_0} 14.2
& 0.0
& 0.0
& 0.3
& 2.3
& \cellcolor{grad_0} 51.1
& \cellcolor{grad_1} 14.0
& \cellcolor{grad_2} 2.8
& \cellcolor{grad_3} 2.5
& 1.3
& 1.4
& 0.0
& 0.2
& \cellcolor{grad_0} 16.3
& \cellcolor{grad_3} 36.9
& \cellcolor{grad_2} 34.8
& \cellcolor{grad_1} 28.1
\\
\cline{2-22}
\hline
\multirow{1}{*}{Analytics Q2}
& 1000
& \cellcolor{grad_3} 1.2
& \cellcolor{grad_2} 1.9
& \cellcolor{grad_1} 10.2
& \cellcolor{grad_0} 14.6
& 0.0
& 0.0
& 0.2
& 1.4
& \cellcolor{grad_0} 49.5
& \cellcolor{grad_1} 14.4
& \cellcolor{grad_2} 2.7
& \cellcolor{grad_3} 2.5
& 1.7
& 1.1
& 0.0
& 0.1
& \cellcolor{grad_0} 17.1
& \cellcolor{grad_3} 36.0
& \cellcolor{grad_2} 35.7
& \cellcolor{grad_1} 27.8
\\
\cline{2-22}
\hline
\multirow{1}{*}{Analytics Q3}
& 1000
& \cellcolor{grad_3} 0.5
& \cellcolor{grad_2} 0.7
& \cellcolor{grad_1} 1.8
& \cellcolor{grad_0} 4.0
& 0.0
& 0.0
& 0.1
& 0.0
& \cellcolor{grad_0} 50.3
& \cellcolor{grad_1} 13.5
& \cellcolor{grad_2} 4.6
& \cellcolor{grad_3} 2.9
& 1.7
& 2.3
& 0.1
& 0.0
& \cellcolor{grad_0} 42.1
& \cellcolor{grad_2} 102.8
& \cellcolor{grad_3} 108.5
& \cellcolor{grad_1} 85.0
\\
\cline{2-22}
\hline
\end{tabular}
}
\end{table*}

Figure~\ref{fig:geth_bd} shows a super-linear increase in the execution time of
\textit{go-ethereum (geth) v1.8.15} with increasing number of miner threads on
three systems under evaluation. Recall that RP3 is unable to run Ethereum. To
investigate the cause of high execution time with more miner threads, we split
break down execution time to three components as described in
BLOCKBENCH~\cite{Dinh_SIGMOD_2017}. We profiled \textit{geth} with Go
\textit{pprof} and analyzing both the callgraph and the cumulative execution
time per routine. From our analysis, the consensus starts with the call of
\texttt{go-ethereum/consensus/ethash.\\(*Ethash).Seal.func1}, while the
execution starts with the invocation,
\texttt{go-ethereum/core/vm.(*EVMInterpreter).Run}. The remaining time is spent
at application layer and data layer of the blockchain stack. We observed that
execution takes longer as the number of miner threads increase. This would
suggest that EVM is inefficient when more miner threads are used.

We observed high variations among the executions of the same benchmark with the
same number of miner threads.
Table~\ref{table:eth_runs_apply} shows five executions of CPUHeavy on NUC with
four miner threads. The high variations are visible in \textit{go-ethereum}
releases starting from \textit{v1.8.14}. Previous releases, including
\textit{v1.4.18} used by BLOCKBENCH~\cite{Dinh_SIGMOD_2017}, exhibit relatively
stable execution. We found that starting with \textit{go-ethereum v1.8.14} a
transaction is started, or \textit{applied}, multiple times and that this number
is inconsistent among different runs. For example, a CPUHeavy transaction is
applied as few as $16$ and as many as $481$ times in \textit{go-ethereum
v1.8.14}. This is explained by the fact that \textit{go-ethereum v1.8.14}
underwent a significant design change. Specifically, whenever a miner thread
receives a new block, it discards any transactions currently being executed and
applies the transactions in the newly received block. In our case, there is a
single transaction, and during its execution the miners keep generating empty
blocks. As a result, the probability of receiving a block during the transaction
execution increases with the number of miners. Therefore, the transaction is
interrupted and restarted many times. As we shall see in the next section, the
same results hold for the cluster setting and for non-CPU heavy workloads.

We note that this design works well when a newly received block contains updates
to states currently used by the transaction being executed. In this case, it
saves time to stop and restart the current transaction until after the new block
is applied. However, interrupting transactions even when receiving empty block
results in unnecessary overhead. A more elegant approach is to restart only
transactions whose states are affected by the new block.

In summary, we make the following observation concerning Ethereum execution. 
\begin{obs}
\label{obs:eth}
\textit{In the latest versions of Ethereum, (i) execution time increases
with the number of miner threads and (ii) there is high execution time variation
among different runs of the same workload, especially when the workload is
computation heavy or when more miner threads are used. These are due to 
new transaction restarting mechanism which restarts execution when receiving a new block, even
if that block is empty.}
\end{obs}

\textbf{Time-energy performance.} The time, power and PPR of Ethereum runnign
with one miner thread are shown in Table~\ref{table:eth-ppr}. Across systems, we
observe the same pattern as that of Hyperledger. In other words,
Observation~\ref{obs:hl} holds.  Even if TX2 exhibits the highest execution time
in general, its energy usage is the lowest and, thus, its PPR is the highest.
For example, the IOHeavy Scan benchmark with $10,000$ key-value pairs is
$3.2\times$ slower on TX2, but uses $4.9\times$ less energy than Xeon.

As expected, Ethereum uses more power than Hyperledger. In particular, for
sorting $1$M values, Xeon, NUC and TX2 use 50.6W, 9W and 2.4W, respectively, in
Hyperledger, as opposed to 81W, 17W and 5W, respectively, in Ethereum.
There are two reasons for this behavior. First, Ethereum use more cryptographic
operations which incur high CPU utilization. Second, Ethereum uses EVM, an
interpreted execution environment which is less efficient than Hyperledger's
Docker execution.  Consequently, the CPU performs more work in Ethereum.

Our evaluation demonstrates high variability, especially for IOHeavy operations,
as indicated by the high standard deviation in Table~\ref{table:eth-ppr}.
Execution profiling of IOHeavy Write shows that much of the time is spent in the
EVM interpreter. For example, the writing of $10,000$ key-value pairs on Xeon
spends $71\%$ of the time inside EVM interpreter, while sorting one million
numbers spends only $10\%$ in the same routine. The root cause is the same as
for running multiple miner threads, namely, the transaction is restarted
multiple times until it manages to finish. Transactions that perform more work
and take longer to finish, have higher chances to be restarted and, thus, take
even longer to finish under \textit{geth v1.8.15}. For example, an execution of
sorting one million numbers on Xeon finishes in 9s and restarts the transaction
2 times. In contrast, the execution of IOHeavy Write of $10,000$ key-value pairs
finishes in 458s and is restarted 63 times.

We note that we were unable to run CPUHeavy with input size of $10$M and $100$M.
While BLOCKBENCH paper~\cite{Dinh_SIGMOD_2017} reports execution times for
Ethereum CPUHeavy on $10$M input, in our experiments the clients never finish
the execution.

\subsection{Parity}
\label{sec:parity}

The time-power results for Parity are presented in Table~\ref{table:parity-ppr}.
Unlike Ethereum, Parity is able to run on the wimpy RP3 system. On the other
hand, all systems are not able to run the CPUHeavy workload with $100$M input.

Recall that RP3 is $2-3\times$ slower than TX2 for Hyperledger. This gap is much
bigger for Parity. In particular, RP3 is $8\times$ slower than TX2 when running
CPUHeavy with $10$M input. Our profiling using Linux \textit{perf} shows that
RP3 spends significant time in \texttt{libarmmem.so} which is a library for
memory operations for ARM-based systems. This, together with a low CPU
utilization of 10\%, suggest that memory is the main bottleneck of Parity
execution on RP3. In contrast, the other systems spend most of the time in the
execution layer, i.e., inside EVM interpreter.

The variability in execution time among different runs is less visible in Parity
compared to Ethereum. Table~\ref{table:parity-ppr} shows high standard
deviations only for IOHeavy workloads with small input size. We attribute this
to the memory hierarchy, especially to CPU caches and memory buffers that need
time to warm-up and may exhibit unpredictable behavior on shorter executions.
Indeed, CPUHeavy and Analytics do not exhibit execution time variability. The
former is not memory or I/O intensive. The latter includes an initialization
step that warms up the caches and memory buffers.

As expected, the power consumption of Parity is lower compared to Ethereum, but
higher when compared to Hyperledger. Taking CPUHeavy workload as example, Xeon,
NUC and TX2 use 57.6W, 12.7W and 3.6W, respectively, to sort one million values
in Parity. For the same amount of work, Ethereum consumes 81W, 17W, and 5W on
Xeon, NUC and TX2, respectively, while Hyperledger consumes 50.6W, 9W and 2.4W,
respectively. This behavior can be explained by the lower power overhead of
Parity's PoA consensus protocol compared to Ethereum's PoW. Parity also has an
interpreted EVM which is not as efficient as Hyperledger's Docker execution
engine, and which draws additional power.

Observation~\ref{obs:hl} holds also for Parity. In particular, Xeon and NUC are
the fastest systems, while TX2 uses the smallest amount of energy due to its
shorter execution than RP3 and lower power usage than Xeon and NUC.

\subsection{Impact of Storage Subsystem}
\label{sec:storage}
In this section, we analyze the impact of different types of storage subsystems
on blockchain performance using the IOHeavy benchmarks. We select the TX2 system
which has interfaces for SD card, SATA storage and USB3 devices. While the
baseline is the system with a 64GB SD card (\textbf{TX2+SDC}), we separately
connect a 1TB SSD through SATA (\textbf{TX2+SSD}) and an external 2TB HDD
through USB3 (\textbf{TX2+HDD}). The SD card stores the OS in both TX2+SSD and
TX2+HDD.

We first measure the IO performance in terms of raw read/write throughput and
latency. Then, we measure the performance of IOHeavy benchmarks in Hyperledger
and Parity. Ethereum is not included in this analysis because of its
unpredictable behaviour, as discussed in Section~\ref{sec:eth}. In addition, we
evaluate the impact of storage on the total power by measuring the idle power
when the hardware is running only the OS, and the active power during blockchain
execution. The results are summarized in Table~\ref{table:storage}.

\begin{table}[t]
\centering
\caption{Impact of storage subsystem}
\label{table:storage}
\resizebox{0.475\textwidth}{!} {
\begin{tabular}{|l|r|r|r|}
\hline
\multicolumn{1}{|c|}{\multirow{2}{*}{\textbf{Metric}}} &
\multicolumn{3}{c|}{\textbf{System}} \\
\cline{2-4} & \textbf{TX2+SDC} & \textbf{TX2+SSD} & \textbf{TX2+HDD} \\
\hline
\hline
Idle System Power [W] & 2.4 & 2.9 & 5.9 \\
Write Throughput [MB/s] & 16.3 & 206.0 & 87.6 \\
Read Throughput [MB/s] & 88.9 & 277.0 & 93.4 \\
Write Latency [ms] & 17.1 & 2.8 & 13.7 \\
Read Latency [ms] & 2.8 & 1.8 & 1.2 \\
\hline
\hline
\multicolumn{4}{|c|}{IOHeavy Write (10000)} \\
\hline
Hyperledger Time [s] & 18.2 & 22.4 & 24.0 \\
Parity Time [s] & 380.7 & 382.8 & 386.2 \\
Hyperledger Power [W] & 7.2 & 7.7 & 10.0 \\
Parity Power [W] & 4.2 & 4.8 & 7.7 \\
\hline
\hline
\multicolumn{4}{|c|}{IOHeavy Scan (10000)} \\
\hline
Hyperledger Time [s] & 28.8 & 31.7 & 29.8 \\
Parity Time [s] & 30.3 & 32.5 & 43.1 \\
Hyperledger Power [W] & 3.1 & 3.6 & 6.6 \\
Parity Power [W] & 2.8 & 3.3 & 6.4 \\
\hline
\end{tabular}
}
\figvspace
\end{table}

In terms of raw performance, the SSD is the clear winner. It has almost
$13\times$ higher write throughput and $3\times$ higher read throughput than SD
card, while adding only 0.5W to the idle power. In contrast, the HDD adds 3.5W
to the idle power, thus increasing it by $2.5\times$. The HDD has more than
$5\times$ higher write throughput but similar read throughput compared to
TX2+SDC.

Interestingly, Jetson with SD card exhibits slightly better execution time when
running the IOHeavy. We attribute this to the fact that the SD card
stores the OS, libraries and Docker containers in all three configurations,
including the TX2+SSD and TX2+HDD. Hence, the ledger storage subsystem is not a
bottleneck, otherwise, TX2+SDC would exhibit higher execution time due to its
lower raw throughput and higher latency. In fact, our profiling of IOHeavy Write
shows that write operations are sparse, with an average of 1MB/s and a peak of
21.5MB/s across all subsystems. These values are within the capabilities of all
storage subsystems, but switching between the execution context of the SD card
and the ledger storage may induce overhead.

In terms of power, we observe that IOHeavy Scan adds only 0.7W and 0.4W to the
idle power off all system configurations for Hyperledger and Parity,
respectively. IOHeavy Write uses more power, adding between 4.1W and 4.8W for
Hyperledger and around 1.8W for Parity. These results are stable, in
general. The only notable exception is IOHeavy Scan in Parity on the TX2+HDD
which, in general, finishes in 32.5s, but in some cases it finishes in
$64$s or $97$s. 

In summary, we make the following observation.

\begin{obs}
Wimpy nodes can accommodate conventional storage subsystems of large
capacity, therefore they can store large ledgers. The storage subsystem type does
not significantly affect the I/O performance of Hyperledger and Parity.
\end{obs}

\subsection{Bootstrapping Performance}
\label{sec:bootstrap}

\begin{figure*}[t]
\centering
\begin{subfigure}{\subfigsizeb}
\centering
\includegraphics[width=0.99\textwidth]{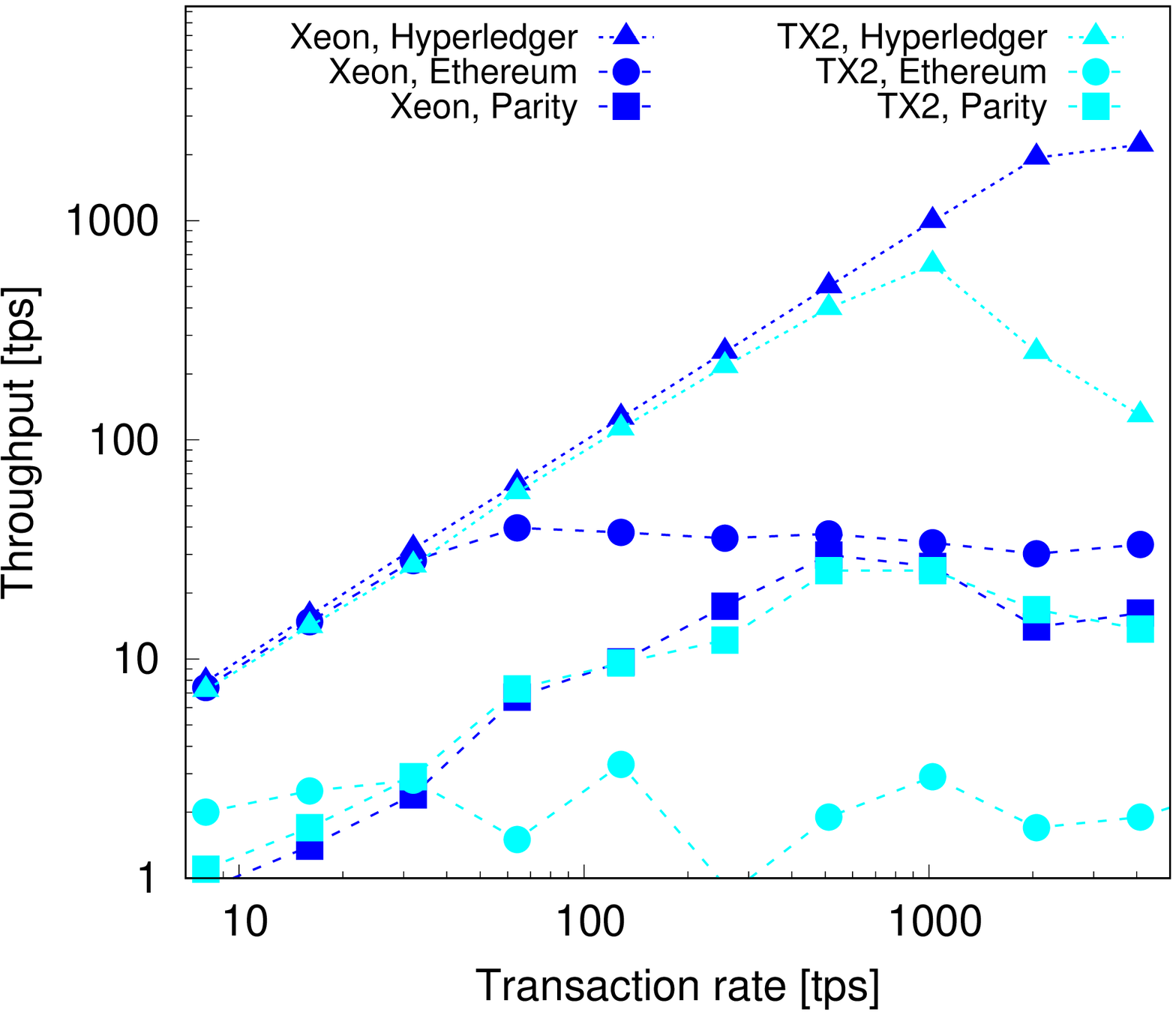}
\caption{Throughput}
\end{subfigure}
\begin{subfigure}{\subfigsizeb}
\centering
\includegraphics[width=0.96\textwidth]{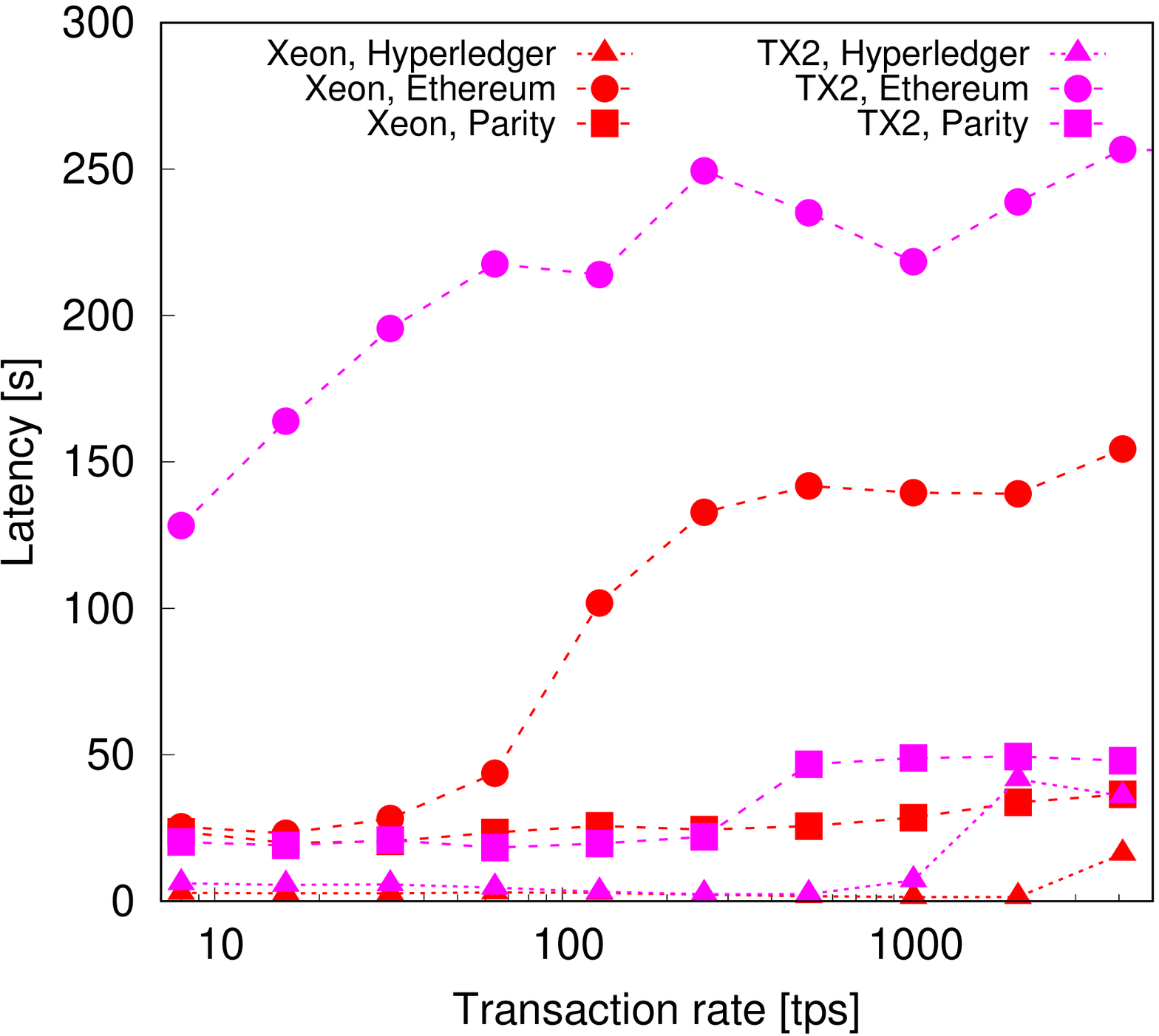}
\caption{Latency}
\end{subfigure}
\begin{subfigure}{\subfigsizeb}
\centering
\includegraphics[width=0.99\textwidth]{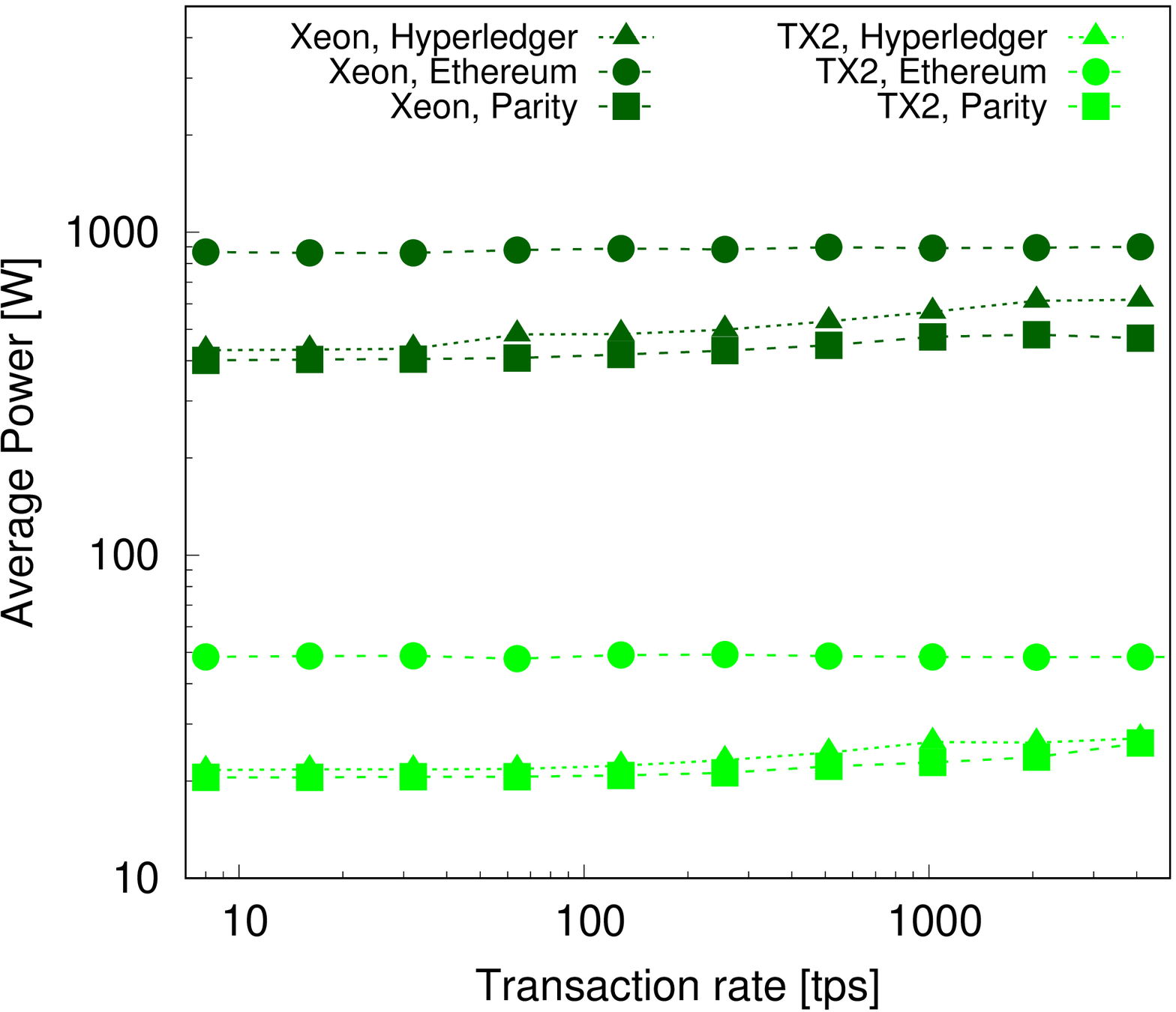}
\caption{Power}
\end{subfigure}
\caption{The performance of YCSB benchmark with increasing transaction rate}
\label{fig:cluster_vary_rate}
\figvspace
\end{figure*}

In this section, we analyze the performance of \textit{bootstrapping} which is
the process of joining a blockchain network and synchronizing the distributed
ledger. We consider one node that joins an existing network of seven other nodes
of the same type. Prior to the bootstrapping process, we generate over $100$
blocks by running the YCSB workload on the 8-node blockchain network. We then
stop one node, delete its ledger and caches, and restart it so that it
synchronizes the ledger with other nodes.

Hyperledger v0.6 adopts a lazy bootstrapping approach where the synchronization
is started when new transactions are submitted. Hence, the execution time and
power of synchronization and of transaction cannot be clearly separated. Here,
we report the time taken by Hyperledger to update its block tip to a certain
value. To synchronize around $2750$ blocks, Hyperledger on Xeon takes $40$s
while using 51.25W. Interestingly, TX2 is faster than Xeon: it takes less than
$20$s while using up to 3W. We attribute this to networking setup. In
particular, the Xeon cluster runs on NFS which adds some overhead.
We note that both systems use a relatively low power compared to their peak
power. This is because the blocks are downloaded from the other peers without
executing all transactions.

Ethereum supports three bootstrapping modes, fast, full and
light~\cite{eth_light}. In light mode, which is intended for wimpy systems, only
the current state is downloaded from other peers. In fast and full mode, all
blocks are downloaded. However, only in full mode are all the transactions
applied, which means it is slower than fast mode. In our experiments, we do not
consider light mode because it is very fast on both wimpy and brawny nodes.
Ethereum takes $14.8$s and $28.8$s to synchronize around $2000$ blocks in fast
and full mode on Xeon, respectively, while using 120W. On TX2, it takes $57.2$s
and 6.7W to synchronize in full mode, and only $4$s and 6W to synchronize in
fast mode.

By default, Parity uses fast (or \textit{warp}) synchronization which skips
``almost all of the block processing''~\cite{parity_sync}. However, we observed
that synchronizing the ledger in Parity takes much longer than Ethereum, even
when \textit{warp syncing} is on. In particular, synchronizing $100$ blocks in
Parity takes over 4 hours on Xeon and over 3 hours on TX2, whereas in Ethereum
it takes $2.6$s and $2.2$s on Xeon and TX2, respectively. This is a well-known
issue in Parity\footnote{For example, users report on StackExchange that
synchronizing with the main network in 2018 took few days
(\url{https://bit.ly/2UvIR1g})}, with some users blaming the I/O subsystem. But
our profiling shows that the peak I/O write rate is around 1MB/s which is much
lower than the available throughput of the storage system. Moreover, the power
during Parity's synchronization is close to the idle power: 51W and 2.4W on Xeon
and TX2, respectively. This shows that Parity is not doing much work during the
synchronization process.  We therefore conclude that the synchronization
inefficiency lies in Parity's implementation rather than in the hardware.

\begin{figure*}[!t]
\centering
\begin{subfigure}{\subfigsizeb}
\centering
\includegraphics[width=0.99\textwidth]{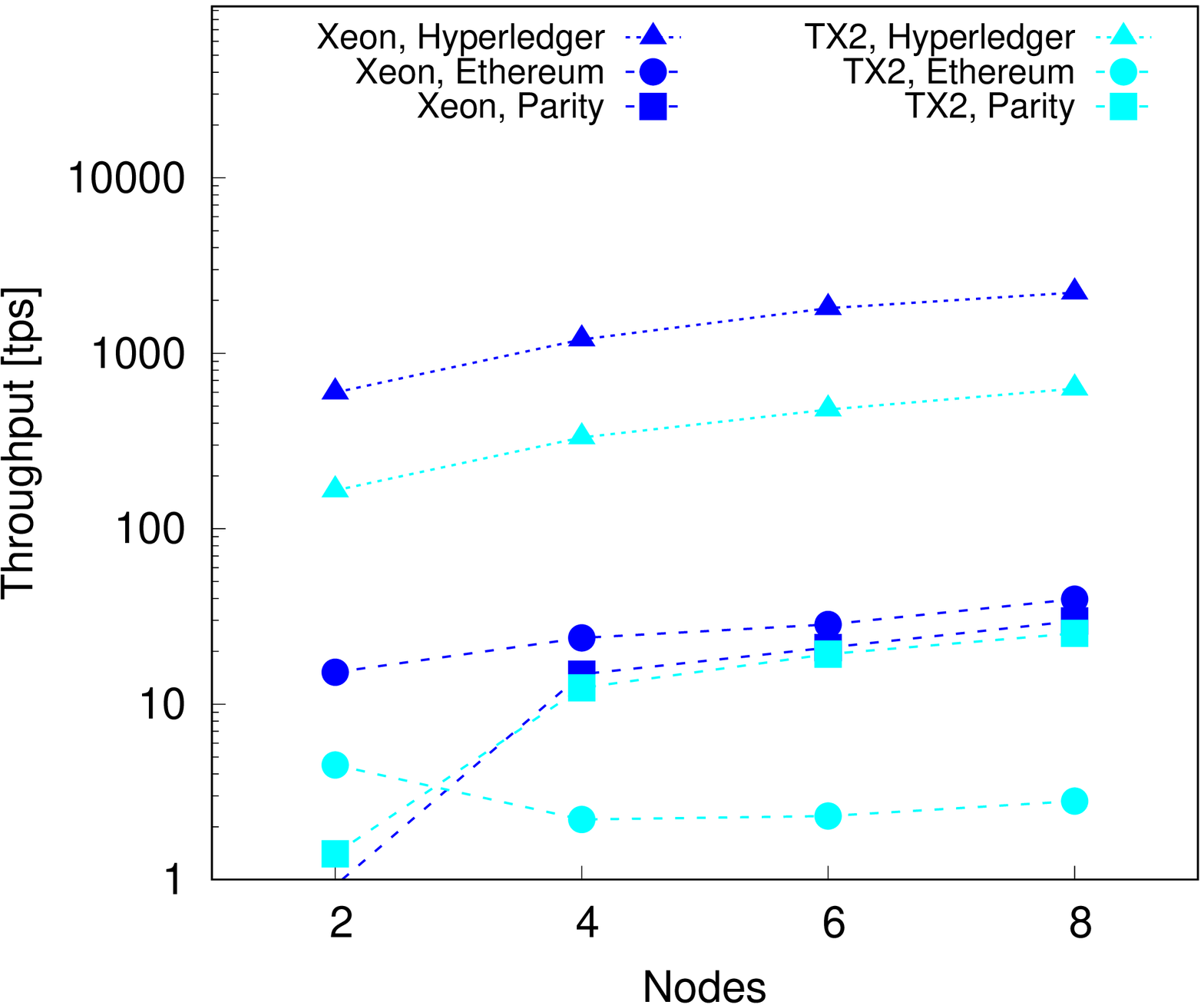}
\caption{Throughput}
\end{subfigure}
\begin{subfigure}{\subfigsizeb}
\centering
\includegraphics[width=0.96\textwidth]{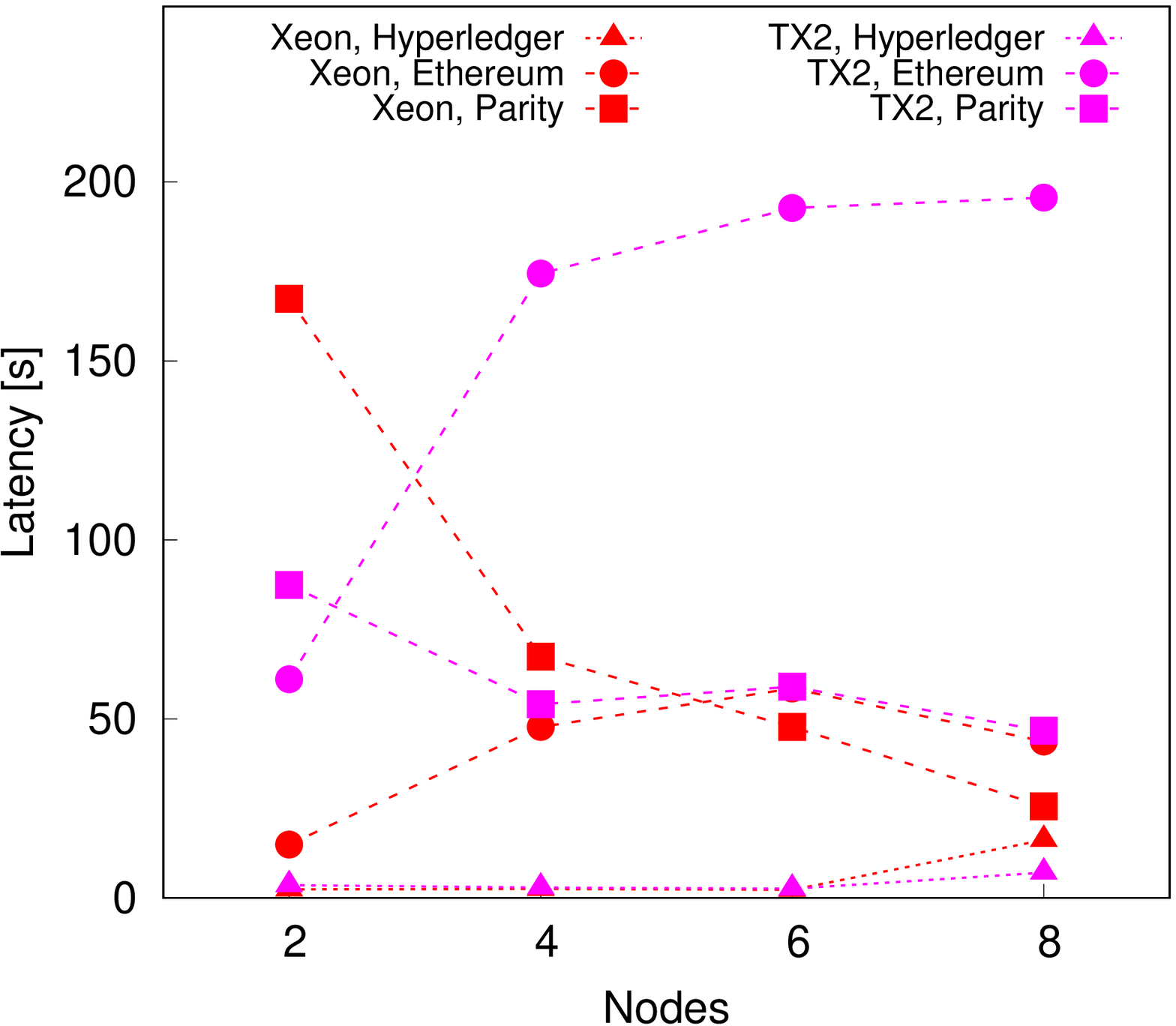}
\caption{Latency}
\end{subfigure}
\begin{subfigure}{\subfigsizeb}
\centering
\includegraphics[width=0.99\textwidth]{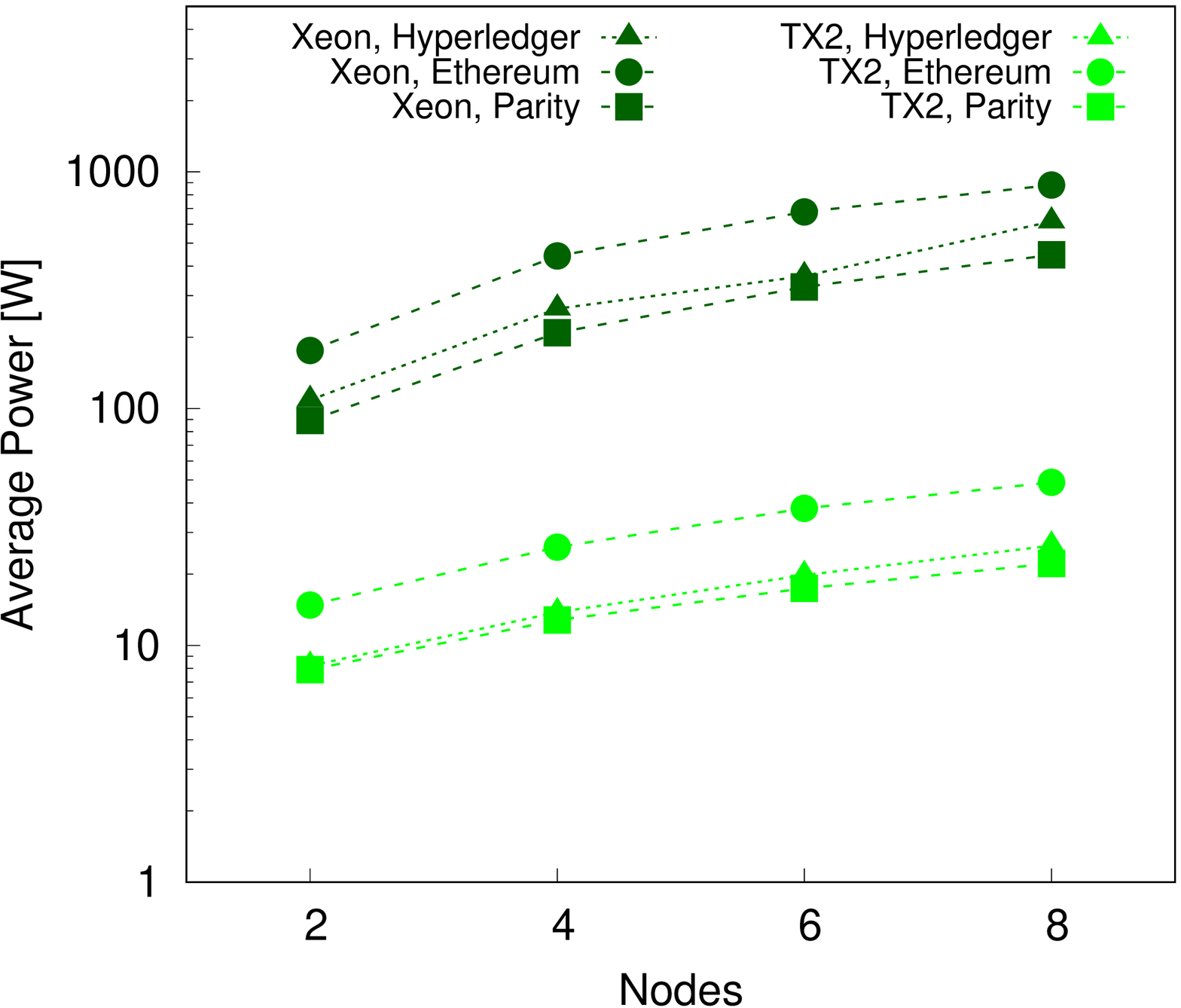}
\caption{Power}
\end{subfigure}
\caption{The performance of YCSB benchmark with increasing number of nodes}
\label{fig:cluster_vary_nodes}
\figvspace
\end{figure*}


\section[Cluster Analysis]{Cluster Analysis}
\label{sec:analysis_mn}

In this section, we analyze the time-energy performance of blockchains on a
cluster. We consider both homogeneous consisting of nodes of the same type, and
heterogeneous cluster consisting of multiple types of nodes.

\subsection{Homogeneous Cluster}
\label{sec:homo}
We consider Xeon-only and TX2-only clusters. The former is the faster, the
latter the most energy efficient.
We vary the cluster size from $2$ to $8$ nodes. The clients that issues requests
run on separate nodes, and unlike the analysis in~\cite{Sankaran_ICDCS_2018},
they are not included in our performance evaluation. Our main focus is on the
blockchain nodes.

\subsubsection{Impact of request rate}
We first examine the throughput, latency and power usage with increasing request
rate. We fix the cluster size to $8$ nodes, and use $8$ to send transactions. We
increase the transaction rate from $8$ to $4096$ transactions per second (tps).
The results, depicted in Figure~\ref{fig:cluster_vary_rate} for the YCSB
benchmark, show that Hyperledger is able to sustain a throughput of up to $2220$
tps and $630$ tps on Xeon and TX2, respectively. Ethereum achieves a throughput
of up to $39.7$ tps and only $3.3$ tps on Xeon and TX2, respectively.  Parity
achieves a maximum throughput of $30$ tps and $25$ tps on Xeon and TX2,
respectively, when the client request rate is 512tps.  Similar patterns are
observed when running Smallbank and Donothing benchmarks.

To achieve peak throughput, Hyperledger uses 618W on Xeon and only 26.4W on TX2.
Parity uses even less power, ranging between 400W and 480W on Xeon, and between
20W and 26W on TX2. In contrast, Ethereum uses the most power, between 860W and
900W on Xeon and around 49W on TX2.

These results can be summarized in the following observation.

\begin{figure}[t]
\centering
\includegraphics[width=0.42\textwidth]{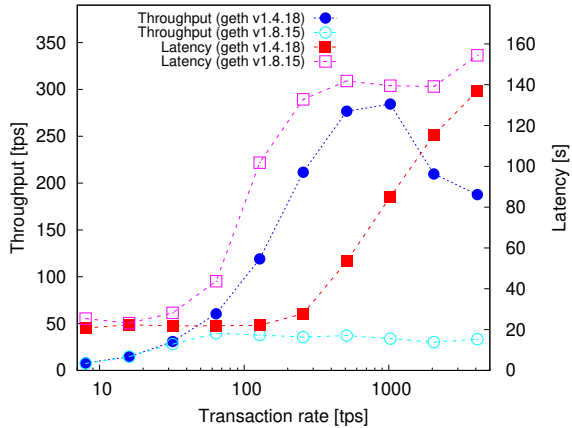}
\caption{Throughput and latency comparison between different versions of
Ethereum}
\label{fig:cluster_vary_rate_geth_comp}
\figvspace
\end{figure}

\begin{obs}
Higher-end wimpy nodes, such as Jetson TX2, achieve around one-third of
Hyperledger throughput and almost the same performance for Parity compared to
brawny Xeon nodes, while using $18\times$ to $23\times$ less power. These nodes
have potential of achieving significant power and cost savings.
\end{obs}

Standard deviation is relatively low in most of the cases, with the highest
outliers being the latency under high request rates. In particular,
Hyperledger's latency exhibits a standard deviation of $111.5$\% and $42.5$\% on
Xeon with $4096$ tps and TX2 with $1024$ tps request rate, respectively. For
throughput, the maximum standard deviation is $101.1\%$ for Ethereum on TX2 and
$30.8\%$ for Parity on Xeon. Power consumptions have low standard deviation:
below $1\%$ on Xeon and $4.5\%$ on TX2.

Ethereum execution on TX2 is irregular, as shown in
Figure~\ref{fig:cluster_vary_rate}, and has higher standard deviation compared
to the other two blockchains. Moreover, Ethereum throughput is much lower and
its latency is higher when using version \textit{v1.8.15} compared to
\textit{v1.4.18} evaluated in BLOCKBENCH~\cite{Dinh_SIGMOD_2017}. As shown in
Figure~\ref{fig:cluster_vary_rate_geth_comp} for YCSB, \textit{v1.4.18} achieves
a maximum of $284.4$ tps for a transaction request rate of $1024$ tps, while
\textit{v1.8.15} achieves only $39.7$ tps. The increase in latency is relatively
smaller, with maximum latencies of $137$ and $154$ seconds for \textit{v1.4.18}
and \textit{v1.8.15}, respectively.

\begin{figure}[t]
\centering
\includegraphics[width=0.4\textwidth]{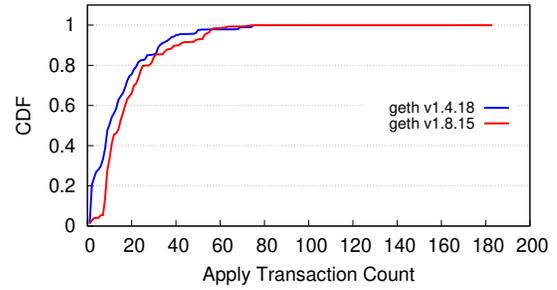}
\caption{Comparison of apply transaction count distribution in two versions of
Ethereum}
\label{fig:cluster_applytx_cdf}
\figvspace
\figvspace
\end{figure}

\begin{figure*}[!t]
\centering
\begin{subfigure}{\subfigsizea}
\centering
\includegraphics[width=0.43\textwidth,angle=270]{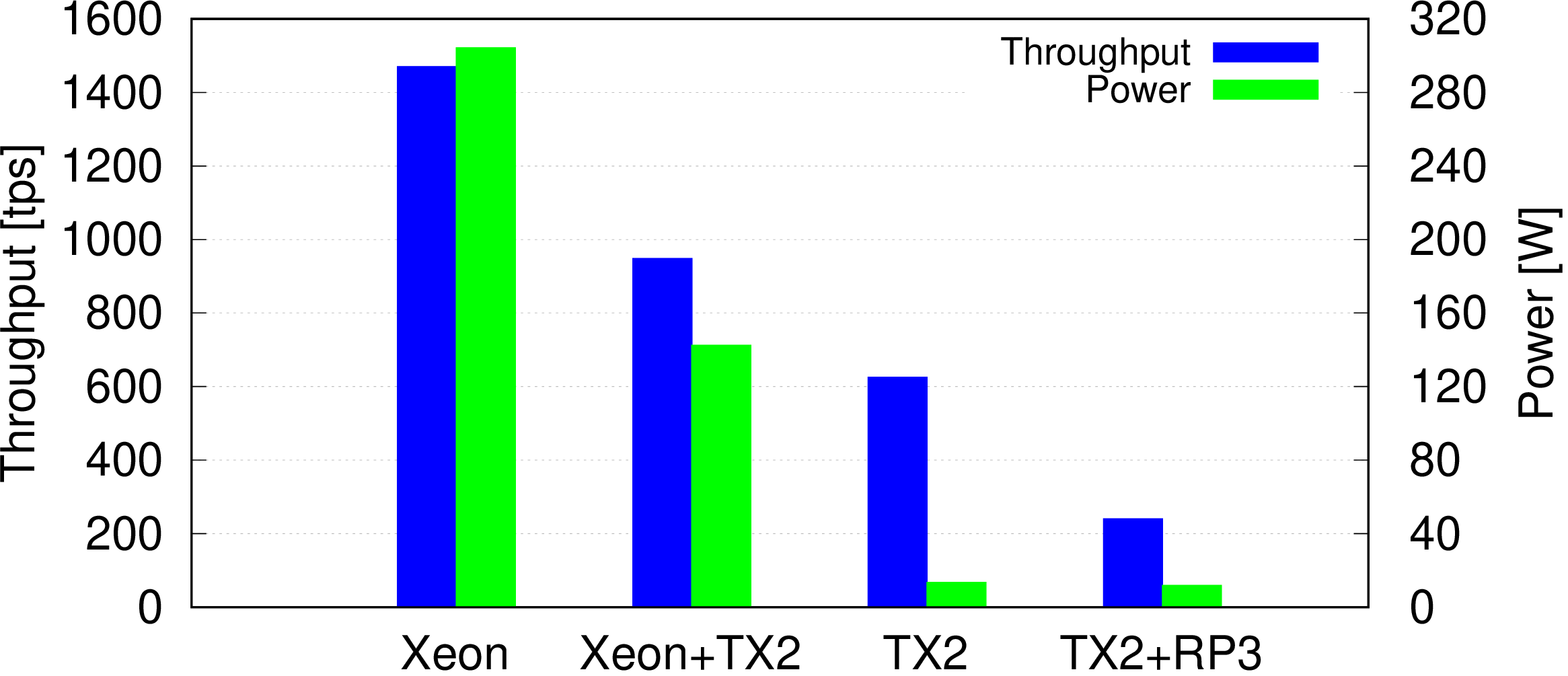}
\caption{Hyperledger}
\end{subfigure}
\begin{subfigure}{\subfigsizea}
\centering
\includegraphics[width=0.43\textwidth,angle=270]{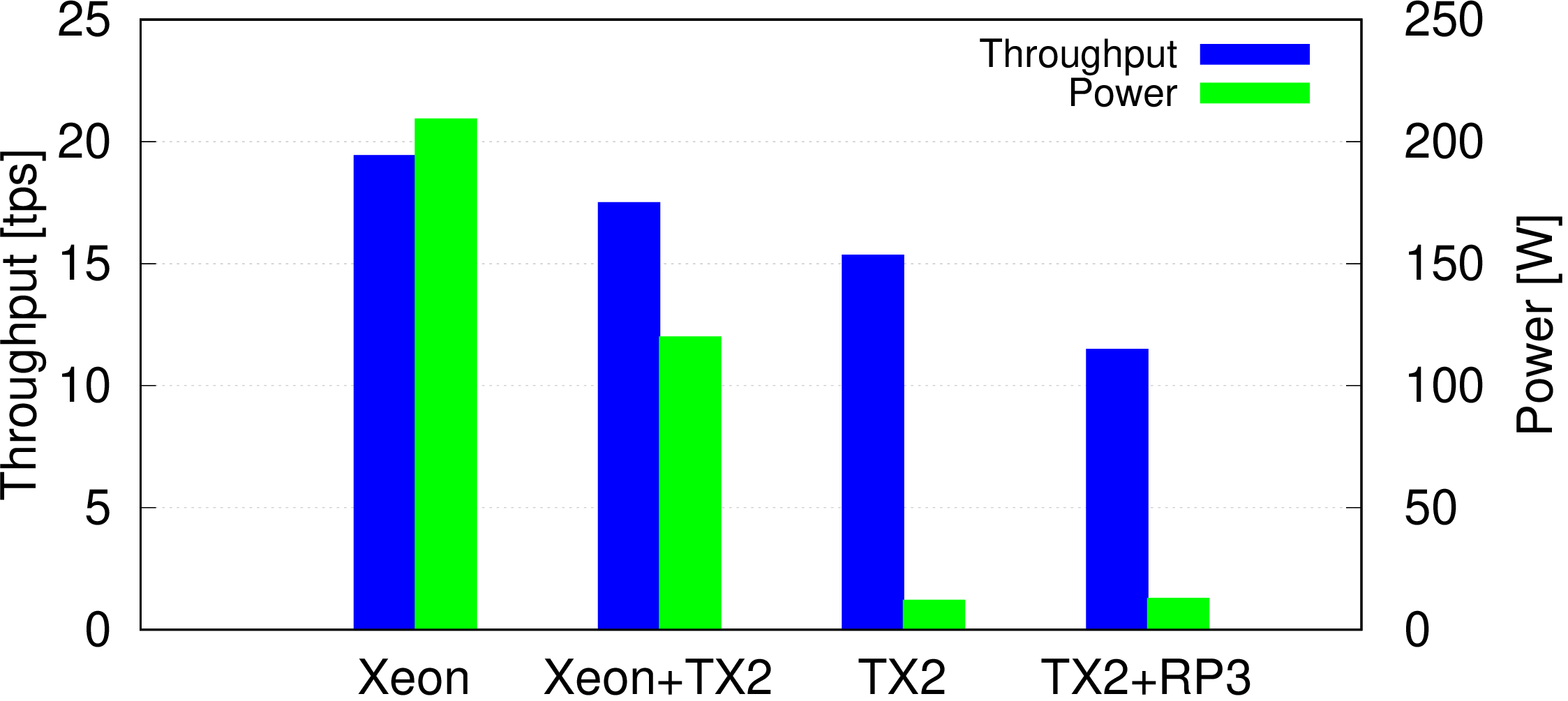}
\caption{Parity}
\end{subfigure}
\caption{The performance of YCSB on heterogeneous clusters with 4 nodes}
\label{fig:cluster_heterogeneous}
\figvspace
\end{figure*}

We note that the higher throughput reported for \textit{v1.4.18} is attributed
to (i) different parameter settings, and more fundamentally to (ii) a design
change in Ethereum. First, there are changes in gas values in the newer
versions. This requires to increase the gas value to $0x10000$ in order to run
YCSB benchmark on \textit{v1.8.15}. Second, a transaction may be restarted
multiple times in \textit{v1.8.15}, as discussed in Section~\ref{sec:eth}.

To understand the second factor contributing to the low throughput, we profile
the code to record the number of times an unique transaction, as represented by
its hash, is restarted, or \textit{applied}. Even though the average number of
times a transaction is applied is similar, which is 20 times, we observed that a
higher number of unique transactions are executed by \textit{geth v1.4.18} than
by \textit{geth v1.8.15}. These unique transactions are reflected in the
throughput. Furthermore, transactions are restarted more often in \textit{geth
v1.8.15}, as shown in Figure~\ref{fig:cluster_applytx_cdf}. The maximum number
of restarts in \textit{v1.8.15} is much higher than \textit{v1.4.18}, namely
$183$ times versus $105$ times.

\subsubsection{Impact of network size}

Next, we examine the scalability with increasing number of blockchain nodes and
clients. We use the same number of clients as the number of nodes. We choose a
request rate that saturates the systems, as identified in the previous section.
In particular, for Xeon we set the rate per client node to $512$, $8$ and $64$
tps for Hyperledger, Ethereum and Parity, respectively. On TX2, we set the rate
per client to $128$, $4$ and $64$ tps.

Figure~\ref{fig:cluster_vary_nodes} shows the throughput for YCSB with
increasing number of nodes. We attribute the fluctuations of Ethereum on TX2 to
the non-deterministic transaction restarting mechanism. The lower throughput,
when compared to Xeon, is due to the compute-intensive PoW consensus protocol.
In fact, the power usage of Ethereum is $2\times$ higher than Hyperledger and
Parity on TX2.  Specifically, $6$ TX2 nodes use, on average, 37.9W, 19.8W and
17.4W when running Ethereum, Hyperledger and Parity, respectively.

The latency of Ethereum increases significantly on TX2, from 46.7s on $2$ nodes
to 195.6s on $8$ nodes. This is $4.5\times$ higher than the latency on $8$ Xeon
nodes. On the other hand, Parity's latency decreases with the number of nodes:
from $87.4$s on $2$ nodes to $46.7$s on $8$ nodes.  In summary, Ethereum is
virtually unusable on wimpy systems due to (i) low throughput and high latency
caused by PoW consensus, and (ii) unstable performance due to transaction
restarting.

\subsection{Heterogeneous Cluster}
\label{sec:hetero}

In this section we examine the effects of heterogeneous nodes on the overall
blockchain performance. The baselines of homogeneous clusters are represented by
(i) $4$ Xeon nodes and (ii) $4$ TX2 nodes. From the homogeneous Xeon cluster, we
replace two nodes with TX2 (\textbf{Xeon+TX2}); from the homogeneous TX2 cluster
we replace two nodes with RP3 (\textbf{TX2+RP3}). We run the distributed
benchmarks for Hyperledger and Parity. Ethereum is left out because it cannot be
run on RP3.

As shown in Figure~\ref{fig:cluster_heterogeneous} for the peak throughput of
YCSB, the performance degrades when lower-performance nodes are introduced. But
the power consumption improves because the heterogeneous cluster uses less
power. In particular, Xeon+TX2 has a performance drop of 35\% but uses 53\% less
power than the homogeneous Xeon cluster, when running Hyperledger.
The results are better for Parity, where a $43\%$ power savings causes only
$10\%$ loss of throughput. However, adding RP3 nodes to a TX2 cluster does not
yield satisfactory results. For Hyperledger, the throughput drops 62\%, while
the power decreases only slightly from 13.4W to 11.8W (or only 12\% power
savings). For Parity, the power consumption of the heterogeneous cluster is even
higher than the homogeneous cluster, 12.8W versus 12W, while the throughput
drops from $15.3$ tps to $11.5$ tps.

Similar to the analysis of homogeneous clusters, the results here demonstrate
that higher-end wimpy nodes have the potential of reducing power usage, while
achieving reasonable performance. However, heterogeneous clusters with wimpy
nodes may not always achieve the best PPR. More specifically, if the performance
gap between different types of nodes is too large, the low-power profile of the
wimpy nodes does not lead to better energy efficiency due to lower throughput
and increasing latency.

\section{Conclusions}
\label{sec:concl}

In this paper, we performed an extensive time-energy analysis of representative
blockchain workloads on low-power, wimpy nodes in comparison with traditional
brawny nodes. The wimpy nodes used in our analysis cover the low-end and
high-end performance spectrum, and both x86/64 and ARM architectures.

We found that higher-end wimpy nodes achieve reasonable performance with
significantly lower energy than brawny nodes. In particular, a Jetson TX2
cluster with eight nodes achieves more than 80\% and almost 30\% of Parity and
Hyperledger throughput, respectively, while using $18\times$ and $23\times$ less
power, respectively, than an 8-node Xeon cluster.

We also found that wimpy nodes with well-balanced PPR achieve higher energy
efficiency compared to extremely low-power nodes. For example, a TX2 is more
energy efficient than a Raspberry Pi 3, even though the former has an idle power
of 2.4W and a peak power of more than 10W, while the latter has 2W and 5W,
respectively. The better energy efficiency of TX2 compared to RP3 is due to its
higher performance while keeping a low power profile at subsystem level,
including the CPU, memory and storage.

Finally, we found that recent versions of Ethereum suffer from low and unstable
performance. It is due to the transaction restarting mechanism that stops and
discards transaction execution whenever new blocks are received, even if those
blocks are empty. This fact, together with the high cost of the PoW consensus
protocol, make Ethereum unusable on wimpy nodes.

\bibliographystyle{abbrv}

\balance

\clearpage

\begin{appendix}

\section{Additional Results}

\begin{figure}[t]
\centering
\begin{subfigure}{\subfigsizeb}
\centering
\includegraphics[width=0.99\textwidth]{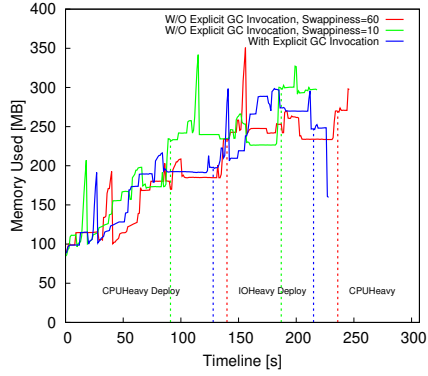}
\caption{Run 1}
\label{fig:gc_mem_rp3_extra_r1}
\end{subfigure}
\begin{subfigure}{\subfigsizeb}
\centering
\includegraphics[width=0.99\textwidth]{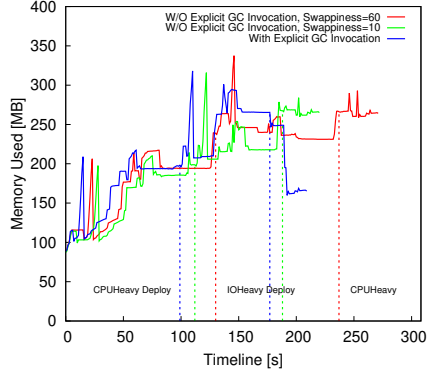}
\caption{Run 2}
\end{subfigure}
\begin{subfigure}{\subfigsizeb}
\centering
\includegraphics[width=0.99\textwidth]{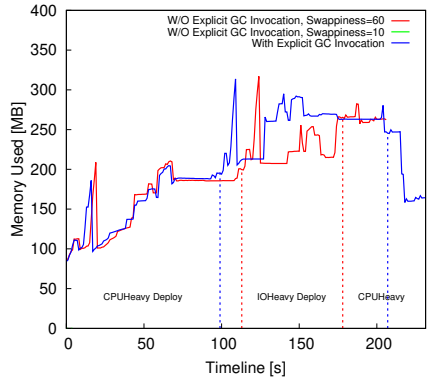}
\caption{Run 3}
\end{subfigure}
\caption{Hyperledger memory usage on RP3 (multiple runs)}
\label{fig:gc_mem_rp3_extra}
\end{figure}

\textbf{Hyperledger on RP3.} Figure~\ref{fig:gc_mem_rp3_extra} compares three
different runs of Hyperledger with and without explicitly calling Go's garbage
collector on RP3. Figure~\ref{fig:gc_mem_rp3_extra_r1} represents the same
execution plotted in detail in Figure~\ref{fig:gc_mem_rp3}. In almost all cases,
Hyperledger with explicit GC invocation uses less memory and is as fast, if not
faster, than Hyperledger without explicit GC invocation. On the one hand, the GC
incurs more \texttt{mmap/munmap} system calls. On average across our
experiments, Hyperledger with explicit GC incurs 70 \texttt{mmap} and 4
\texttt{munmap} calls, respectively, while Hyperledger without GC invocation
incurs 50 \texttt{mmap} and 2 \texttt{munmap} calls, respectively.

\begin{figure}[t]
\centering
\includegraphics[width=0.18\textwidth,angle=270]{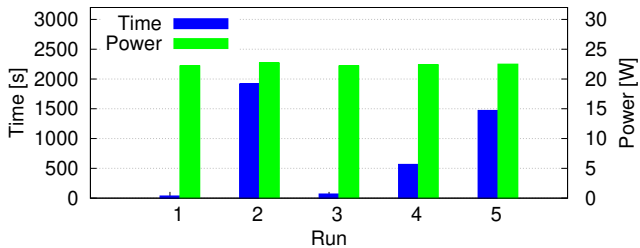}
\caption{Difference in Ethereum CPUHeavy execution time among different runs}
\label{fig:eth_runs}
\end{figure}

\textbf{Ethereum.} Figure~\ref{fig:eth_runs} compares the execution time and
power usage of different runs of the same CPUHeavy workload on Ethereum with
four miner threads, when running on the NUC node. We observe significant
execution time differences, while the power is roughly constant at around 23W.
Compared to the idle power of 9W and the CoreMark power of 18.6W, Ethereum's
power usage is higher, suggesting that the system is doing heavy work not only
at CPU level, but also at memory and I/O.

Table~\ref{table:eth_cluster_runs_apply} compares the number of times
transactions are re-started (applied) in two versions of Ethereum on a cluster
setup with varying request rate. We present the minimum, maximum and average
times, with standard deviation, across all unique transactions. We also show how
many unique transactions are executed and the total number of times
\texttt{ApplyTransaction()} method is called.

\begin{figure*}[!t]
\centering
\begin{subfigure}{\subfigsizeb}
\centering
\includegraphics[width=0.99\textwidth,angle=270]{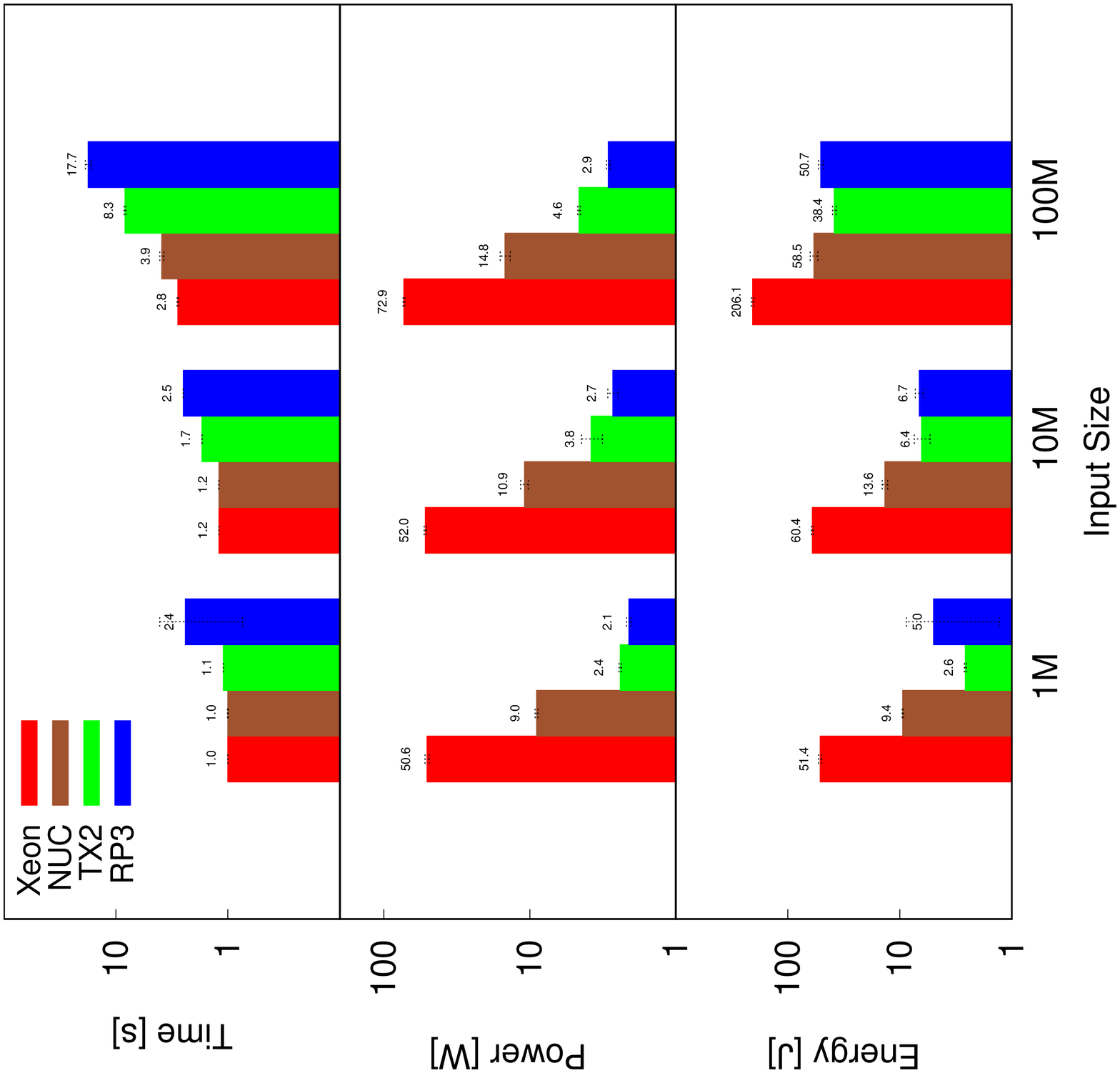}
\caption{CPUHeavy}
\end{subfigure}
\begin{subfigure}{\subfigsizeb}
\centering
\includegraphics[width=0.99\textwidth,angle=270]{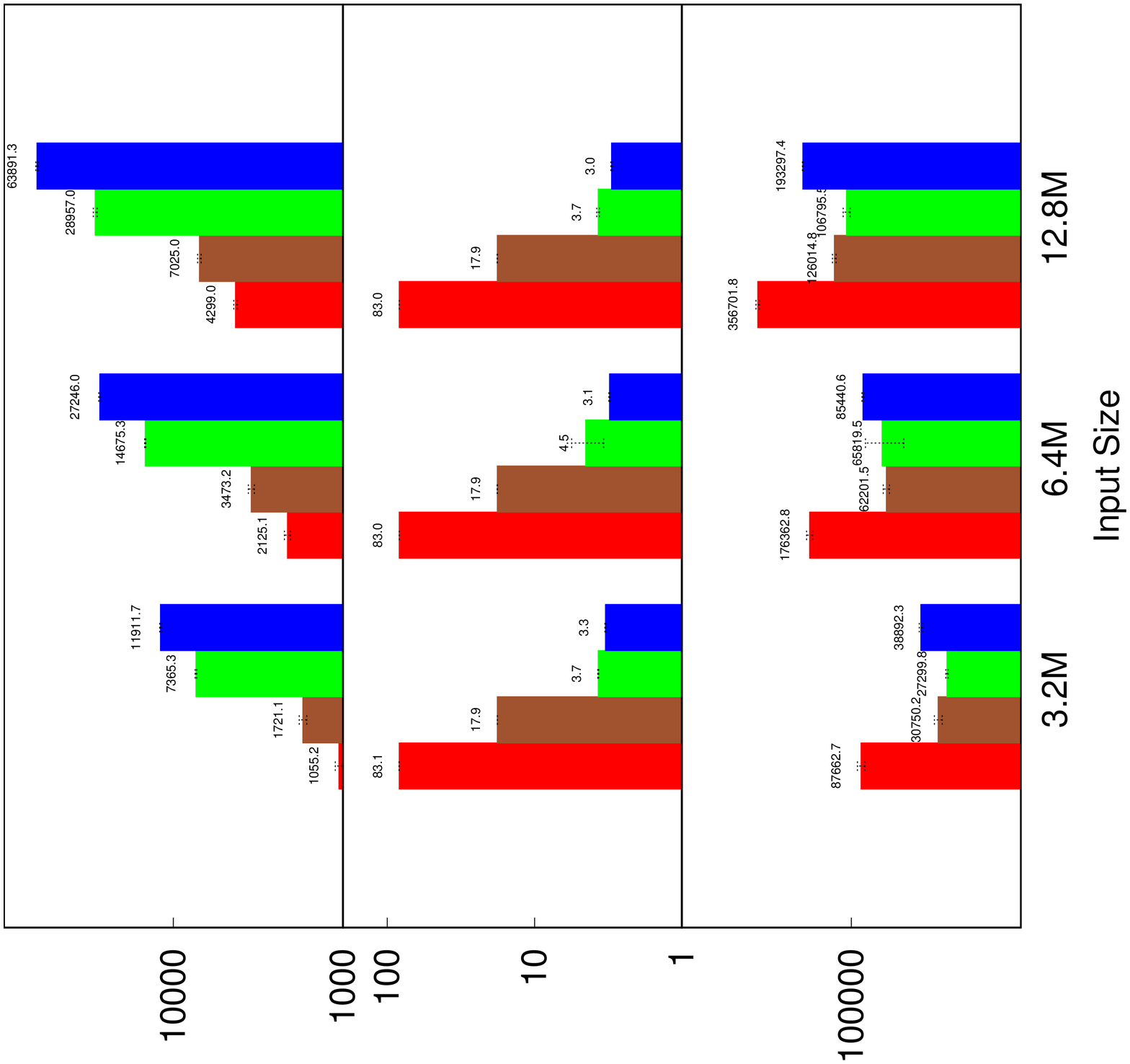}
\caption{IOHeavy Write}
\end{subfigure}
\begin{subfigure}{\subfigsizeb}
\centering
\includegraphics[width=0.99\textwidth,angle=270]{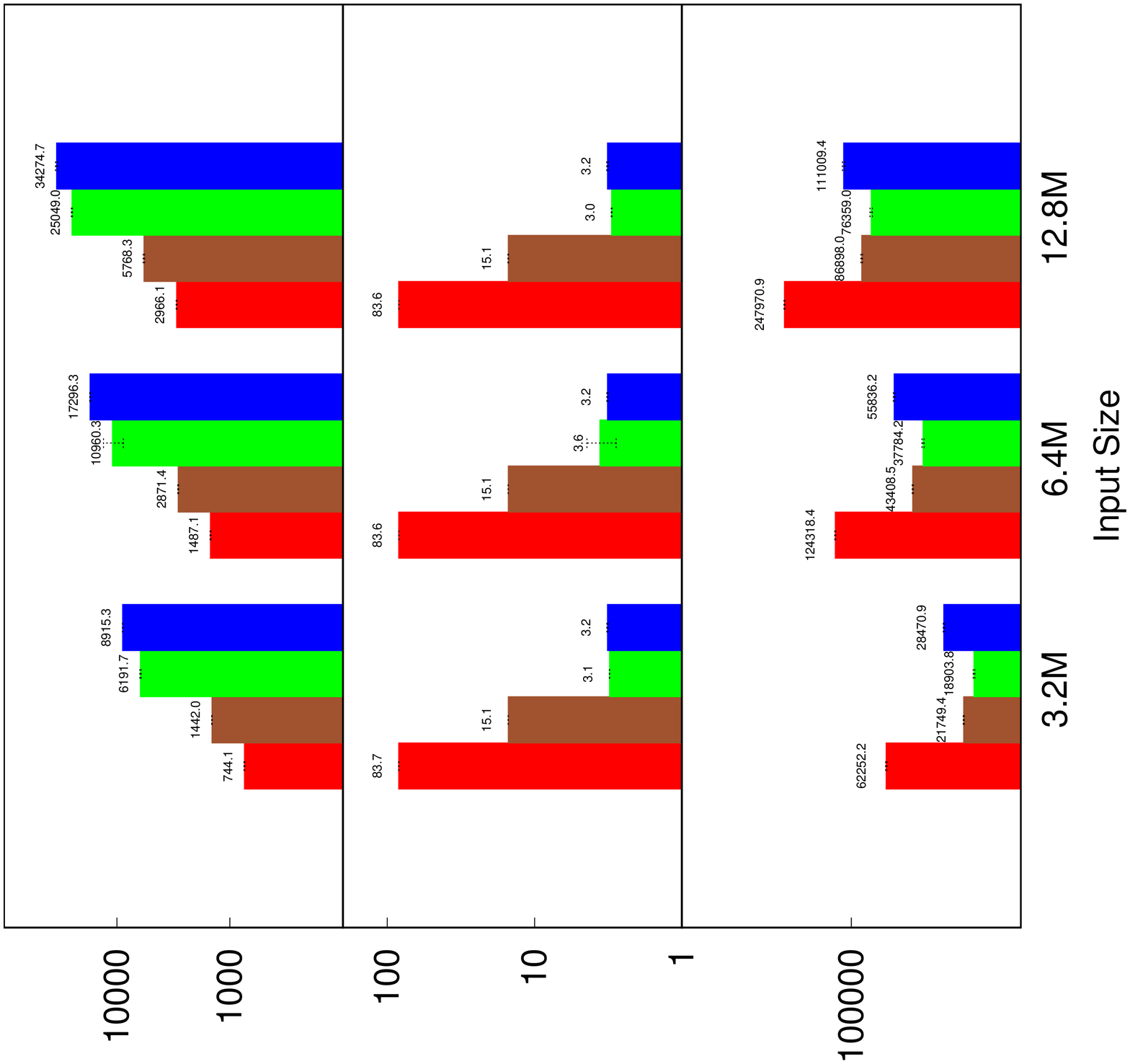}
\caption{IOHeavy Scan}
\end{subfigure}
\caption{The time-power-energy of Hyperlegder}
\label{fig:hl_tpe}
\end{figure*}

\begin{figure*}[!t]
\centering
\begin{subfigure}{\subfigsizeb}
\centering
\includegraphics[width=0.99\textwidth,angle=270]{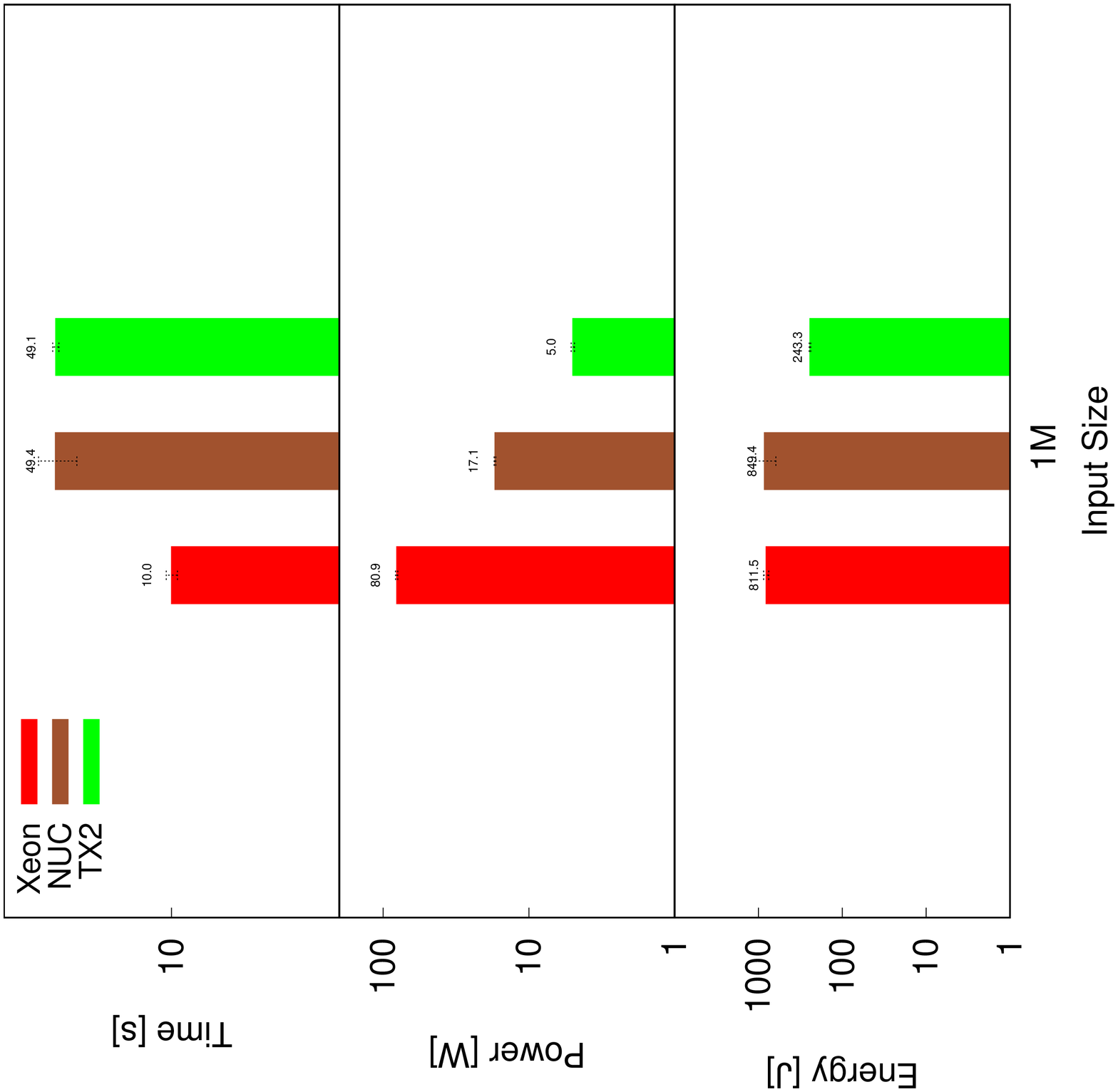}
\caption{CPUHeavy}
\end{subfigure}
\begin{subfigure}{\subfigsizeb}
\centering
\includegraphics[width=0.99\textwidth,angle=270]{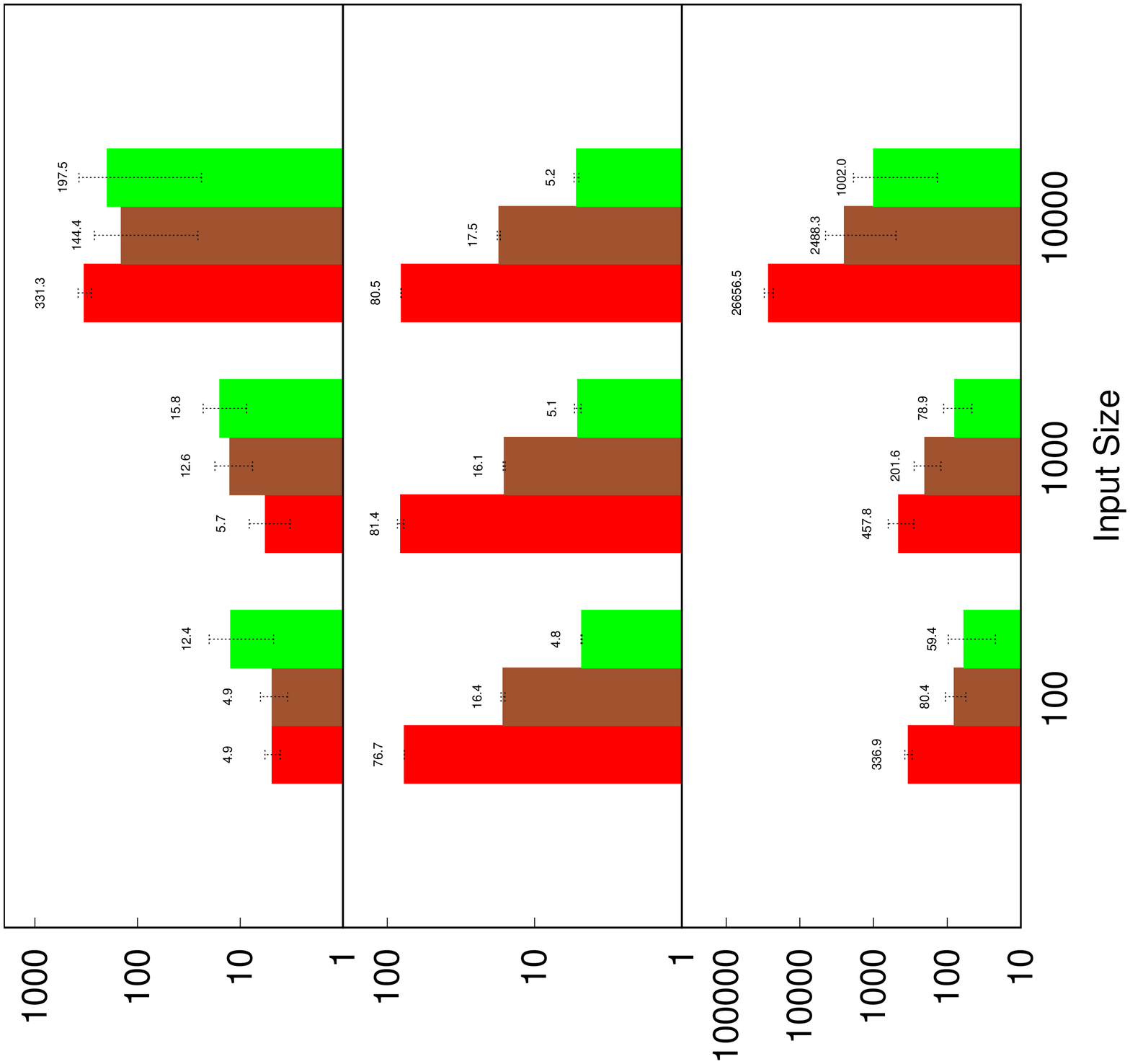}
\caption{IOHeavy Write}
\end{subfigure}
\begin{subfigure}{\subfigsizeb}
\centering
\includegraphics[width=0.99\textwidth,angle=270]{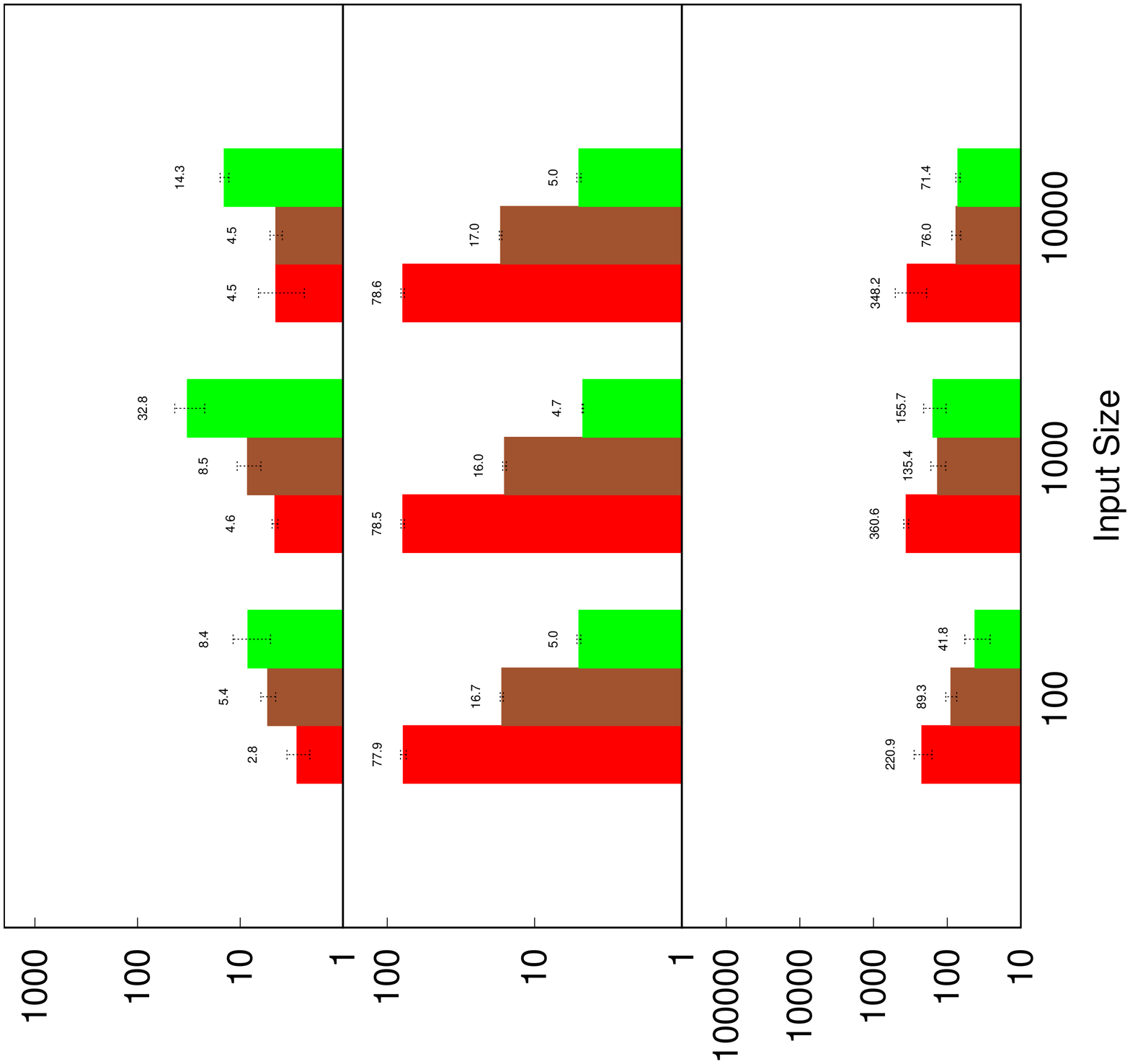}
\caption{IOHeavy Scan}
\end{subfigure}
\caption{The time-power-energy of Ethereum (with one miner thread)}
\label{fig:eth_tpe}
\end{figure*}

\begin{figure*}[!t]
\centering
\begin{subfigure}{\subfigsizeb}
\centering
\includegraphics[width=0.99\textwidth,angle=270]{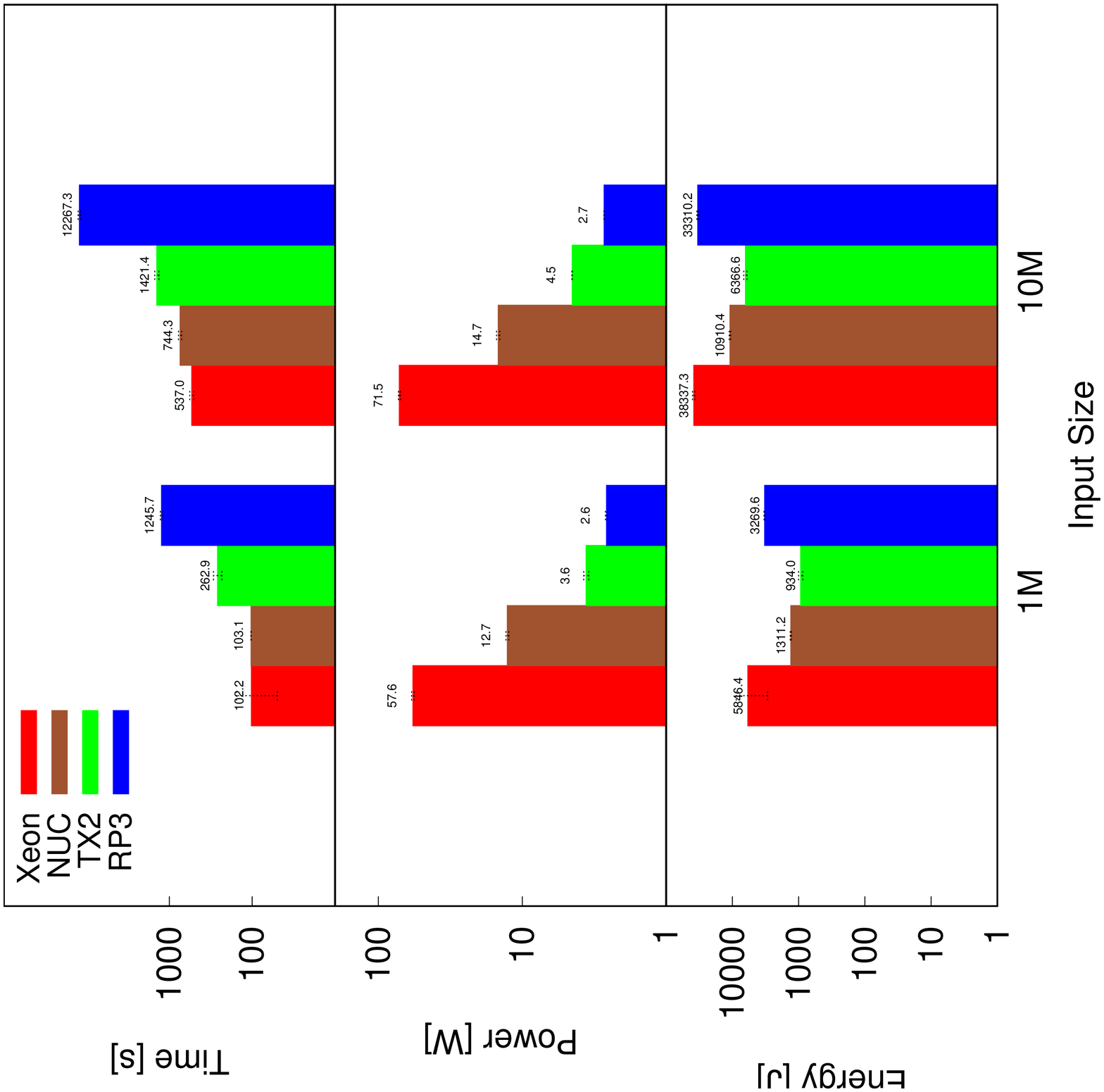}
\caption{CPUHeavy}
\end{subfigure}
\begin{subfigure}{\subfigsizeb}
\centering
\includegraphics[width=0.99\textwidth,angle=270]{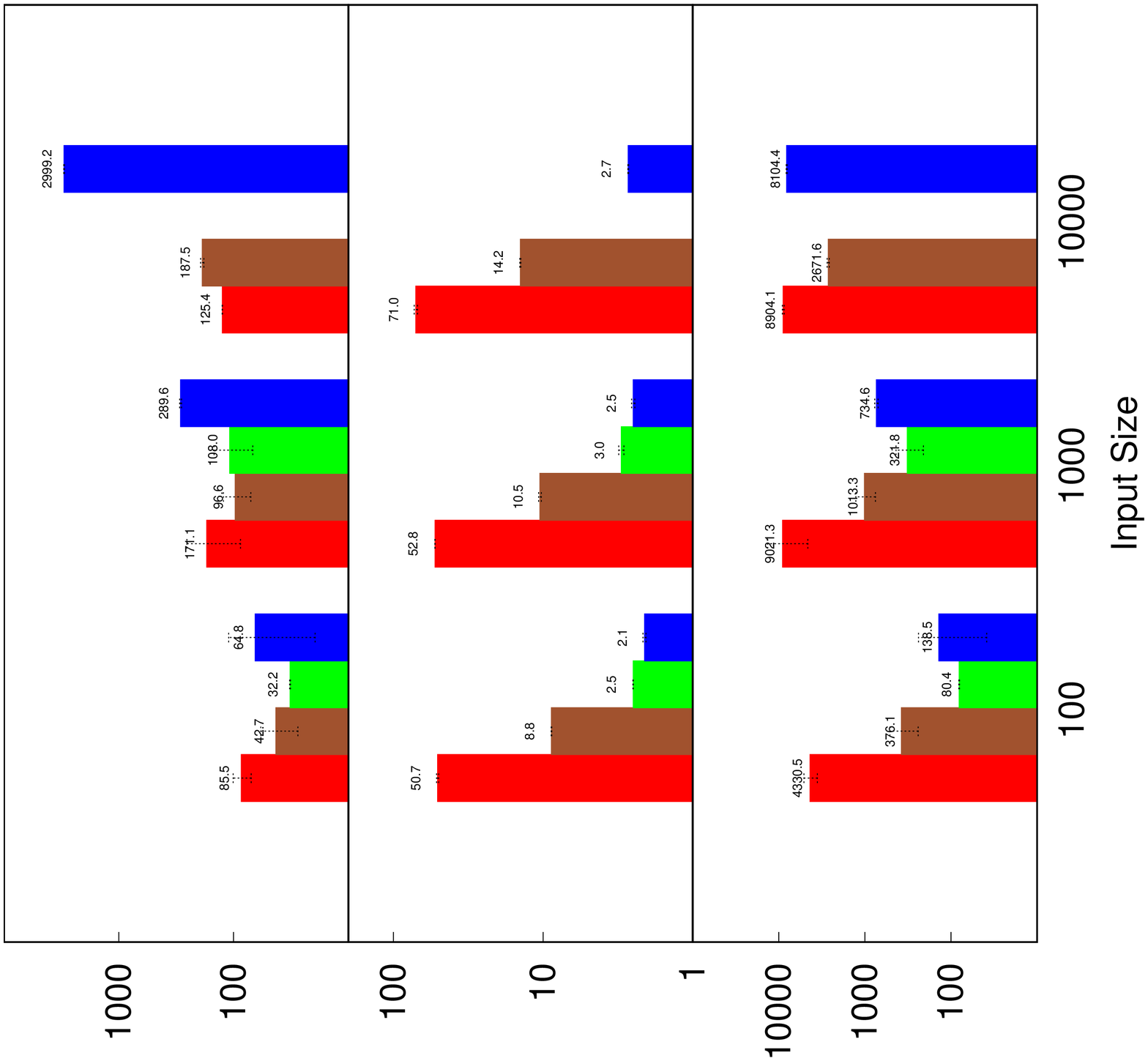}
\caption{IOHeavy Write}
\end{subfigure}
\begin{subfigure}{\subfigsizeb}
\centering
\includegraphics[width=0.99\textwidth,angle=270]{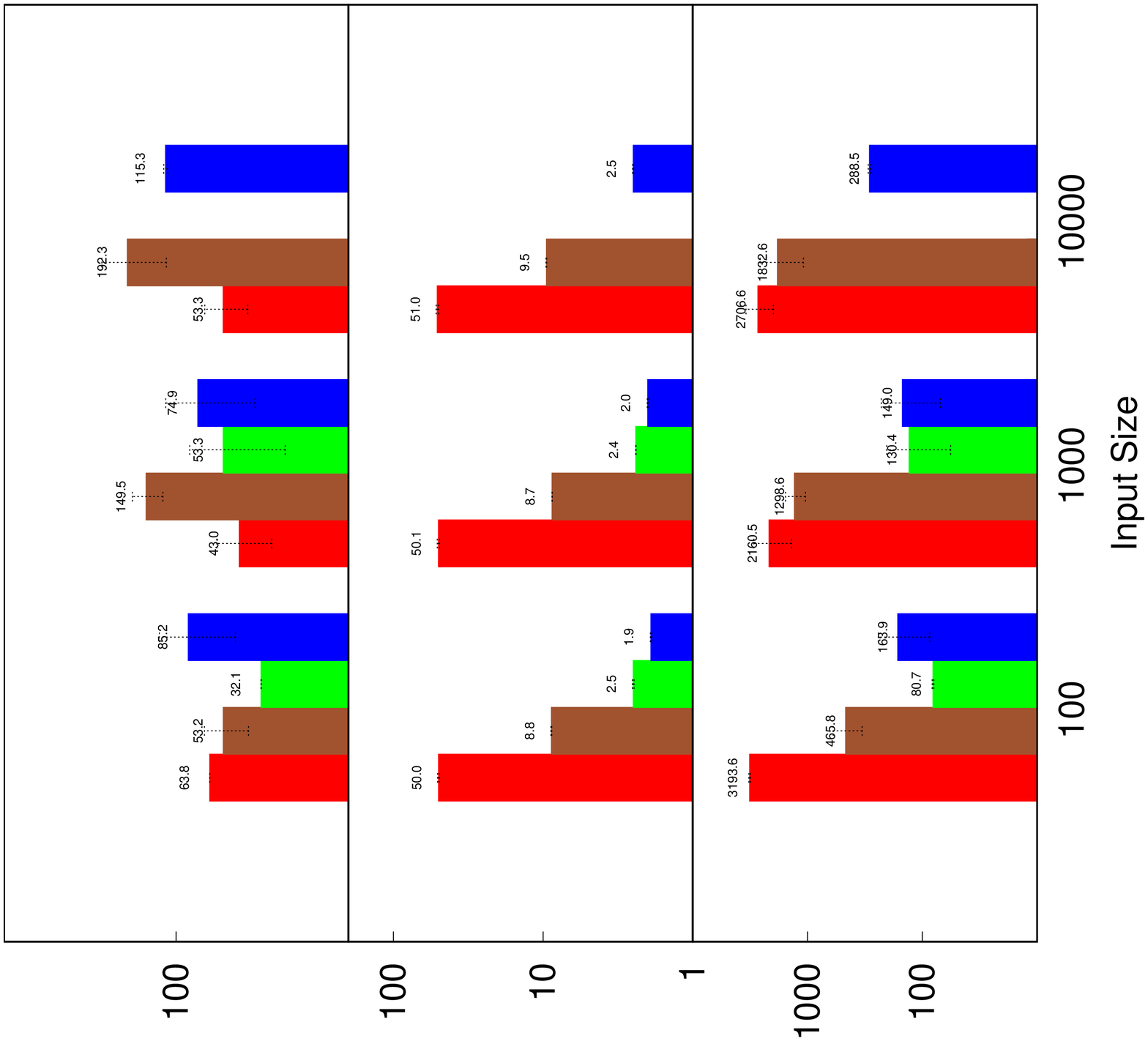}
\caption{IOHeavy Scan}
\end{subfigure}
\caption{The time-power-energy of Parity}
\label{fig:parity_tpe}
\end{figure*}

\textbf{Single-node Time-Power-Energy.}~~~Execution time, power usage and total
energy of CPUHeavy and IOHeavy workloads are plotted in
Figures~\ref{fig:hl_tpe}, \ref{fig:eth_tpe} and \ref{fig:parity_tpe} for
Hyperledger, Ethereum and Parity, respectively.

\begin{figure*}[t]
\centering
\begin{subfigure}{\subfigsizeb}
\centering
\includegraphics[width=0.99\textwidth]{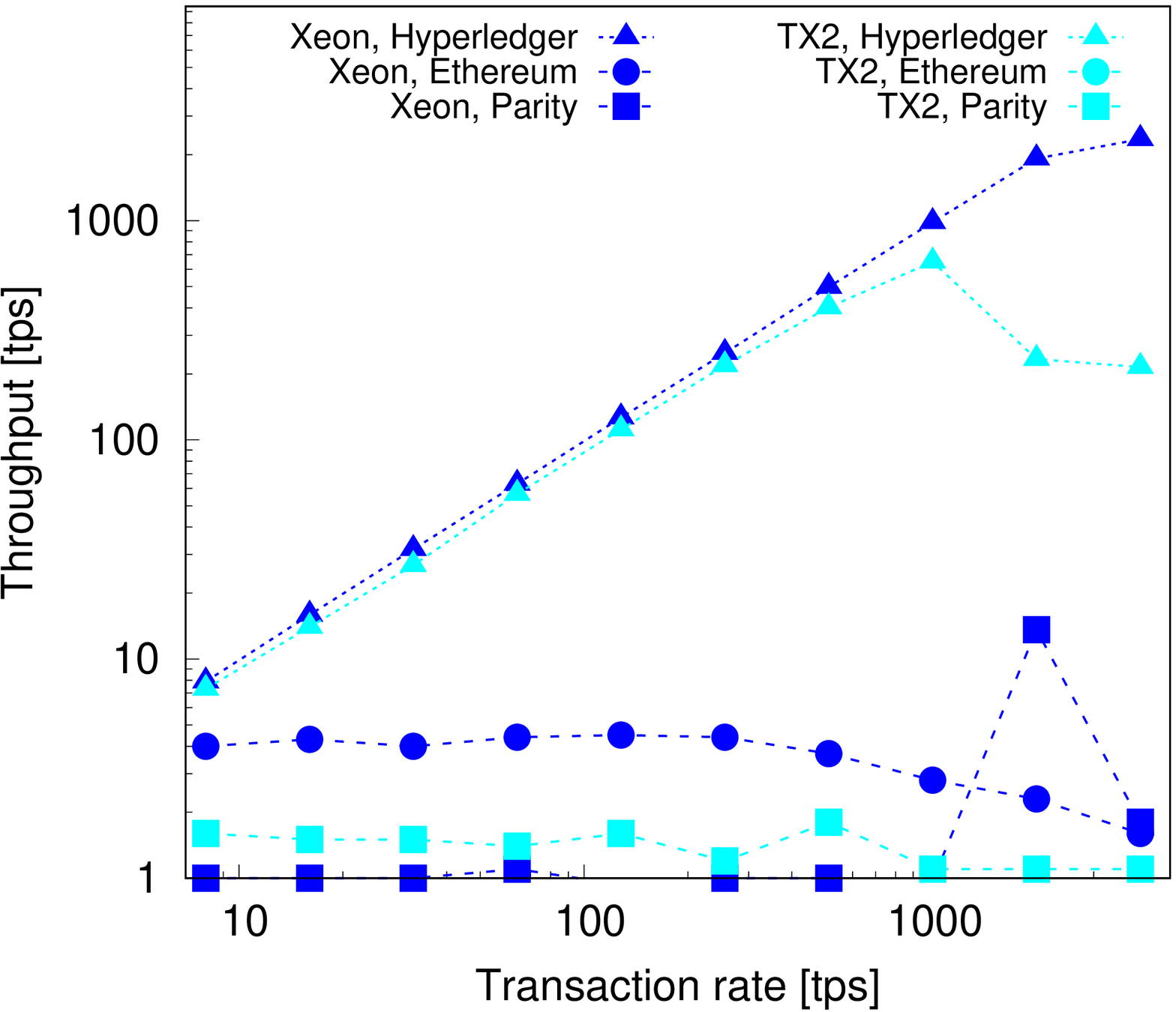}
\caption{Throughput}
\end{subfigure}
\begin{subfigure}{\subfigsizeb}
\centering
\includegraphics[width=0.99\textwidth]{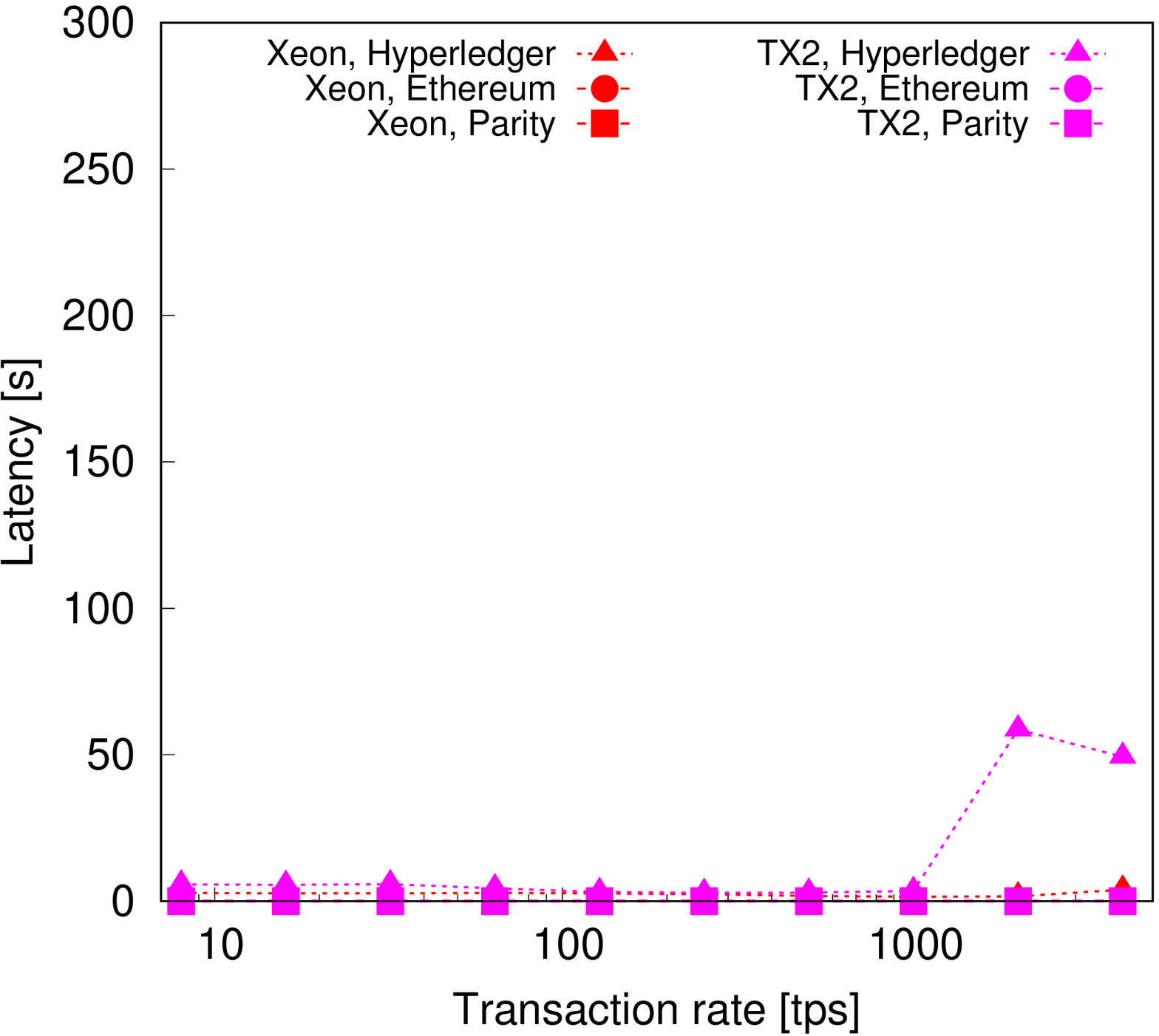}
\caption{Latency}
\end{subfigure}
\begin{subfigure}{\subfigsizeb}
\centering
\includegraphics[width=0.99\textwidth]{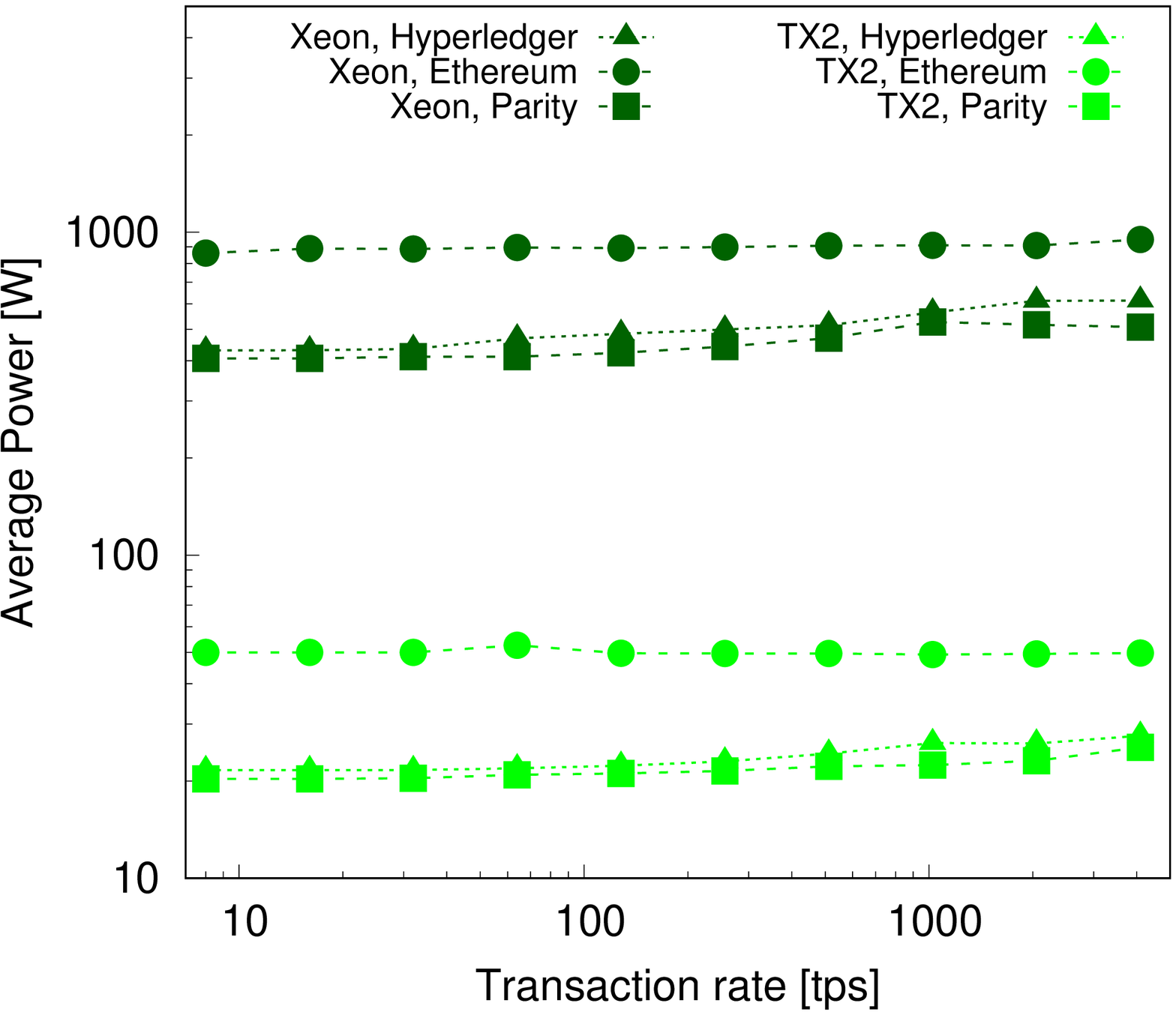}
\caption{Power}
\end{subfigure}
\caption{The performance of Smallbank benchmark with increasing transaction
rate}
\label{fig:cluster_smallbank_vary_rate}
\end{figure*}

\begin{figure*}[t] \centering
\begin{subfigure}{\subfigsizeb}
\centering
\includegraphics[width=0.99\textwidth]{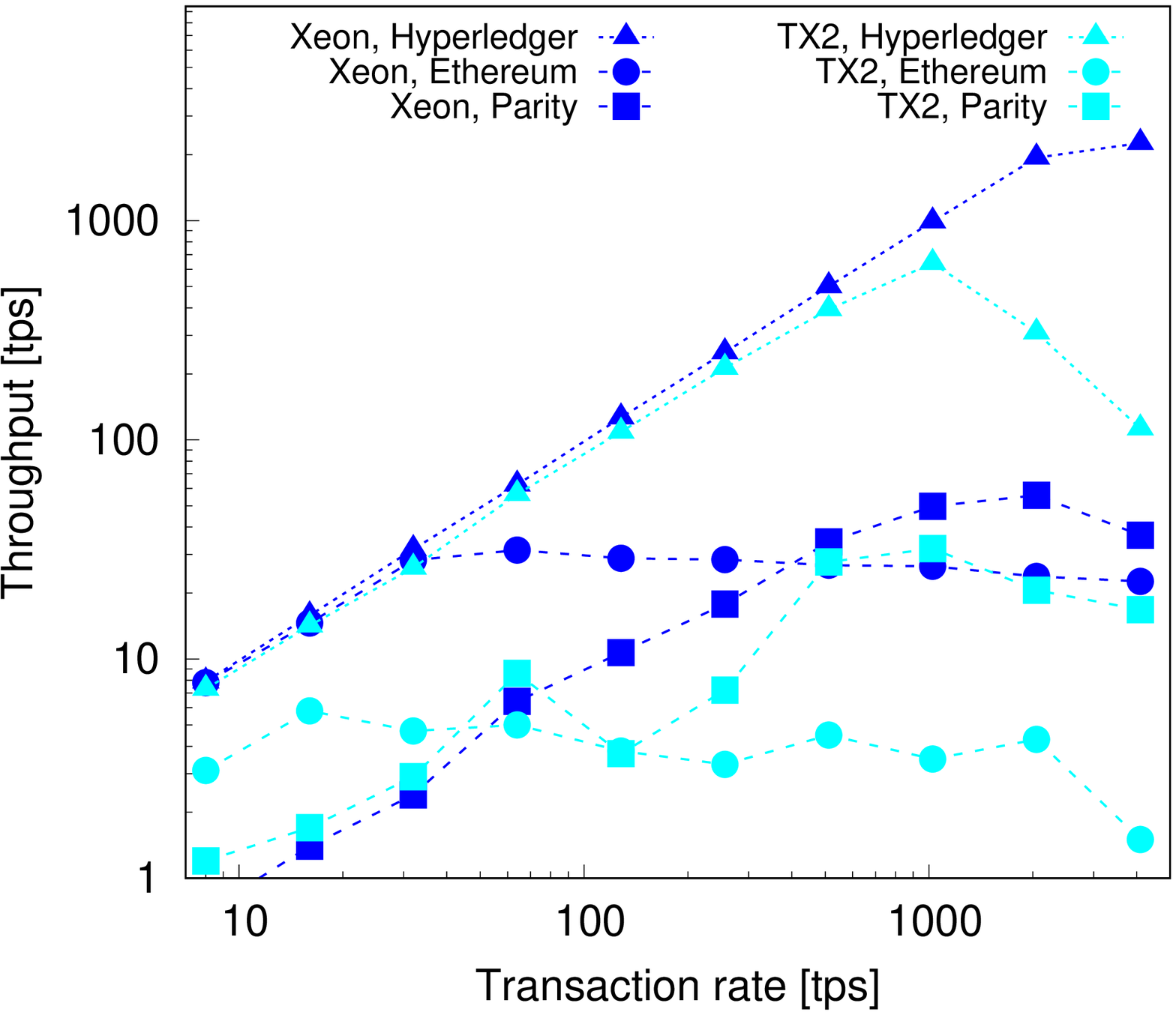}
\caption{Throughput}
\end{subfigure}
\begin{subfigure}{\subfigsizeb}
\centering
\includegraphics[width=0.99\textwidth]{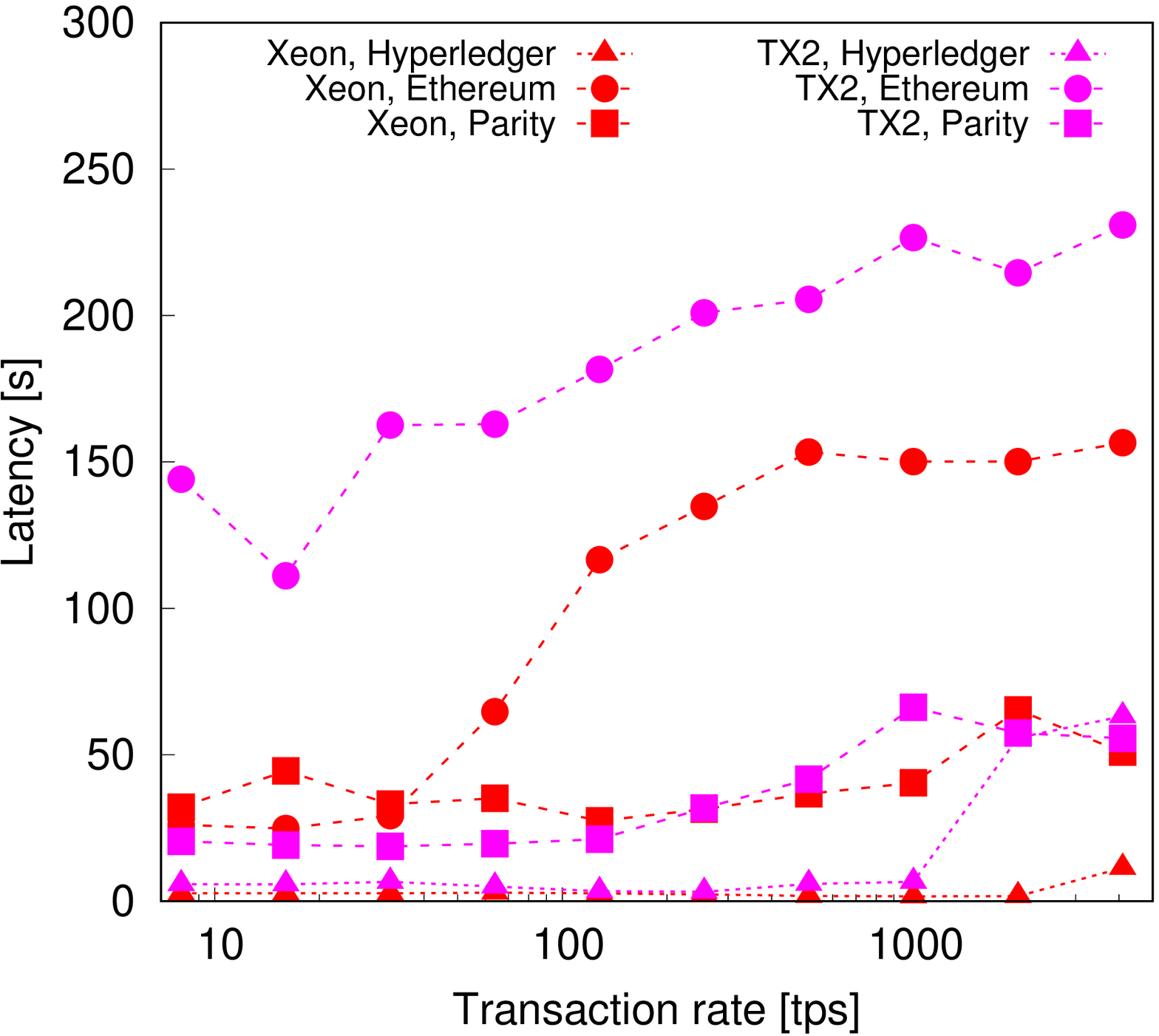}
\caption{Latency}
\end{subfigure}
\begin{subfigure}{\subfigsizeb}
\centering
\includegraphics[width=0.99\textwidth]{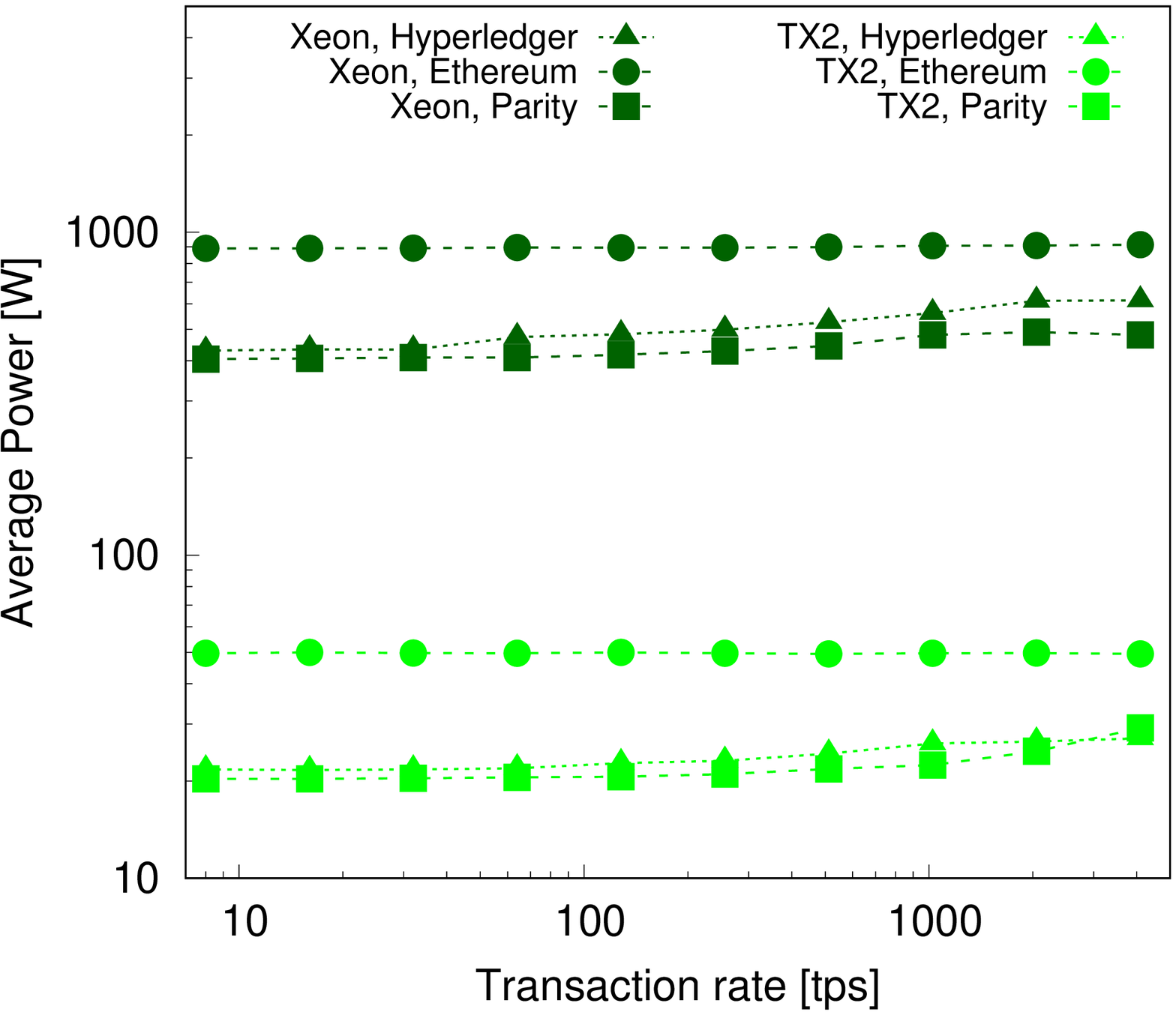}
\caption{Power}
\end{subfigure}
\caption{The performance of Donothing benchmark with increasing transaction
rate}
\label{fig:cluster_donothing_vary_rate}
\end{figure*}

\begin{figure*}[t]
\centering
\begin{subfigure}{\subfigsizeb}
\centering
\includegraphics[width=0.99\textwidth]{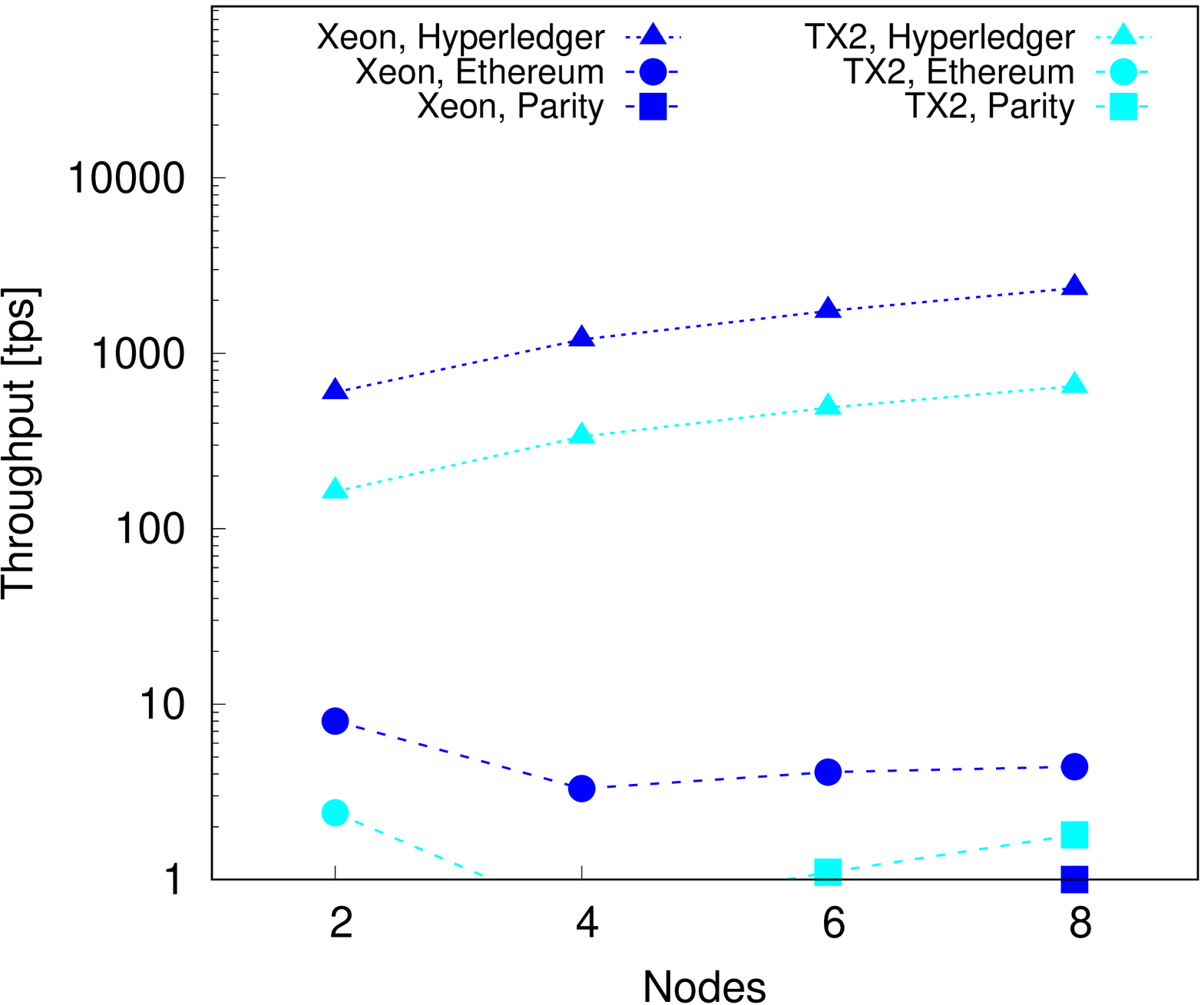}
\caption{Throughput}
\end{subfigure}
\begin{subfigure}{\subfigsizeb}
\centering
\includegraphics[width=0.96\textwidth]{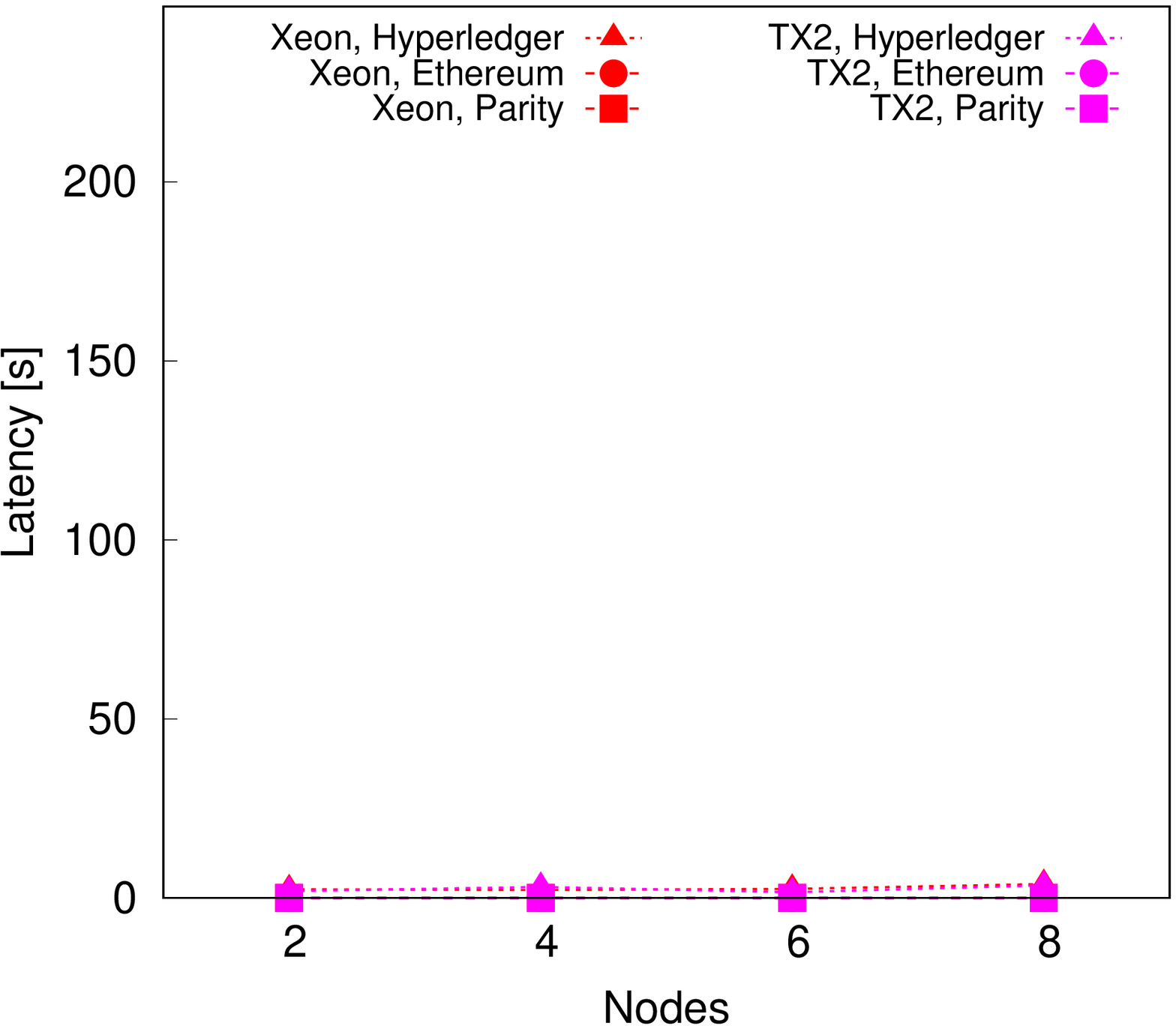}
\caption{Latency}
\end{subfigure}
\begin{subfigure}{\subfigsizeb}
\centering
\includegraphics[width=0.99\textwidth]{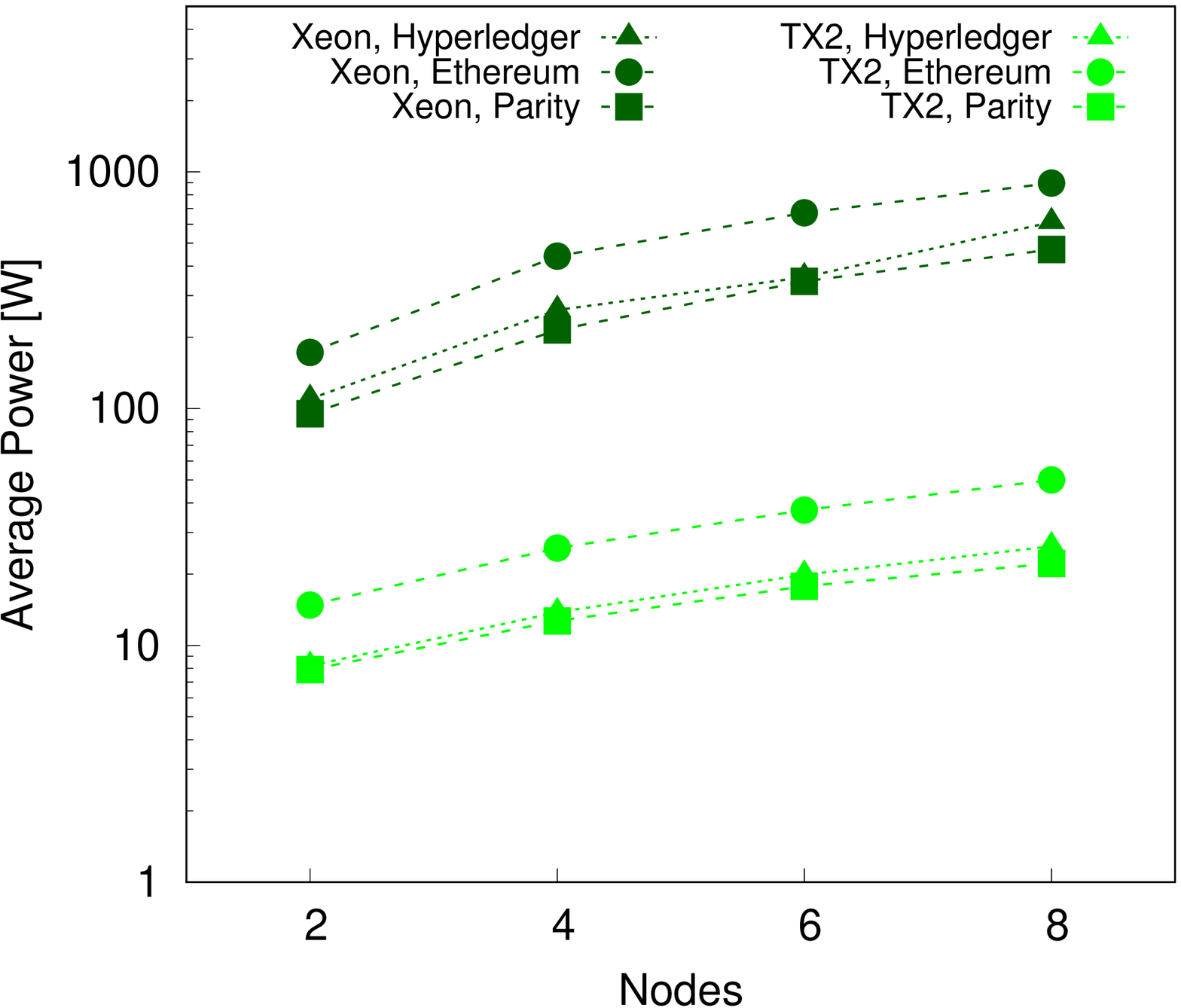}
\caption{Power}
\end{subfigure}
\caption{The performance of Smallbank benchmark with increasing number of nodes}
\label{fig:cluster_smallbank_vary_nodes}
\end{figure*}

\begin{figure*}[t]
\centering
\begin{subfigure}{\subfigsizeb}
\centering
\includegraphics[width=0.99\textwidth]{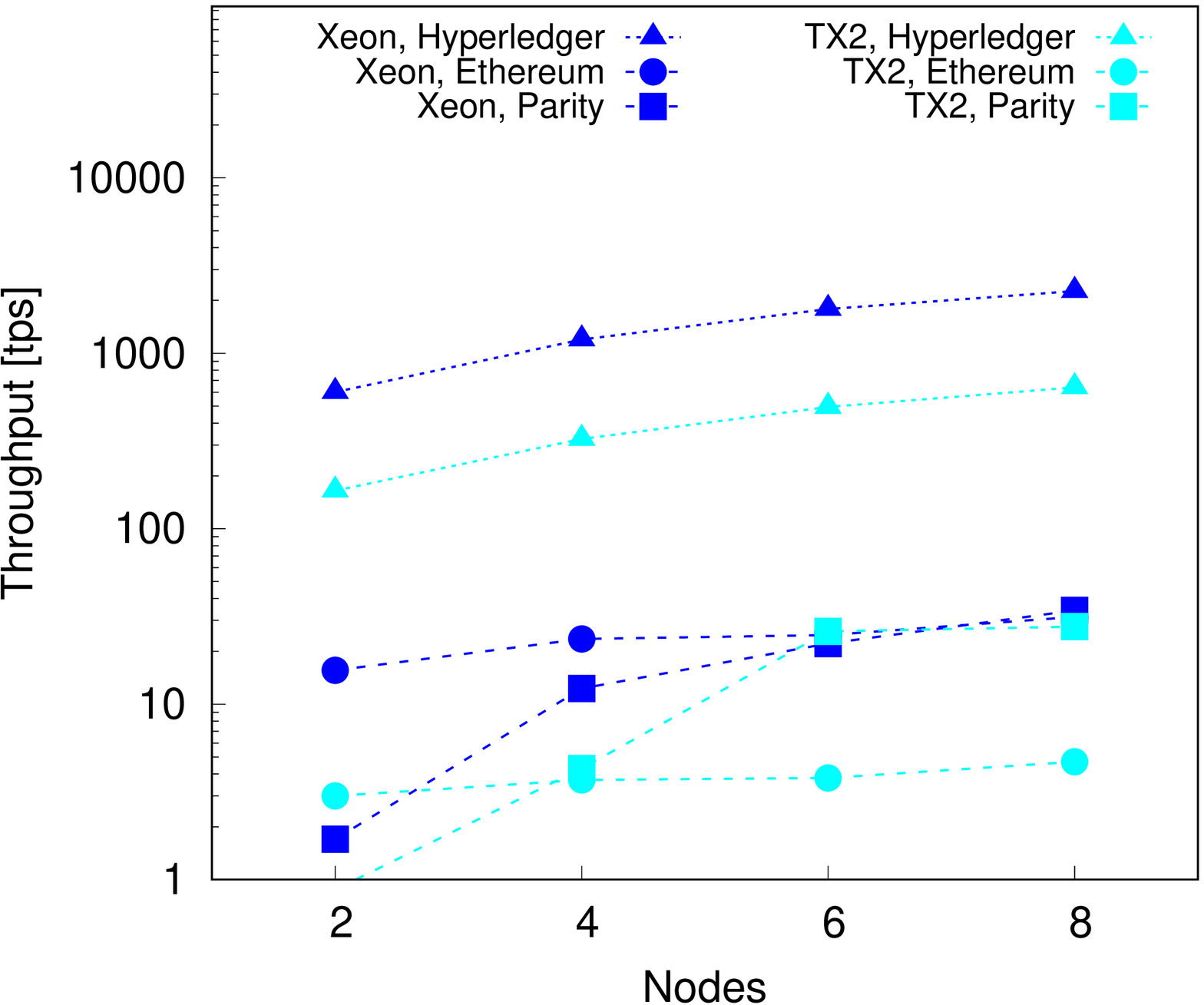}
\caption{Throughput}
\end{subfigure}
\begin{subfigure}{\subfigsizeb}
\centering
\includegraphics[width=0.96\textwidth]{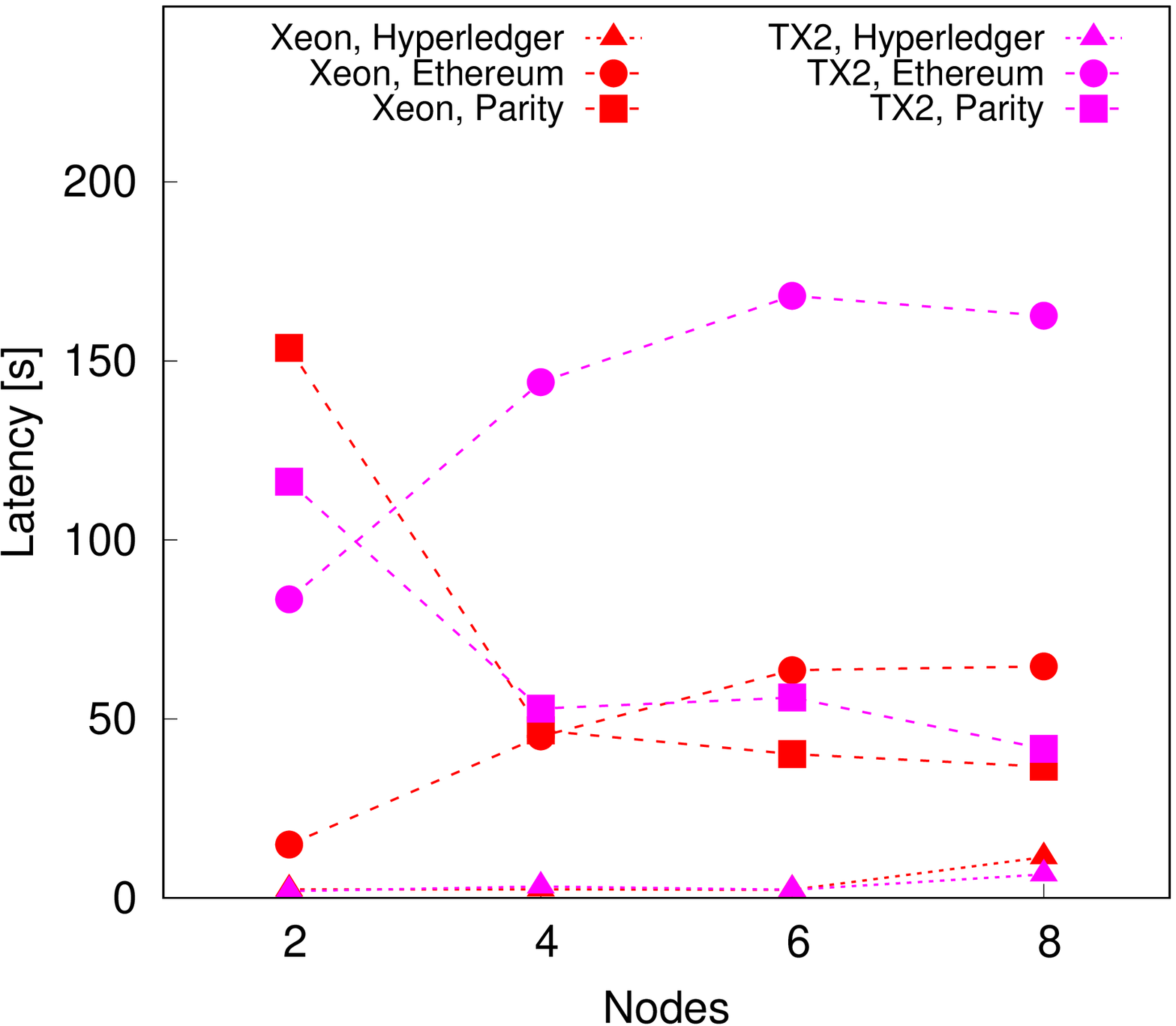}
\caption{Latency}
\end{subfigure}
\begin{subfigure}{\subfigsizeb}
\centering
\includegraphics[width=0.99\textwidth]{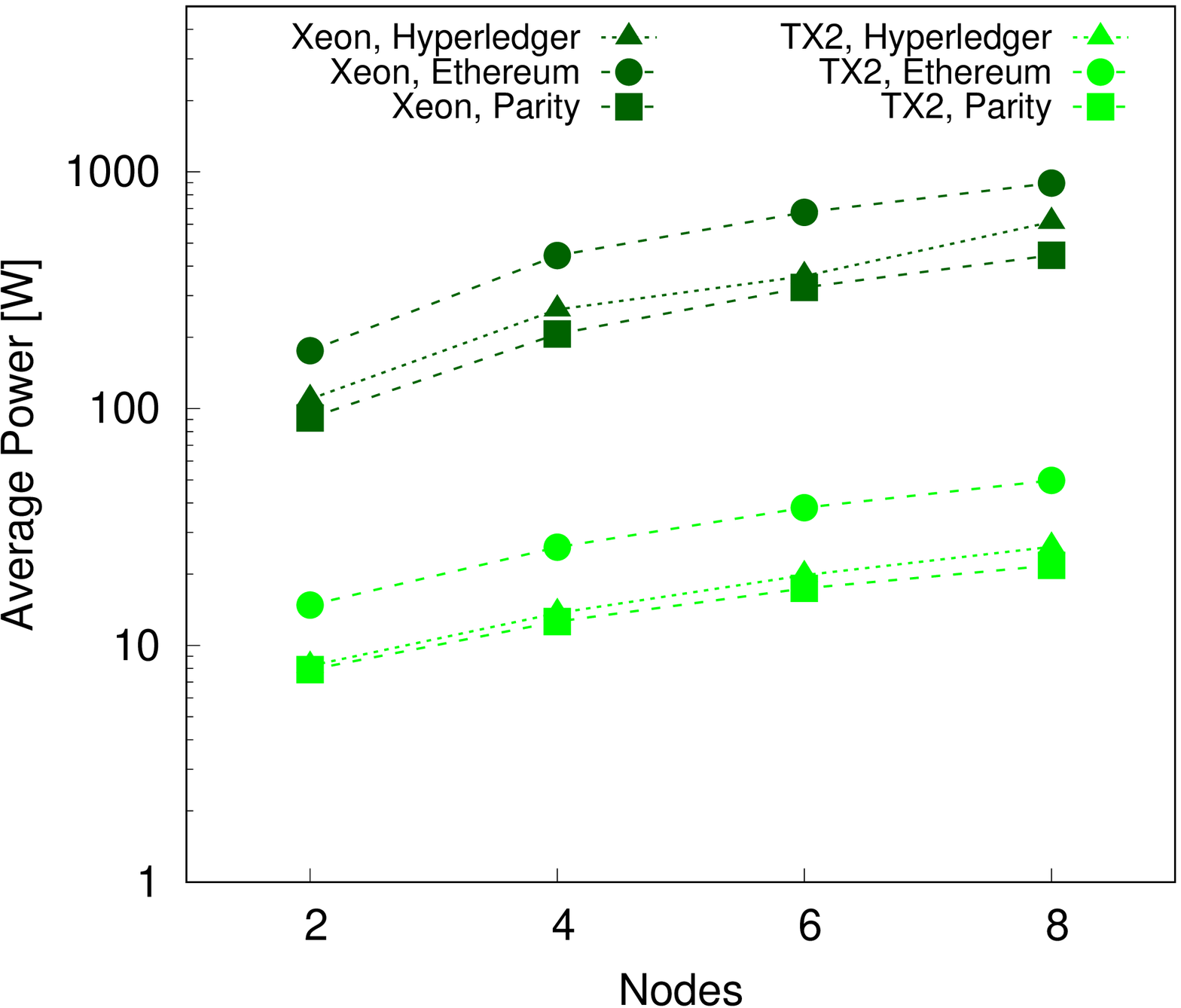}
\caption{Power}
\end{subfigure}
\caption{The performance of Donothing benchmark with increasing number of
nodes}
\label{fig:cluster_donothing_vary_nodes}
\end{figure*}

\begin{figure*}[!t]
\centering
\begin{subfigure}{\subfigsizea}
\centering
\includegraphics[width=0.43\textwidth,angle=270]{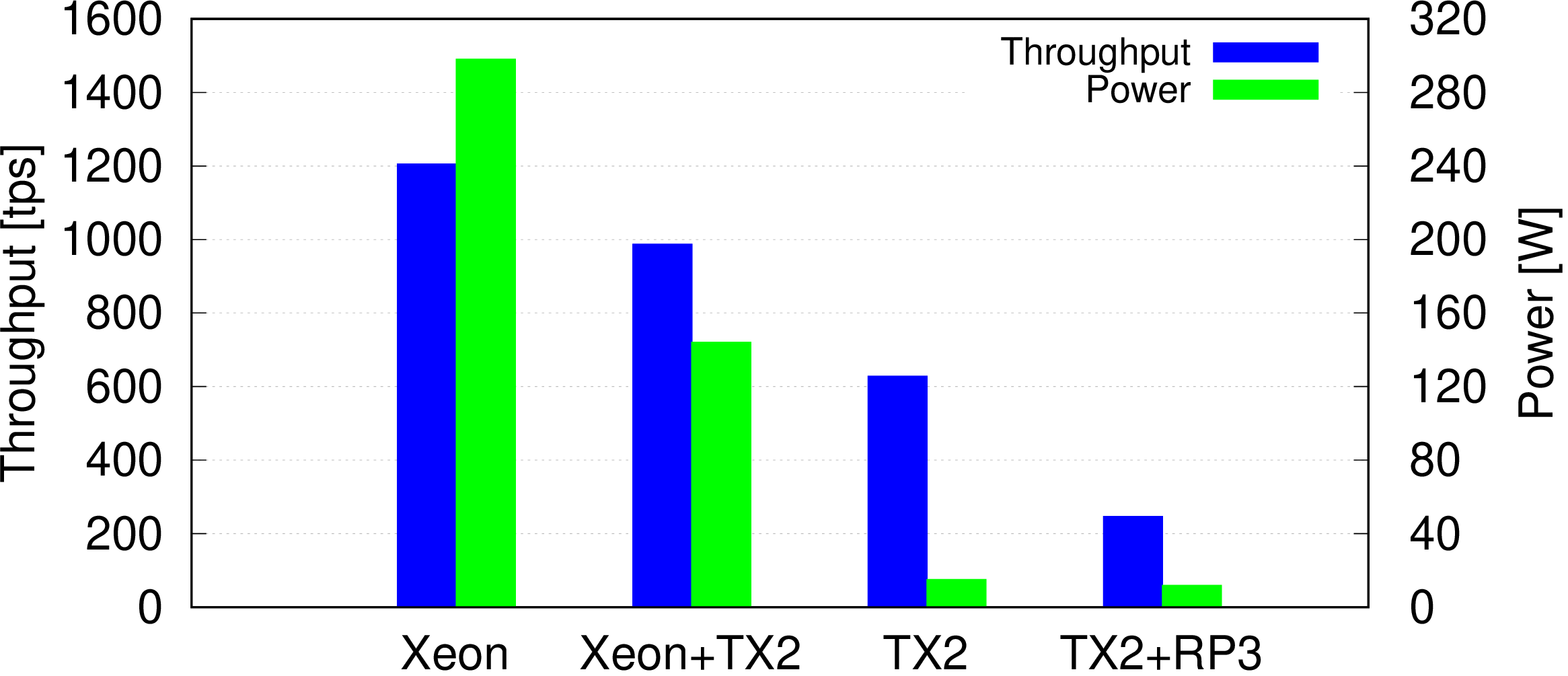}
\caption{Hyperledger}
\end{subfigure}
\begin{subfigure}{\subfigsizea}
\centering
\includegraphics[width=0.43\textwidth,angle=270]{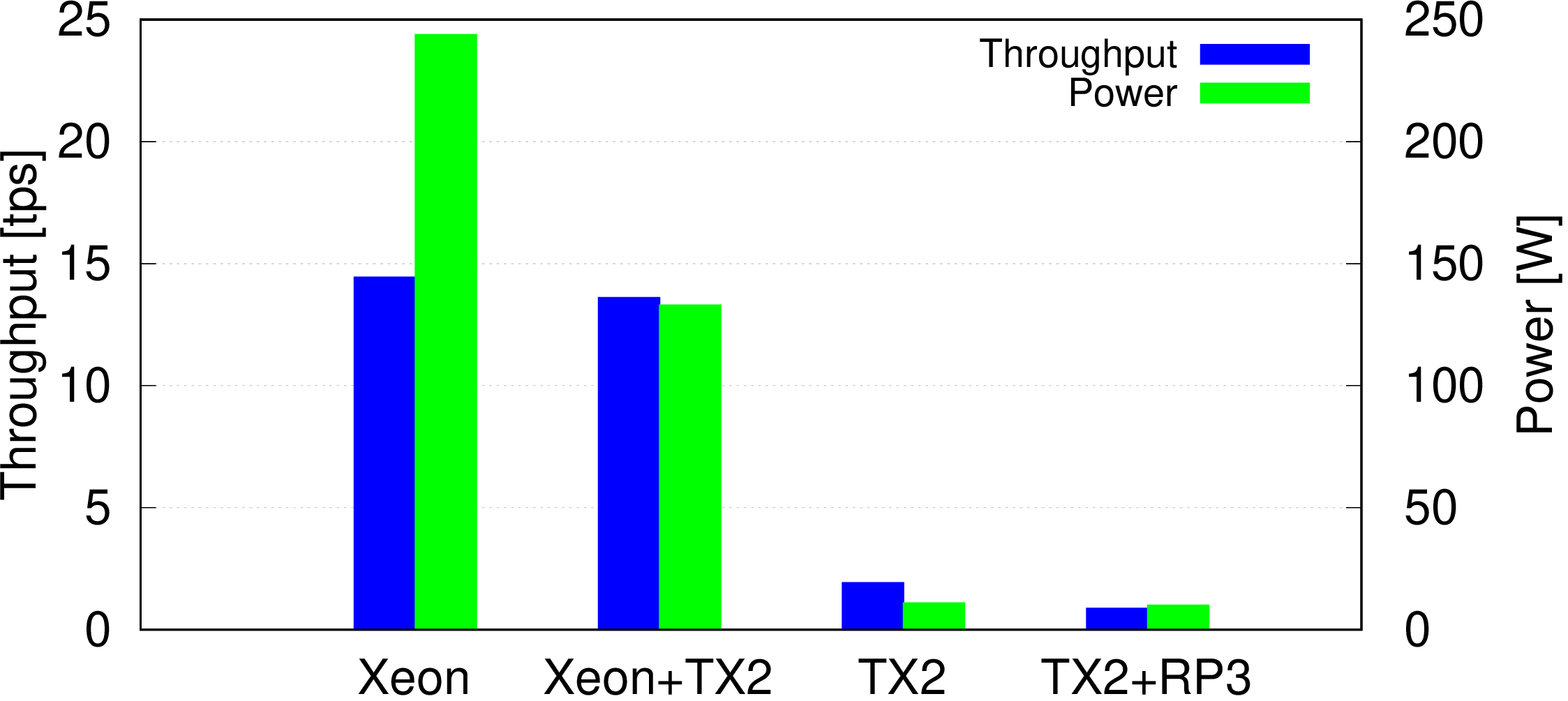}
\caption{Parity}
\end{subfigure}
\caption{The performance of Smallbank on heterogeneous clusters with 4 nodes}
\label{fig:cluster_heterogeneous_smallbank}
\figvspace
\end{figure*}

\begin{figure*}[!t]
\centering
\begin{subfigure}{\subfigsizea}
\centering
\includegraphics[width=0.43\textwidth,angle=270]{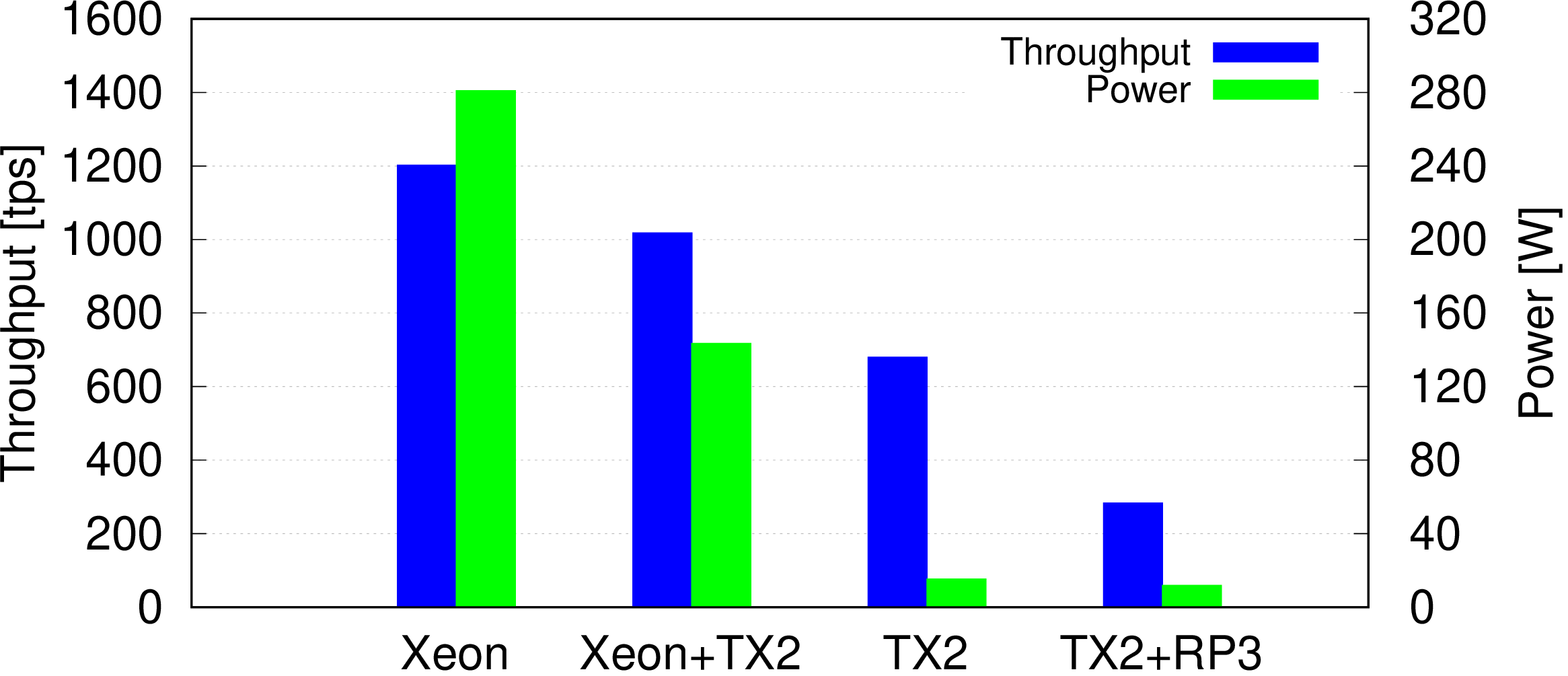}
\caption{Hyperledger}
\end{subfigure}
\begin{subfigure}{\subfigsizea}
\centering
\includegraphics[width=0.43\textwidth,angle=270]{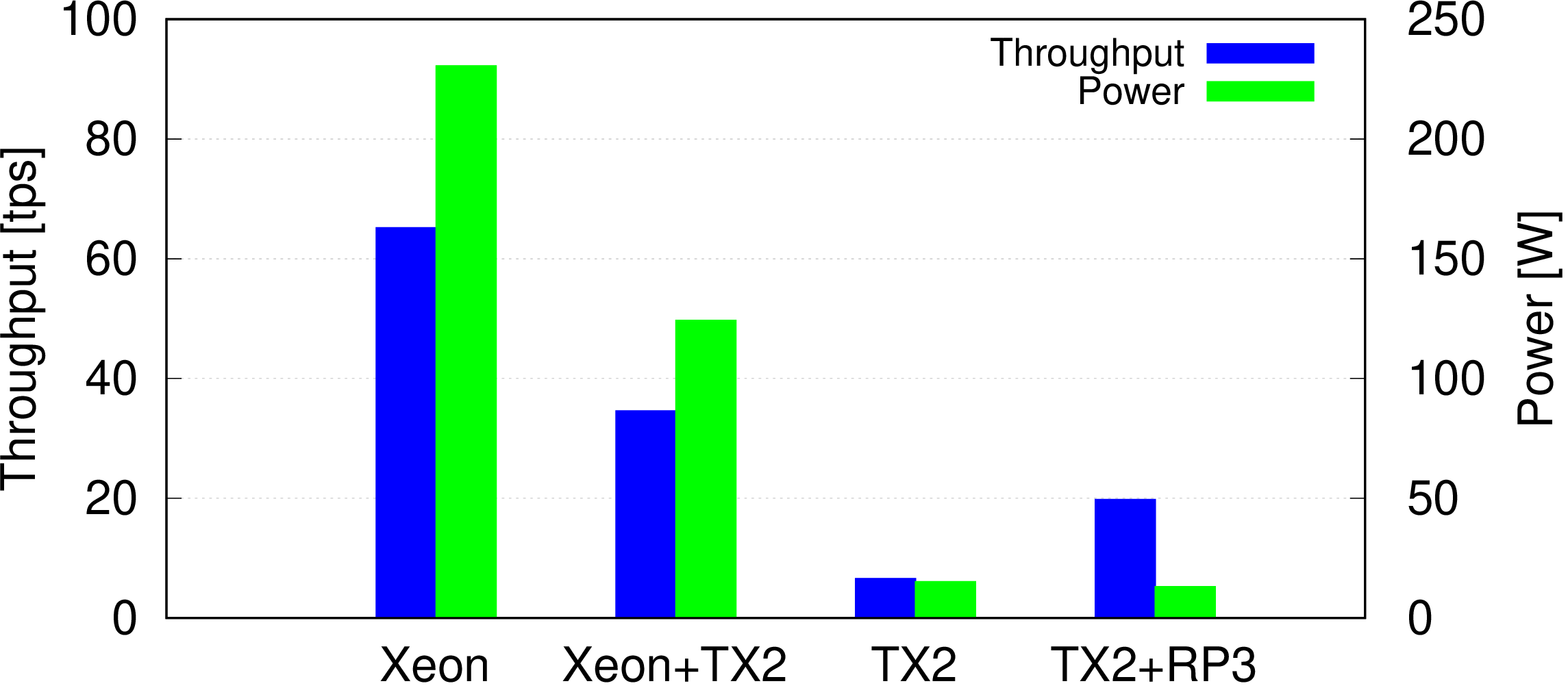}
\caption{Parity}
\end{subfigure}
\caption{The performance of Donothing on heterogeneous clusters with 4 nodes}
\label{fig:cluster_heterogeneous_donothing}
\figvspace
\end{figure*}

\textbf{Cluster Performance.} The throughput, latency and power usage of
Smallbank and Donothing workloads at cluster level are plotted in
Figures~\ref{fig:cluster_smallbank_vary_rate},
\ref{fig:cluster_donothing_vary_rate}, \ref{fig:cluster_smallbank_vary_nodes},
\ref{fig:cluster_donothing_vary_nodes},
\ref{fig:cluster_heterogeneous_smallbank} and
\ref{fig:cluster_heterogeneous_donothing}.
Figures~\ref{fig:cluster_smallbank_vary_rate} and
\ref{fig:cluster_donothing_vary_rate} reflect the performance with varying
transaction request rate for Smallbank and Donothing, respectively.
Figures~\ref{fig:cluster_smallbank_vary_nodes},
\ref{fig:cluster_donothing_vary_nodes} show the performance on increasing number
of nodes. Figures~\ref{fig:cluster_heterogeneous_smallbank} and
\ref{fig:cluster_heterogeneous_donothing} show the performance of Smallbank and
Donothing, respectively, on heterogeneous clusters, as discussed in
Section~\ref{sec:hetero}.

\begin{table}[b] \centering \caption{Comparison of YCSB transaction
(re-)execution count between two versions of Ethereum}
\label{table:eth_cluster_runs_apply}
\resizebox{0.475\textwidth}{!} {
\begin{tabular}{|c|l|r|r|r|r|r|r|}
\hline
\multirow{3}{*}{\textbf{geth}} & \multicolumn{1}{c|}{\textbf{Apply}} &
\multicolumn{6}{c|}{\textbf{Txn Request Rate [tps]}} \\
\cline{3-8} & \multicolumn{1}{c|}{\textbf{Txn}} & & & & & & \\
& \multicolumn{1}{c|}{\textbf{Count}} & 64 & 128 & 256 & 512 & 1024 & 2048 \\
\hline
\hline
\multirow{6}{*}{v1.4.18} & Min & 8 & 1 & 1 & 1 & 1 & 1 \\
& Max & 71 & 59 & 92 & 106 & 105 & 143 \\
& Average & 29.7 & 27.8 & 22.2 & 19.1 & 14.9 & 13.0 \\
& Std.dev. & 9.8 & 7.6 & 14.4 & 14.6 & 14.3 & 11.8 \\
& Unique & 19,531 & 38,832 & 74,535 & 122,500 & 125,093 & 126,455 \\
\cline{2-8} & Total & 580,507 & 1,080,921 & 1,653,870 & 2,342,661 & 1,863,023 &
1,641,687 \\
\hline
\multirow{6}{*}{v1.8.15} & Min & 1 & 1 & 1 & 1 & 1 & 1 \\
& Max & 528 & 385 & 497 & 246 & 183 & 114 \\
& Average & 21.0 & 18.4 & 20.8 & 18.8 & 19.4 & 21.1 \\
& Std.dev. & 17.4 & 14.9 & 16.3 & 14.5 & 14.5 & 18.4 \\
& Unique & 10,935 & 12,289 & 10,713 & 10,957 & 10,826 & 10,725 \\
\cline{2-8} & Total & 229,754 & 226,488 & 222,503 & 206,357 & 209,560 & 226,455
\\
\hline
\end{tabular}
}
\end{table}

\end{appendix}

\end{document}